\documentclass[11pt]{article}

\usepackage{latexsym}
\usepackage{amsfonts}
\usepackage[usenames,dvipsnames]{xcolor}
\usepackage{graphicx,amssymb,amsmath,epsfig}
\usepackage[english]{babel}
\usepackage{graphicx}
\usepackage{dcolumn}
\usepackage{bm}
\usepackage{caption}
\definecolor{myblue}{rgb}{0,0,0.8}
\usepackage[colorlinks=true
,urlcolor=myblue	
,anchorcolor=myblue
,citecolor=myblue
,filecolor=myblue
,linkcolor=myblue
,menucolor=myblue
,pagecolor=myblue
,linktocpage=true  
]{hyperref}

\catcode`\@=11
\def\marginnote#1{}

\newcount\hour
\newcount\minute
\newtoks\amorpm
\hour=\time\divide\hour by60 \minute=\time{\multiply\hour by60
\global\advance\minute by-\hour}\edef\standardtime{{\ifnum\hour<12
\global\amorpm={am}%
        \else\global\amorpm={pm}\advance\hour by-12 \fi
        \ifnum\hour=0 \hour=12 \fi
        \number\hour:\ifnum\minute<10
0\fi\number\minute\the\amorpm}}
\edef\militarytime{\number\hour:\ifnum\minute<10 0\fi\number\minute}

\def\draftlabel#1{{\@bsphack\if@filesw {\let\thepage\relax
   \xdef\@gtempa{\write\@auxout{\string
      \newlabel{#1}{{\@currentlabel}{\thepage}}}}}\@gtempa
   \if@nobreak \ifvmode\nobreak\fi\fi\fi\@esphack}
        \gdef\@eqnlabel{#1}}
\def\@eqnlabel{}
\def\@vacuum{}
\def\draftmarginnote#1{\marginpar{\raggedright\scriptsize\tt#1}}
\def\draft{\oddsidemargin -.5truein
        \def\@oddfoot{\sl preliminary draft \hfil
        \rm\thepage\hfil\sl\today\quad\militarytime}
        \let\@evenfoot\@oddfoot \overfullrule 3pt
        \let\label=\draftlabel
        \let\marginnote=\draftmarginnote

\def\@eqnnum{(\theequation)\rlap{\kern\marginparsep\tt\@eqnlabel}%
\global\let\@eqnlabel\@vacuum}  }

\def\numberbysection{\@addtoreset{equation}{section}
        \def\theequation{\thesection.\arabic{equation}}}

\def\underline#1{\relax\ifmmode\@@underline#1\else
 $\@@underline{\hbox{#1}}$\relax\fi}

\catcode`@=12 \relax

\numberbysection

\topmargin by -\headsep
\def\nonu{\nonumber}
\def\br{\begin{eqnarray}}
\def\er{\end{eqnarray}}

\def\({\left(}
\def\){\right)}
\def\[{\left[}
\def\]{\right]}

\def\a{\alpha}
\def\b{\beta}

\def\bc{{\beta\chi}}

\def\g{\gamma}
\def\G{\Gamma}

\def\l{\lambda}

\def\c{\chi}
\def\p{\phi}

\def\pa{\partial}

\def\s{\sigma}

\def\sech{\mathrm{sech}}

\def\th{\theta}

\def\tp0{\Theta_{+}^{(0)}}
\def\tm0{\Theta_{-}^{(0)}}

\def\wt{\widetilde}
\def\wh{\widehat}
\def\z{\zeta}

\def\vp{\varphi}

\def\cl{{\cal L}}

\def\l{\lambda}
\def\th{\theta}
\def\nonu{\nonumber}

\def\bi{\begin{itemize}}
\def\ei{\end{itemize}}
\def\e{\epsilon}

\def\act{\tilde}

\def\wl{\widetilde{\cal L}}
\def\ww{\widetilde{W}}

\def\gp{\gamma \psi}
\def\bw{{\bf W}}

\begin{document}

\vspace*{1cm}
\noindent

\vskip 1 cm
\begin{center}
{\Large\bf  Extended multi-scalar field theories in (1+1) dimensions}
\end{center}
\normalsize
\vskip 1cm
\begin{center}
{A. R. Aguirre}\footnote{\href{mailto:alexis.roaaguirre@unifei.edu.br}{alexis.roaaguirre@unifei.edu.br}}, and 
{E. S. Souza}\footnote{\href{mailto:edsons_souza@unifei.edu.br}{edsons\_souza@unifei.edu.br}}
\\[.7cm]

\par \vskip .1in \noindent
\emph{Institute of Physics and Chemistry,\\ Federal University of Itajub\'a \!\!,\\
Av. BPS 1303, Itajub\'a - MG, 37500-903, Brazil.}
\vskip 2cm

\end{center}

\begin{abstract}

We present the explicit construction of some multi-scalar field theories in (1+1) dimensions supporting BPS (Bogomol'nyi--Prasad--Sommerfield)  kink solutions. The construction is based on the ideas of the so-called  extension method. In particular, several new interesting two-scalar and three-scalar field theories are explicitly constructed from non-trivial couplings between well-known one-scalar field theories. The BPS solutions of the original one-field systems will be also BPS solutions of the multi-scalar system by construction, and therefore we will analyse their linear stability properties for the constructed models. 
\end{abstract}

\newpage
\tableofcontents

\hspace{-0.73cm}
\rule{\linewidth }{1pt}
\vskip .4in

\section{Introduction}

Recently, there has been a great deal of interest in investigating multi-scalar field theories in \mbox{($1+1$)} dimensions, which support kink solutions obeying first-order BPS equations  \cite{santos13}--\cite{Alonso2019}. Interesting applications have been found mainly in the context  of cosmological models \cite{Moraes14}--\cite{Brito19},  in the study of some aspects of self-duality in generalised BPS theories \cite{Luiz1}--\cite{Klimas19}, and also in connections with several others relevant subjects \cite{livrophi4}--\cite{Adam2018}. Among others topological defects \cite{Rajaraman,Vilenkin}, the study of kink solutions in scalar field theories are of great importance in several areas of the modern theoretical physics \cite{Manton,Shnir}.  They are non-trivial static solutions of the non-linear field equations with finite energy satisfying especially boundary conditions, which usually describe models that exhibit spontaneous symmetry breaking. However, finding out  analytically such kind of solutions for a given field theory is in general a quite difficult task, especially for the case of multi-fields systems.  In some cases thought,  it is possible to make use of some indirect methods to find analytical solutions, e.g  the so-called trial orbit  method \cite{orb1}--\cite{orb4}, which allow us decouple the field equations by introducing very specific orbit equations of the form $\mathcal{O}(\p_1,\dots,\p_n)=0$, that is constraints in the target space.  Although useful, this method has been shown not so efficient when one is looking for new analytical multi-scalar models.

In that scenario, a simplifier tool in searching for kink solutions is provided by the so-called BPS (Bogomolnyi--Prasad--Sommerfield) method \cite{Bogo, PS}, which allows to find solutions from first-order differential (BPS) equations instead of second-order Euler-Lagrange equations. BPS solutions correspond to static configurations of minimal energy. Although simpler, the problem of solving analytically first-order differential coupled equations is still not easy, and then it is necessary to use additional procedures to sort it out. 

In this work, we will use the extension method, originally proposed in \cite{santos13, santos2018}, to systematically construct  several new multi-scalar field theories in ($1+1$) dimensions supporting BPS states, starting from a system of several one-scalar models. The basic ingredients in the construction are the so-called deformation functions and its inverses \cite{Bazeia1}--\cite{Bazeia2005}, which provide suitable links between the fields to be coupled. In addition, the method has the nice advantage that the BPS solutions of the one-field systems are also solutions for the multi-scalar system. 

We aim that the new models constructed in the present work could improve the knowledge and understanding of the analytical solutions of multi-scalar systems, and believe that they have potential applications to cosmological models, to the study of kink scattering process of multi-solitons, and also to analyse integrability and self-duality properties in the multi-scalar-models. For that reason, special attention will be given to the theories with periodic potentials with infinitely degenerate vacua, as is the case of sine-Gordon model, or even more exotic models as the one studied in \cite{Gabriel,susykink}.

This paper is organized as follows. In section \ref{general}, we  briefly review some basics aspects of scalar BPS theories, introducing the superpotential function, or sometimes called the prepotential function \cite{Luiz1}, as the key ingredient of the whole construction. In section \ref{deforming}, we present  the main ideas of the deformation procedure, and then we apply it to obtain several examples of deformed theories. In section \ref{2escalares}, we introduce the extension method and  construct several new interesting two-scalar field theories.  The linear stability of the BPS solutions for these new models will be discussed in section \ref{linear}. In section \ref{3field},  we construct some new three-scalar fields extended models by applying a straightforward generalization of the extension method for three-field systems \cite{santos2018}, and also analyse the linear stability of their BPS solutions. Final remarks and comments of our work are presented in section \ref{remarks}. In appendix \ref{exatos}, we have summarized some basics features of the underlying exactly solvable potentials which appear in the linear stability analysis. Finally, appendix \ref{3campos} contains some explicit calculations on the derivation of the superpotentials for the three-field systems.

\section{General settings}
\label{general}

Let us start considering theories with $n$ real scalar fields $\p_a(x,t)$, $a=1,\dots n$, in $(1+1)$ dimensions described by the following Lagrangian, 
\br
 \cl=\frac{1}{2}\sum_{a=1}^{n}(\pa_\mu\p_a)^2 - V(\p),
\er
where $\mu=\{0,1\}$, with metric convention  $\eta_{\mu\nu}=diag(+1,-1)$, $\pa_\mu \equiv \frac{\pa}{\pa x^{\mu}}$, $x^0 = t, x^1 =x$, in natural units. The corresponding field equations for $\p_a(x,t)$ are given by
\br 
 \pa_t^2\p_a - {\pa_x^2\p_a} +\frac{\pa V}{\pa \p_a}=0, \label{gfe}
\er
and for the static configurations ($\pa_t \p_a=0$), we get
\br
\p_a''(x) = 	\frac{\pa V}{\pa \p_a}, \label{sta1}
\er
where we are using standard conventions $\p'_a\equiv \frac{d}{dx}\p_a$.  These equations can be rewritten as follows,
\br
 \frac{1}{2}(\p'_a)^2 = \int\frac{\pa V}{\pa \p_a}d\phi_a. \label{eqn2.4}
\er
It is worth pointing out that there is no summation assumed in eq. (\ref{eqn2.4}). The corresponding energy functional for the static configurations reads,
\br
 E[\phi] = \int _{-\infty}^{\infty} dx \Big(\frac{1}{2} \sum_{a=1}^{n}(\p'_a)^2 + V(\p) \Big) =2 \int _{-\infty}^{\infty} dx \,V(\p).
\er
Finite energy configurations require existence of the boundary condition $\p_a'(\pm\infty) \to 0$, and a potential possessing at least one vacuum value, $V(\bar{\p})=0$, such that $\p_a(\pm\infty) \to \bar{\p}_a^\pm$. When  two or more minima exist, then the potential supports topological configurations connecting two adjacent minima $\bar{\p}^{-}$ and $\bar{\p}^{+}$.
Now, by introducing a smooth function of the scalar fields $W^{(n)}(\p)$, sometimes named {superpotential} or {pre-potential} \cite{Luiz1}, the potential $V$ can be written as,
\br
 V(\p) =\frac{1}{2}\sum_{a=1}^{n}\(W_{\p_a}^{(n)}\)^2, \label{potentialncampos}
\er
where $W_{\p_a}^{(n)}$ stands for $\frac{\pa W^{(n)}}{\pa{\p_a}}$. Then, the field equations can be rewritten as a set of coupled first-order differential equations,
\br
 \p'_a= \pm W^{(n)}_{\p_a}, \label{primeiraordem}
\er
with energy given by,
\br 
 E =\big| W^{(n)} ({\bar{\p}^{+}}) - W^{(n)} ({\bar{\p}^{-}})\big| = E_{\mbox{\tiny BPS}}.  \label{energiabps}
\er
The solutions of the first-order differential equations (\ref{primeiraordem}) with non-zero energy (\ref{energiabps}) are named BPS states. These minimum energy static configurations are also solutions of the second-order differential equations (\ref{sta1}), which can be understood from the self-duality properties of the BPS theories as claimed in \cite{Luiz1}. In fact, for a given  field theory, the Bogomolnyi bound (\ref{energiabps}) only depends on the boundary conditions, and not on the field configuration, which means that $E_{\mbox{\tiny BPS}}$ is a homotopy invariant, that is invariant under any smooth deformation of the field configurations. These interesting properties makes BPS states so attractive, and it will be the main goal of our work to look for them.

\section{Deforming one-scalar field theories}
\label{deforming}

Recently, it has been proposed an interesting procedure to generate infinite families of one-field theories with topological (kink-like) or non-topological (lump-like) solutions, which is now referred as {deformation procedure} \cite{Bazeia1, Bazeia2}. The main idea is to start from a given ``seed'' one-scalar field theory possessing static solutions, and then perform a field transformation on the target space to obtain a new one-scalar field theory that also supports static solutions. In particular, we will focus in theories supporting BPS solutions.

\noindent Let us  start from a one-scalar field model described by the Lagrangian,
\br
 \cl = \frac{1}{2}\pa_\mu\p \pa^\mu \p - V(\p),  \qquad \mbox{with} \qquad
 V(\p) = \frac{1}{2}\left(W_\p^{(1)}\right)^2,
 \er
which supports BPS solutions satisfying the first-order differential equation,
\br
 \p' &=& W_\p^{(1)} (\p).
\er
Now, we introduce an invertible smooth function $f$ on the target space, called {the deformation function}, such that 
\br 
\p(x) = f(\vp(x)),\label{def1}
\er
where $\vp$ is a new (deformed) scalar field. This function also allows us to introduce a new (deformed) one-field model described by the following Lagrangian,
\br
 \wl = \frac{1}{2}\pa_\mu\vp \pa^\mu \vp - \widetilde{V}(\vp), \qquad \mbox{with}\qquad \widetilde{V}(\vp)=\frac{1}{2}\left(\widetilde{W}_\vp^{(1)}\right)^2,
\er
which satisfies the first-order equation,
\br
 \vp'&=& \ww_\vp^{(1)} (\vp),
\er
providing that the two potentials are related to each other through the deformation function as follows,
\br
\widetilde{V}(\vp) = \frac{1}{f_\vp^2}\, V(\p\to f(\vp)),
\er
where $f_\vp = \frac{df}{d\vp}$. This also implies that the two superpotentials are related in the following form,
\br
 W_\p^{(1)} (\p\to f(\vp) )= W_\p^{(1)}(\vp) =   f_\vp\, \ww_\vp^{(1)}(\vp).
\er
It is worth noting that the static solutions for both scalar fields are related by eq. (\ref{def1}), and then by replacing in the first-order differential equation, we find that they also satisfy the following important constraint,
\br
 \frac{d\p}{d\vp} = \frac{W_\p^{(1)}(\p)}{\ww_\vp^{(1)}(\vp)}. \label{e2.15}
\er
This relation between the fields will play a central role in constructing multi-scalar field theories supporting BPS kink-like solutions. In fact, this relation has been already used in \cite{Dutra1} for studying systems of two coupled fields in (1+1)-dimensions through orbit equation deformations.

Let us now consider a few interesting examples to illustrate the deformation procedure. First of all, we start with the standard $\p^4$ model \cite{Bazeia2005}, whose potential can be written as
\br
 {V}(\p) = \frac{\a^2}{2}  \(1-{\p^2}\)^2, \label{eq2.16}
\er
where $\a>0$,  is a real dimensionless parameter. This potential satisfies the first-order differential equation
\br
 \p'= W_\p^{(1)}(\p) = \a ( 1- \p^2), \label{eq2.17}
\er
and supports the following static solution,
\br
 \p(x) =  \tanh(\a x). \label{eq2.18}
\er
Now, in order to obtain the deformed model, we consider the following function,
\br
 \p= f(\vp) = |\vp|-1. \label{eq2.23}
\er
After using the deformation function, we obtain that the deformed potential 
\br
\wt{V}(\vp)=\frac{\a^2}{2}\vp^2\left(2-|\vp|\right)^2, \label{eq2.24}
\er
describes the so-called  $\vp^6$-like model \cite{Bazeia2004}.  The corresponding  the first-order differential equation is given by,
\br
\vp' = \wt{W}_{\vp}^{(1)}=\a \,\vp\left(2-|\vp|\right),  \label{eq2.25}
\er
with the following topological solutions
\br 
\vp_{\pm}(x)=\pm(1+\tanh(\a x)), \label{eq2.26}
\er
which are quite similar to the solutions of the standard $\p^6$ model \cite{Bazeia2005}. This model possesses three minima at the values $\bar{\vp} = \{0,\pm 1\}$,  and supports two symmetric BPS sectors \cite{Bazeia2004}. Interestingly, this example shows us that the deformation procedure can change the number of vacua of the seed model, and consequently changing the number of topological sectors.

As a second example, let us consider again the $\p^4$ model as the seed model, with superpotential  given  by (\ref{eq2.17}), and introduce the following  periodic deformation
\br 
 \p = f(\chi) = \sin (\b\chi), \label{eq2.27}
\er
where $\chi$ is the new deformed field. The corresponding deformed model describes the sine-Gordon model given by the following first-order equation
\br
 \chi'= \ww_\chi^{(1)}(\chi)= \frac{\a}{\b}\cos(\b\chi).\label{eq2.28}
\er
This very well-known model has infinite degenerate vacua at the values $\bar{\chi}_k=\left(k-\frac{1}{2}\right)\frac{\pi}{\b}$, with $k \in \mathbb{Z}$, and correspondingly an infinite number of equivalent topological sectors.  Connecting the minima $\bar{\chi}_0$ and $\bar{\chi}_1$, its static solution can be written as follows,
\br 
 \chi(x) =\frac{1}{\b}\arcsin(\tanh(\a x)). \label{eq2.29}
\er

As a last example, let us consider the bosonic exotic scalar model (E-model) investigated in \cite{Gabriel,susykink} as our seed model, which is described by the following first-order field equation,
\br
\eta' = W_\eta^{(1)} = \a(1+\eta) \cos\Big(\frac{1}{2}\ln(1+\eta)^2\Big).\label{eq2.30}
\er
This model also has infinitely degenerate trivial vacua at the points $\bar{\eta}_k=-1+e^{\left(k-\frac{1}{2}\right)\pi}$, with $k\in \mathbb{Z}$. However, in this case the infinite number of BPS sector are not equivalent, since the BPS energy depends on the topological sector. A simple kink-like solution for this model connecting the vacua $\bar{\eta}_0$ and $\bar{\eta}_1$, can be written as follows,
\br 
 \eta(x) =  \exp \left( \arctan(\sinh(\a x)) \right) -1.\label{eq2.31}
\er
Now, by considering the following deformation function,
\br
 \eta = f(\chi) = (e^{\b\chi}-1),  \label{eq2.32}
\er
we get the sine-Gordon model, which has been already described  in eq. (\ref{eq2.28}).



\section{Constructing two-scalar fields models}
\label{2escalares}

Let us now describe the method to construct two-scalar field theories from one-scalar field theories. To do that, we will use the deformation method introduced in the last section. The starting point is the first-order equation for the seed one-scalar field model, 
\br
 \p' = W_\p^{(1)}(\p), \label{e4.8}
\er 
which supports static solutions. Now,  by introducing a deformation function, i.e.  $\p = f(\vp)$, we can rewrite eq. (\ref{e4.8}) in two different (but equivalent) ways, namely
\br
\p' = W_\p^{(1)}(\vp), \qquad \mbox{and} \qquad \p'=W_\p^{(1)}(\p,\vp),
\er 
where we have made full use of the function $\p \to f(\vp)$ in the first expression in order to make $W_\p^{(1)}$ a function depending only on $\vp$, while  in  the second expression we have made partial use of this function in order to make $W_\p^{(1)}$ a function depending on both fields $\p$ and $\vp$. Of course, there is an ambiguity in obtaining the last expression since it would depend on how this ``lifting'' from $\p$-space to $(\p,\vp)$-space is made. Then, there will be an infinite number of resulting models once we chose the form of $W_\p^{(1)}(\p,\vp)$. Some of these models would be trivial, and some of them even do not longer support kink-like solutions. However, our main goal here is to construct models that do support BPS solutions, and that will be reach by choosing carefully that form.

Taking into account that the deformed field $\vp$ also satisfied a first-order differential equation,
\br
  \vp'=\ww_\vp^{(1)}(\vp),
\er 
the same procedure can be applied in order to rewrite the equation as follows,
\br
\vp' = \ww _\vp^{(1)}(\p), \qquad \mbox{and} \qquad \vp' = \ww_\vp^{(1)}(\p, \vp),
\er 
 where again there have been both a full as well as a partial use of the deformation (inverse) function, $\vp= f^{-1}(\p)$. Note also that now the eq. (\ref{e2.15}) can be rewritten in several different ways. In fact, in order to proceed with the extension method, we define the new two-fields superpotential through the following ansatz,
\br
\label{modeloextendidoA}
 W_\p^{(2)}(\p,\vp) \!\!&=&\!\! a_1 W_\p^{(1)}(\vp)+a_2 W_\p^{(1)}(\p,\vp)+a_3 W_\p^{(1)}(\p)+p_1\,g(\vp)+p_2\,g(\p,\vp)+p_3\,g(\p),\qquad \,\,\mbox{}\\[0.1cm]
 W_\vp^{(2)}(\p,\vp) \!\!&=&\!\! b_1 \widetilde{W}_\vp^{(1)}(\vp)+b_2 \widetilde{W}_\vp^{(1)}(\p,\vp)+b_3 
\widetilde{W}_\vp^{(1)}(\p)+q_1\,\act{g}(\vp)+q_2\,\act{g}(\p,\vp)+q_3\,\act{g}(\p),\qquad \,\,\mbox{}
\label{modeloextendidoB}
\er
where for consistency, the parameters $a_i$, $b_i$, $p_i$, and $q_i$ with $i=1,2,3$, must satisfy the following constraints 
\br
\sum_{i=1}^{3} a_{i}= \sum_{i=1}^{3} b_{i}  =1,  \qquad \mbox{and}\qquad 
\sum_{i=1}^{3} p_{i}  =\sum_{i=1}^{3} q_{i} =0, 
\er
and $g$ and $\tilde{g}$ are arbitrary functions required for the  consistency conditions, which can be written as follows,
\br
 W_{\p\vp}^{(2)}(\p,\vp) = W_{\vp\p}^{(2)}(\p,\vp).
\label{condicaodecontinuidade}
\er
Thus, by substituting eqs. (\ref{modeloextendidoA}) and (\ref{modeloextendidoB}) in eq.(\ref{condicaodecontinuidade}), we get the following constraint
\br
 0 &=&p_1\,g_\vp(\vp) +p_2\,g_\vp(\p,\vp) -q_2\,\act{g}_\p(\p,\vp) -q_3\,\act{g}_\p(\p)+ a_1\,W_{\p\vp}^{(1)}(\vp)+a_2\,W_{\p\vp}^{(1)}(\p,\vp)\nonu\\ &&- b_2 \,\widetilde{W}_{\vp\p}^{(1)}(\p,\vp)-b_3\,\widetilde{W}_{\vp\p}^{(1)}(\p). \label{eqcontinuidadegeral}
\er
The above constraint allows us to obtain the specific form of the functions $g$ and $\tilde{g}$. After doing that, we can go back to the system given by eqs. (\ref{modeloextendidoA}) and (\ref{modeloextendidoB}), and perform simple integrations to finally determine the form of $W^{(2)}(\p,\vp)$. In what follows, we will illustrate the extension method by explicitly constructing new interesting two-scalar fields models.

\subsection{$\p^4$ model coupled with $\vp^{6}$-like model}

Now, we will consider a model constructed through the coupling of the standard $\p^4$ model and the $\vp^6$-like model \cite{Bazeia2004}. Let us start from eq. (\ref{eq2.17}), namely
\br
\label{superpotencialvp}
W_\p^{(1)}(\p)=\a (1-\p^2),
\er
with the deformation function given in (\ref{eq2.23}), namely,
\br
\label{funcaodevp4tipoz6}
 \p= f(\vp) = |\vp|-1.
\er
The deformed model is the $\vp^6$-like model, whose superpotential satisfies eq.  (\ref{eq2.25}), 
\br
\label{superpotencialdeformadoz6}
\wt{W}_{\vp}^{(1)}(\vp)=\a \,\vp\left(2-|\vp|\right).
\er
Now, we will use the deformation function  to write eq. (\ref{superpotencialvp}) in three equivalent arbitrary forms, where one of these must be a function of only $\p$, another function of $\p$ and $\vp$, and the last function only $\vp$, that is
\begin{subequations}
\br
     W_\p^{(1)}(\vp)&=& \a  |\vp|(2-|\vp|),\\
     W_\p^{(1)}(\p,\vp)&=& \a  \(1+\p(1-|\vp|)\),\\
	 W_\p^{(1)}(\p) &=&\a  (1-\p^2).
\er
\end{subequations}
Similarly, we can use the inverse deformation function to write eq.(\ref{superpotencialdeformadoz6}) in the following three different forms,
\begin{subequations} 
\br
	 \wt{W}_\vp^{(1)}(\vp)&=& \a   \vp(2-|\vp|),\\
     \wt{W}_\vp^{(1)}(\p,\vp)&=& \a  \vp(1-\p),\\	 
	 \wt{W}_\vp^{(1)}(\p) &=& \a  \e (1-\p^2),
\er
\end{subequations}
where the constant parameter $\e = \pm 1$ for solutions $\vp_\pm$  (\ref{eq2.26}), respectively.  Now, by substituting these expressions directly into the constraint (\ref{eqcontinuidadegeral}), we get
\begin{subequations}
\br
\label{gvpz}
g(\p,\vp)&=&-\frac{\a}{p_2}\left(\frac{b_2}{2}\vp^2 +2b_3\p |\vp|\right),\\
\label{actgvpz}
\act{g}(\p,\vp)&=&\frac{\e \a }{q_2}\left(2a_1\p (1-|\vp|)-\frac{a_2}{2}\p^2\right),
\er
\end{subequations}
where we have chosen  $p_1=0$ and $q_3=0$, so $p_3=-p_2$ and $q_1=-q_2$, for simplicity. In addition, we can use the deformation function, and its inverse, to write  
\begin{subequations}
\br
\label{gvp}
p_3 g(\p)&=&\frac{\a  b_2}{2}(1+\p)^2+2\a b_3\p (1+\p),\\
\label{actz}
q_1 \act{g}(\vp)&=&\epsilon \a \left(2a_1+\frac{a_2}{2}\right) (1-|\vp|)^2.
\er
\end{subequations}
By substituting the above results in eqs. (\ref{modeloextendidoA}) and (\ref{modeloextendidoB}), we obtain respectively,
\begin{eqnarray}
\nonumber
W_\p^{(2)}(\p,\vp) &=& \a\left[a_1 |\vp|(2-|\vp|)-\frac{ b_2}{2}\vp^2+ a_2 \left(1+\p (1-|\vp|)\right)-2 b_3\p|\vp|+a_3  (1-\p^2)\right. \\
\label{derivadasuperpotencialvpparavpz}
&&\left. \quad +\frac{ b_2}{2}(1+\vp)^2+2 b_3\vp(1+\vp)\right], \\
\nonumber
W_\vp^{(2)}(\p ,\vp) &=& \a\left[ b_1 \vp (2-|\vp|)+ \epsilon \left(2a_1+\frac{a_2}{2}\right) (1-|\vp|)^2+  b_2 \vp (1-\p)+2a_1 \epsilon \p (1-|\vp|)\right.\\
\label{derivadasuperpotencialzparavpz}
&&\left. \quad -\epsilon \left(b_3+\frac{a_2}{2}\right)\p^2+\epsilon b_3\right],
\end{eqnarray}
which upon being integrated results in the following superpotential, 
\br 
\nonumber
W^{(2)}(\p,\vp) &=&\a\left[ \frac{1}{2}\left(a-b+c+1\right)|\vp| -\frac{1}{2}\left(a+c-1\right)\vp^2 +\frac{1}{6}\left(a+b+c-1\right){|\vp|\vp^2}+  a \p|\vp|\right. \\
\nonumber
&&\quad -\frac{ 1}{2}\(a+b\)\p\vp^2 +\frac{1}{2}\left(a+b-c-1\right)\p^2|\vp|+ \frac{1}{2}(2-a+b){\p}+\frac{1}{2}(1+c-a){\p^2}\\
&&\left.\quad  -\frac{1}{6}\left(a+b-2c\right)\p^3-\frac{1}{6}\left(a-2b+c-1\right)\right],\label{Wphi4zeta6}
\er
where we have just renamed the parameters:  $a\equiv 2a_1$, $b\equiv b_2$, and $c\equiv 2a_1+a_2-b_2-2b_1+1$. This superpotential describes the coupling between the $\p^4$ model and the $\vp^6$-like model, and therefore from now on we will name it as the {extended $(\p^4+\vp^6_l)$ model.} 
Note that there are several models that can be considered depending on the choice of these parameters. 
This model contains three minima at the following values: $m_1=(-1,0)$, $m_2=(1,2)$, and $m_3=(1,-2)$. It supports then three topological sectors, with only two of them are BPS, they are: the sector connecting $m_1$ and $m_2$, and the one connecting $m_1$ to $m_3$,  with the explicit solutions given by  eq.(\ref{eq2.18}) and (\ref{eq2.26}), namely
\br
 \p(x) = \tanh(\a x), \qquad \vp_\pm(x) = \pm(1+\tanh(\a x)), \label{eqbpsphi4zeta6}
\er
which energy is $E_{\mbox{\tiny BPS}}=8\a/3$. On the other hand, the non-BPS configuration, connecting the minima $m_2$ and $m_3$, do not satisfy the first-order equation. In this case, we can write an explicit solution for the specific values $a =1 $,  $b=-1$, and $c=0$,
\br
\p =1, \qquad \vp_\pm(x) = \pm 2\tanh(\a x),
\er
with energy $E = 16\a/3$,  which is twice the energy of the BPS sectors, as it was already expected. It is worth noting that the corresponding anti-kink configurations,
\br
 \p(x) = - \tanh(\a x), \qquad \vp^{(-)}_\pm(x) = \pm (1- \tanh(\a x)),
\er
are also in the BPS sectors connecting  $m_2$ and $m_3$ minima to $m_1$, respectively. There are several others topological sectors that appear after chosing the values of  the parameters. For instance, for $a=b=0$ and $c=-1$, we recovery the BPS sector associated with the $\p^4$ model, connecting the two minima $(\pm 1,0)$, with energy $E_{\mbox{\tiny BPS}}=4\a/3$. On the other hand, we can also verify that the trivial configuration $\p =0$ does not belong to the minima space of this potential. Finally, we would like to pointing out that for the values $a=c=1$, and  $b=0$, the superpotential $W^{(2)}(\p,\vp)$ becomes harmonic, and consequently all the solution will be BPS solutions \cite{Menezes1, Menezes2}.

\subsection{$\p^4$ model coupled to sine-Gordon model}

Our starting point will be again the  $\p^4$ model. Now, by using the deformation function (\ref{eq2.27}) we will write the right-hand side of eq. (\ref{eq2.17}) in  the following forms,
\begin{subequations}
\br
	 W_\p^{(1)}(\p)&=& \a(1-\p^2),\\
	 W_\p^{(1)}(\chi)&=& \a \cos^2(\b \chi),\\
	 W_\p^{(1)}(\p,\chi) &=& \a\(1-\p\sin(\b\chi)\),
\er
\end{subequations}
and similarly eq. (\ref{eq2.28}) as
\begin{subequations} 
\br
	 \widetilde{W}_\chi^{(1)}(\chi)&=& \frac{\a}{\b}\cos(\b\chi),\\
	 \widetilde{W}_\chi^{(1)}(\p)&=& \frac{\a}{\b} \sqrt{1-\p^2},\\
	 \widetilde{W}_\chi^{(1)}(\p,\chi) &=& \frac{\a}{\b}\sqrt{1-\p\sin(\b\chi)}.
\er
\end{subequations}
We see that the last two forms for $\widetilde{W}_\chi^{(1)}$ contain square root functionals, and then it is convenient to consider the following choice of parameters $b_2=b_3=0$. Also, without loss generality we choose $p_1=p_2=q_3=0$.  This implies immediately that $b_1=1$, $p_3=0$, and  $q_2=-q_1$. Then, by solving the constraint (\ref{eqcontinuidadegeral}) to determine the $\tilde{g}$-function, we get
\br
 \tilde{g}(\p,\chi) = -\Big(\frac{\a\b}{q_2} \Big)\left(2a_1 \p\sin(\b\chi) + \frac{a_2}{2}\p^2\right) \cos(\b\chi),
\er
which can be rewritten, by using the deformation function, as follows
\br
\tilde{g}(\chi) =  \Big(\frac{\a\b}{q_1} \Big) \Big(2a_1+{a_2 \over 2}\Big) \sin^2(\b\chi) \cos(\b\chi),
\er
By substituting the above results in eqs. (\ref{modeloextendidoA}) and (\ref{modeloextendidoB}), we get
\br
W^{(2)}_\p(\p,\chi) &=& \a\left[a_1 \cos^2(\b\chi) +a_2\(1-\p\sin(\b\chi)\)+(1-a_1-a_2)\(1-\p^2\) \right],\\
W^{(2)}_\chi(\p,\chi) &=& \a\left[\frac{1}{\b}\cos(\b\chi)  +{\b} \Big(2a_1+{a_2 \over 2}\Big) \sin^2(\b\chi) \cos(\b\chi) - a_1 \b \p\sin(2\b\chi)  \right.\nonumber \\
&& \left. \quad - \frac{a_2\b}{2}\p^2 \cos(\b\chi)\right].
\er
Integrating out these expressions we finally obtain the two-fields superpotential, which is given by the following form
\br
 W^{(2)}(\p,\chi) 
 &=& \a\left[\p-(1-a_1-a_2)\frac{\p^3}{3}- a_1 \p \sin^2(\b\chi)-\frac{ a_2}{2} \p^2\sin(\b\chi) +\frac{1}{\b^2}\sin(\b\chi) \right.\nonumber \\&& \left. \quad + \frac{1}{3}\Big(2a_1+\frac{a_2}{2}\Big)\sin^3(\b\chi) \right].
\label{potencialfsin}
\er
This superpotential describes the coupling of the  $\p^4$ and  sine-Gordon models, which from now on will be named as the {extended} $(\p^4$+sG) {model}. The static kink-like solutions,
\br
 \p(x) = \tanh(\a x), \qquad  \chi(x) =\frac{1}{\b} \arcsin\(\tanh(\a x)\), \label{eqs3.56}
\er
\begin{figure}[t!]
\begin{center}
 \includegraphics[scale=0.4]{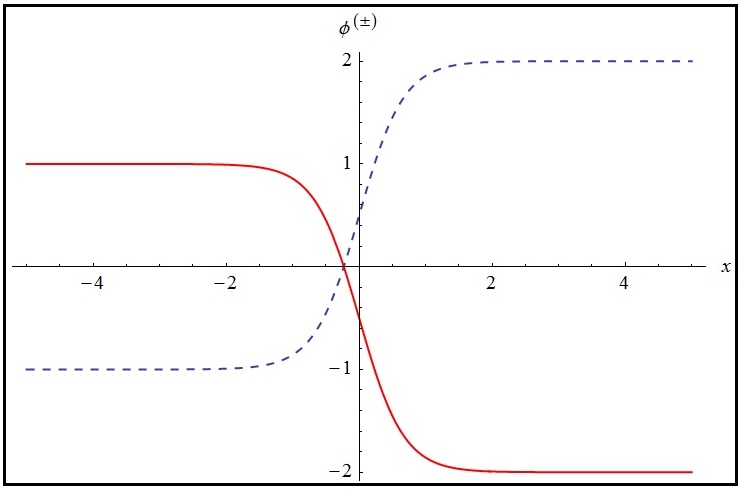}
 \caption{The solid (red) line is the plot of the $\phi^{(+)}$ solution for the parameter values $a_1=3$ and $a_2=-1$. The dashed (blue) line is the plot of the $\phi^{(-)}$ solution for the parameter values $a_1=-1$ and $a_2=1$, with $\a=1$.}\label{outsol}
\end{center}
\end{figure}
are BPS solutions of this model connecting the minima $m_1 =(-1,-\frac{\pi}{2\b})$ and $m_2=(1,\frac{\pi}{2\b})$, with BPS energy given by,
\br
E_{\mbox{\tiny{BPS}}} =  2\a\Big(\frac{2}{3}+\frac{1}{\b^2}\Big).
\er
In general, we can verify that this potential possesses minima at the points\mbox{ $(-1, (2k-\frac{1}{2})\frac{\pi}{\b})$}, and \mbox{ $(+1, (2k+\frac{1}{2})\frac{\pi}{\b})$}, with $k\in \mathbb{Z}$. It is also worth highlighting the existence of other BPS solutions. For instance, a particular solution is
\br 
 \p^{(-)}(x) = \frac{(1-a_1) - e^{(-2(1-a_1)+a_2)\a x} }{((1-a_1)-a_2) + e^{(-2(1-a_1)+a_2)\a x}},  \qquad \chi^{(-)}(x) = \frac{\pi}{\b}\Big(2k -\frac{1}{2}\Big), \label{sol3.58}
\er
providing that the parameters $a_1$ and $a_2$ are restricted to satisfy $a_2< (1-a_1)$, and \mbox{$a_2\neq 2(1-a_1)$}, otherwise $\phi^{(-)}(x)$ becomes an exponential or a constant solution,  respectively. By chosing $k=0$, we notice that this solution connects the minimum $m_1$ to a new minimum \mbox{$m_3=(\frac{1-a_1}{1-a_1-a_2},-\frac{\pi}{2\b})$}. We see that the BPS energy of this solution is given by,
\br 
 E_{\mbox{\tiny{BPS}}}= \frac{\a}{6}\frac{\big|2(1-a_1)-a_2\big|^3}{\(1-a_1-a_2\)^2}. \label{eq3.59}
\er
Another possible solution is,
\br
 \phi^{(+)}(x) = \frac{(1-a_1) + e^{ (2(1-a_1)-a_2)\a x} }{(a_2 -(1-a_1)) + e^{ (2(1-a_1)-a_2)\a x} }, \qquad \chi^{(+)}(x) = \frac{\pi}{\b}\Big(2k +\frac{1}{2}\Big). \label{sol3.60}
\er
In this case $a_2>(1-a_1)$, and again $a_2\neq 2(1-a_1)$. For $k=0$, this solution connects the minimum $m_2$ to a new minimum $m_4 = (-\frac{1-a_1}{1-a_1-a_2}, +\frac{\pi}{2\b})$. We also note that the solution (\ref{sol3.60}) possesses the same BPS energy as the solution (\ref{sol3.58}), given by eq.  (\ref{eq3.59}). We have plotted $\phi^{(\pm)}(x)$ in figure \ref{outsol}.


\subsection{E-model coupled to sine-Gordon model}

Let us consider now the sine-Gordon model and the E-model described by the first-order field equations (\ref{eq2.28}) and  (\ref{eq2.30}), respectively, together with the deformation function (\ref{eq2.32}). Then, we write the following expressions,
\br
 W_\eta^{(1)}(\eta) &=&\a (1+\eta )\cos\Big( \frac{1}{2} \ln (1+\eta)^2 \Big),\\
 W_\eta^{(1)}(\chi) &=& \a\, e^{\b \chi} \cos(\b \chi),\\
 W_\eta^{(1)}(\eta,\chi) &=& \a (1+\eta) \cos(\b\chi),
\er
and
\br 
 \wt{W}_\chi ^{(1)}(	\chi) &=& \frac{\a}{\b}\cos(\b \chi),\\
 \wt{W}_\chi^{(1)} (	\eta) &=& \frac{\a}{\b}\cos\Big( \frac{1}{2} \ln (1+\eta)^2 \Big),\\
  \wt{W}_\chi^{(1)} (	\eta,\chi) &=& \frac{2\a}{\b}\left[\cos\Big( \frac{\b \chi}{2} \Big) \cos\Big( \frac{1}{4} \ln (1+\eta)^2 \Big) - \frac{1}{2}\right].
\er
By choosing $p_1=q_3=0$ in the constraint (\ref{eqcontinuidadegeral}), we get
\br
p_2 g_\chi ( \eta,\chi)-q_2\act{g}_\eta ( \eta,\chi) &=&-\a\b a_1 e^{\bc}\big(\cos(\bc)-\sin(\bc)\big)+\a\b a_2(1+\eta)\sin(\bc)\qquad\mbox{}\nonu\\
&&-\frac{\a}{\b}\frac{b_2}{(1+\eta)}\cos\Big(\frac{\bc}{2}\Big)\sin\Big(\frac{1}{4} \ln (1+\eta)^2\Big)\nonu\\&&-\frac{\a}{\b}\frac{b_3}{(1+\eta)}\sin\Big(\frac{1}{2} \ln (1+\eta)^2\Big).\qquad\mbox{}
\er
After integrating, we find the following solution, 
\br
g(\eta,\chi) &=& -\a\frac{a_1}{p_2}e^{\bc}\cos(\bc)-\frac{\a}{\b}\frac{b_3}{p_2}\frac{\chi}{(1+\eta)}\sin\Big(\frac{1}{2} \ln (1+\eta)^2\Big),\\[0.1cm]
\act{g}(\eta,\chi) &=&-\frac{2\a}{\b}\frac{b_2}{q_2}\cos\Big(\frac{\bc}{2}\Big)\cos\Big(\frac{1}{4} \ln (1+\eta)^2\Big)-\a\b\frac{a_2}{q_2}\left(\eta+\frac{\eta^2}{2}\right)\sin(\bc).
\er
As it was done before, we can use the deformation function to write
\br
g(\eta) &=& -\frac{\a a_1}{p_2}(1+\eta)\cos\Big(\frac{1}{2} \ln (1+\eta)^2\Big)-\frac{\a b_3}{\b^2p_2}\frac{\ln (1+\eta)}{(1+\eta)}\sin\Big(\frac{1}{2} \ln (1+\eta)^2\Big),\\[0.1cm]
\act{g}(\chi) &=& -\frac{2\a b_2}{\b q_2}\cos^2\Big(\frac{\bc}{2}\Big)-\frac{\a\b a_2}{2q_2}\left(e^{2\bc}-1\right)\sin(\bc).
\er
Using these results, we obtain 
\br
\nonumber
 W^{(2)}_\eta(\eta,\chi) &=&\a a_2(1+\eta)\cos(\bc)+\a(1-a_2)(1+\eta)\cos\Big(\frac{1}{2} \ln (1+\eta)^2\Big)\\
 &&+\frac{\a b_3}{\b^2}\left(\frac{\ln(1+\eta)-\bc}{(1+\eta)}\right)\sin\Big(\frac{1}{2} \ln (1+\eta)^2\Big),\\[0.1cm]
 W^{(2)}_\chi(\eta,\chi) &=& \frac{\a}{\b}(1-b_3)\cos(\bc) +\frac{\a b_3}{\b}\cos\Big(\frac{1}{2} \ln (1+\eta)^2\Big)\nonu\\
 &&+\frac{\a\b a_2}{2}\left(e^{2\bc} - (1+\eta)^2\right)\sin(\bc).
\er
We then can construct the corresponding two-fields superpotential,
\br
 W^{(2)}(\eta,\chi) &=& \frac{\a}{5}(1-a_2)(1+\eta)^2\left[2\cos\Big(\frac{1}{2} \ln (1+\eta)^2\Big)+\sin\Big(\frac{1}{2} \ln (1+\eta)^2\Big)\right]\nonu\\&&+\frac{\a b_3}{\b^2}\left[\big(\bc-\ln (1+\eta)\big)\cos\Big(\frac{1}{2} \ln (1+\eta)^2\Big)+\sin\Big(\frac{1}{2} \ln (1+\eta)^2\Big)\right] \nonu \\ && + \a a_2\left(\eta+\frac{\eta^2}{2}\right)\cos(\bc) +\frac{\a a_2}{10}\left[\left(5-e^{2\bc}\right)\cos(\bc)+2e^{2\bc}\sin(\bc)\right]\nonu\\
 &&+\frac{\a}{\b^2}(1-b_3)\sin(\bc).\label{ESGpot}
\er
This two-parameters superpotential leads us to a potential $V(\eta,\chi)$ describing the coupling of the sine-Gordon model and the E-model, which from now on we will named as the {extended (sG+E) model}. This superpotential supports the static kink-like solutions (\ref{eq2.29}) and (\ref{eq2.31}), 
\br
 \eta(x) = \Big[ \exp \left( \arctan(\sinh(\a x)) \right) -1\Big], \qquad  \chi(x) =\frac{1}{\b} \arctan\(\sinh(\a x)\),\label{eqs3.78}
\er
connecting the minima $m_1=\big(e^{-\pi/2}-1, -\pi/2\b\big)$ and $m_2=\big(e^{\pi/2}-1, \pi/{2\b}\big)$, with BPS energy given by
\br
E_{\mbox{\tiny{BPS}}} =  \frac{2\a}{5 \b^2}\left(5+\b^2\cosh\pi\right).
\er
We notice that for the particular values of the parameters $a_2=0$ and $b_3=1$, it is possible to obtain other BPS solutions, at least numerically. In this case, we have that $\eta(x)=-1$, and $\chi(x)$ has to satisfy
\br 
\chi'(x) = \frac{\a}{2\b}\big(2\cos(\b\chi) + \b^2 e^{2\b\chi }\sin(\b\chi)\big). \label{kinkesg2}
\er
\begin{figure}[h]
\begin{center}
\includegraphics[scale=0.3]{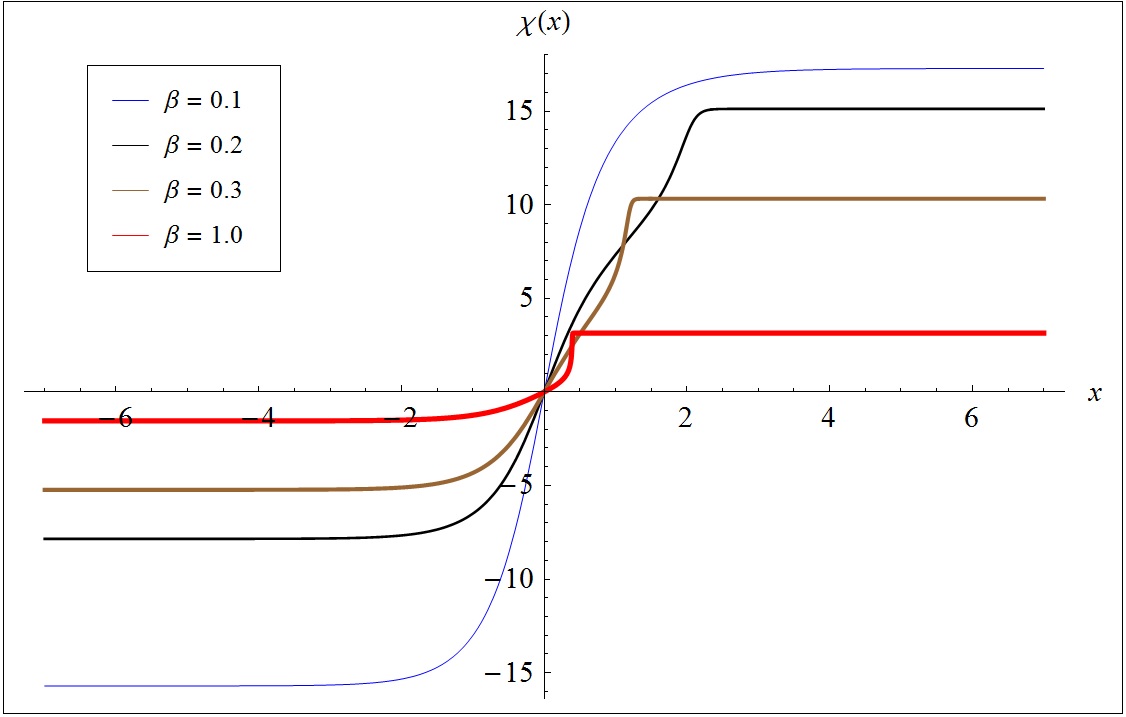}
\caption{Plot of the kink-like numerical solution of the eq. (\ref{kinkesg2}) for $\a=2$, and different values of $\b$. The thickness  of the curves increases as $\b$ also increases. This solutions interpolates between the minima $\bar{\chi}_{0}$ and $\bar{\chi}_{+1}$.}\label{kinkesg}
\end{center}
\end{figure}
\noindent It is interesting to see that the associated potential  $V(-1,\chi)$ represents a modification of the sine-Gordon model. In fact, we have verified that it has infinite minima, and supports BPS solutions. The minima are located approximately at the following points,
\br 
 \bar{\chi}_{k} \approx \left\{ \begin{array}{l c l} \frac{1}{2\b}\left[(2k-1)\pi+\frac{\b^3}{e^{(1-2k)\pi}-\b^3} \]& , &  k \leq 0, \\[0.2cm] \frac{\pi k}{\b} - \frac{2}{\b^3} e^{-2 k \pi} &,& k>0.\end{array}  \right.
\er
Then, there are at least three type of topological sectors, a small one for $k< 0$, a medium one for $k=0$, and the large one for $k>0$. The  corresponding BPS energies are given by 
\br 
E_{\mbox{\tiny BPS}}^{(k, k+1)}&\approx & \left\{\begin{array}{l l}
  \frac{2\a}{5\b^2} \(5 + \b^2 \, e^{2k\pi} \cosh(\pi)\) +\frac{\a \b^3}{2}\left[\frac{\cosh (2 \pi )-\b^3\,e^{2 \pi  k} \cosh (\pi ) }{\b^6-2 \b^3 \,e^{-2 \pi  k}\cosh (\pi ) +e^{-4 \pi  k}}\right], &  k<0, \\[0.3cm] \frac{\a}{20}\left[\(2e^{2\pi}-e^{-\pi}\) +\frac{20}{\b^4} \(\b^2+2e^{-2\pi}\)+ \frac{5}{e^\pi - \b^3}\right], & k=0, \\[0.3cm]   \frac{\a \cosh(\pi)}{5\b^4}\left[20\, e^{-(2k+1)\pi} +\b^4 \,e^{(2k+1)\pi} \right],   & k>0.
 \end{array} 
  \right.\qquad\quad \mbox{}
\er
Unfortunately, we have not  been  able to obtain the corresponding analytical solutions of the first-order equation (\ref{kinkesg2}) for the kink solutions associated to each topological sector. However, we did construct them numerically\footnote{Specifically, we have used the NDSolve package of the Wolfram Mathematica Software, and chosen as initial condition $\chi(0)=0$ for solving the differential equation.}. For instance,  we have plotted in figure \ref{kinkesg} the numerical kink solutions connecting the minima ${\bar\chi}_{0}$ 
to $\bar{\chi}_{+1}$, 
for several values of the parameter $\b$. We can see that for $\b \lesssim 0.1$, the profile tends to fit the sine-Gordon kinks. For greater values it undergoes a rapid deformation.  It is worth noting that this first-order approximation fails  when $\b= e^{(1-2k)\pi/3}$, for $k\leq 0$. However, there is nothing special about those values, but in that case a second-order approximation would be necessary.


\section{Linear stability of the BPS configurations}
\label{linear}
Let us now discuss the linear stability for the two-scalar fields models we have constructed. The main issue is basically to analyse the spectrum of the corresponding Schr\"odinger-like operator associated with the normal modes of the classical model. The stability will be ensured when this Schr\"odinger-like is positive semi-definite,  implying that negative eigenvalues will be absent from its spectrum, and the zero mode will correspond to  the lowest bound state \cite{sta1}--\cite{Nascimento97}.

First of all,  it is well-known for one-field models that the static configurations of the $\phi^4$ model (\ref{eq2.18}), the $\vp^6$-like model (\ref{eq2.26}), the sine-Gordon model (\ref{eq2.29}), and the E-model  (\ref{eq2.31}) are all stable \cite{Rajaraman, Bazeia2004, Gabriel, Flugge}, with the corresponding Schr\"odinger-like operators related to the so-called Rosen-Morse II potential (or modified Po\"schl-Teller potential) for the first three models, and to the so-called Scarf-II (hyperbolic) potential in the latter case \cite{susyqm} (see more details in appendix \ref{exatos}).

Now, the stability analysis for multi-fields models is in general a highly non-trivial problem. Here, we will follow the line of reasoning introduced in \cite{Nascimento97}, to study the stability of static solutions in the two-scalar field models constructed in section \ref{2escalares}. The starting point is to consider a pair of static solutions, say $\p_s(x)$ and $\vp_s(x)$, and then introduce small fluctuations around these solutions, given in the following form
\br
\p(x,t)&=&\p_s(x)+\sum_k\rho_k(x)\cos(w_kt),\label{camposcomflutuacoes1}\\
\label{camposcomflutuacoes2}
\vp(x,t)&=&\vp_s(x)+\sum_k\s_k(x)\cos(w_kt),
\er
where $\rho_k$ and $\s_k$ are the small perturbations, when compared to the static configurations. Now, by substituting the fields $\p(x,t)$ and $\vp(x,t)$ into the second-order equations (\ref{sta1}), and considering only first-order terms in the fluctuations, we obtain the Schr\"odinger-like equation $H\Psi_k(x)=w^2_k\Psi_k(x)$, where
\br
&&H=-\frac{d^2}{dx^2}+\left(\begin{array}{c}
V_{\p\p} \ \ \ V_{\p\vp}\\
\\
V_{\vp\p} \ \ \ V_{\vp\vp}
\end{array}\right),\qquad 
\label{elHV}
\Psi_k(x)=\left(\begin{array}{c}
\rho_k(x)
\\
\s_k(x)
\end{array}\right).
\er
Notice that the derivatives of the potential $V(\p,\vp$) are written in terms of the static fields $\p_s(x)$ and $\vp_s(x)$. In addition, as it can be seen from eqs. (\ref{camposcomflutuacoes1}) and (\ref{camposcomflutuacoes2}), linear stability requires that the eigenvalues of $H$ have to be positive semi-definite, i.e. $w_k^2\geq 0$, with the zero mode $H\Psi_0(x)=0$, being given by
\br
\Psi_0(x)=N_0\left(\begin{array}{c}
\rho_0(x)
\\
\s_0(x)
\end{array}\right)=N_0\left(\begin{array}{c}
\p'(x)
\\
\vp'(x)
\end{array}\right),
\er
where the normalization constant $N_0$ can be chosen to be the unit. When the potential $V(\p,\vp)$ supports BPS states the Hamiltonian in (\ref{elHV}) can be written as follows,
\br
\label{elHS}
H=A^\dag_-A_-=A_+A_-,
\er
where the first-order operators
\br
\label{elS}
A_{\pm}=\pm\frac{d}{dx}+ \bw, \qquad \qquad \bw=\left(\begin{array}{c}
W^{(2)}_{\p\p} \ \ \ W^{(2)}_{\p\vp}\\
\\
W^{(2)}_{\vp\p} \ \ \ W^{(2)}_{\vp\vp}
\end{array}\right),
\er
have been introduced.  Note that $A^{\dag}_\pm=A_\mp$, implies that the Schr\"odinger-like operator $H$  is always positive semi-definite for the BPS case,  thus ensuring the linear stability of the BPS configurations. In this case, the ground state coincides with zero mode, and can be written as
\br
\Psi_0(x)=\left(\begin{array}{c}
W^{(2)}_\p
\\
W^{(2)}_\vp
\end{array}\right).
\er
Here, we will have an inherent difficulty regarding the explicit determination of the eigenvalue spectrum of the associated Schr\"odinger-like operator. As it can be seen, the coupling between static fields results in the coupling of the fluctuations in (\ref{elHV}). However, the problem turns to be more manageable if we take advantage of the first-order operators (\ref{elS}), and diagonalize the matrix $\bw$, to obtain
\br
\label{eldiagS}
A_{\pm}=\pm\frac{d}{dx}+\left(\begin{array}{c}
u_+ \ \ \ 0\\
\\
0 \ \ \ u_-
\end{array}\right),
\er
where the respective eigenvalues are  \cite{Nascimento97}
\br
\label{elv+-}
u_{\pm}=\frac{1}{2}\big(W^{(2)}_{\p\p}+W^{(2)}_{\vp\vp}\big)\pm\sqrt{\frac{1}{4}\big(W^{(2)}_{\p\p}-W^{(2)}_{\vp\vp}\big)^2+\big(W^{(2)}_{\p\vp}\big)^2}.
\er
By substituting (\ref{eldiagS}) in (\ref{elHS}), we obtain two decoupled eigenvalue equations
\br
&&\left[-\frac{d^2}{dx^2}+ U_+ (x)\right]\rho_k(x)=w^2_k\,\rho_k(x),\\
&&\left[-\frac{d^2}{dx^2}+U_-(x)\right]\s_k(x)=w^2_k\,\s_k(x),
\er
where the quantum mechanical potentials are given by
\br
U_{\pm}(x) = u^2_{\pm} +\frac{du_\pm}{dx}. \label{eqpotU}
\er
It is worth pointing out that in general this method would require certain simplifications since the square root term appearing in eq.(\ref{elv+-}) brings some complications for the explicit analytical calculations. In what follows, we will try then to simplify this term whenever is possible to perform the analytical analysis of the stability of the BPS configurations for the two-fields models we have constructed. Otherwise, the corresponding spectral problems should be analysed from a numerical point of view.

\subsection{The extended $(\p^4 +\vp^6_l)$ model}

\begin{figure}[t]
\begin{center}
 \includegraphics[width=\linewidth]{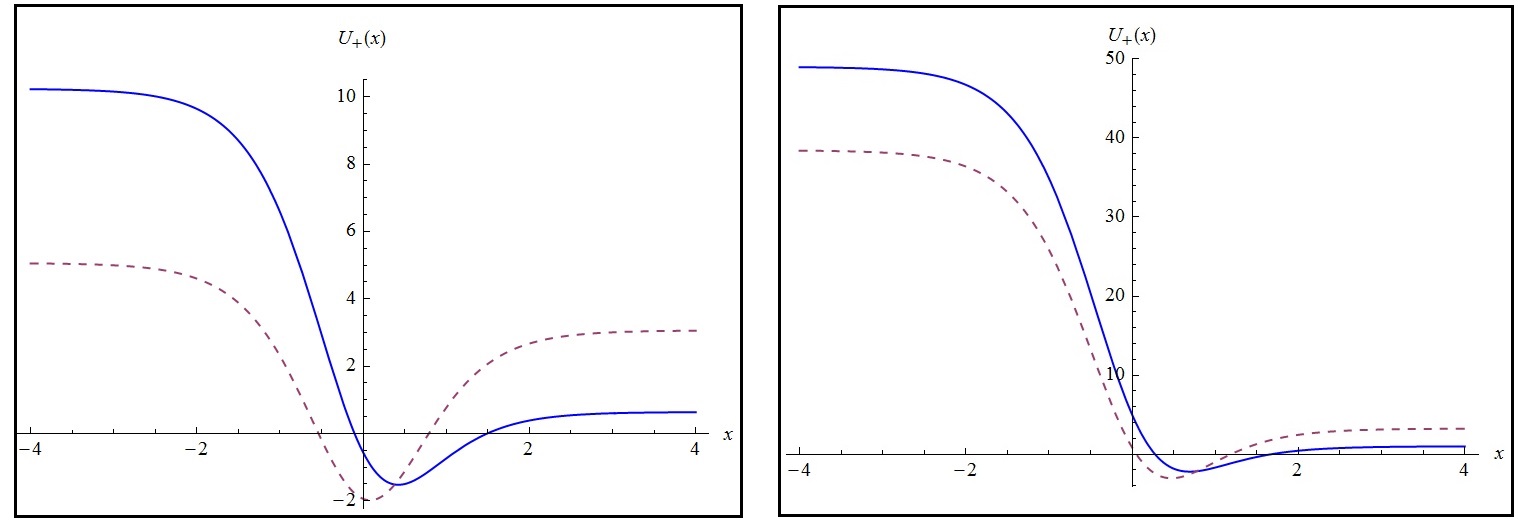}
 \caption{Quantum mechanical potential $U_{+}(x)$ associated to the extended $(\p^4+ \vp^{6}_l)$ model for different values of the parameters. On the left, we have plotted the potential with $c=-1$, and $b=0.6$ (solid line), and $b=0.125$ (dashed line). On the right, we have plotted the potential with $c=-2$, and $b=1.5$ (solid line), and $b=1.1$ (dashed line).}\label{potphi4zeta6}
 \end{center}
\end{figure}

Let us first study the stability of the BPS solutions (\ref{eqbpsphi4zeta6}) of the extended $(\p^4 + \vp^6_l)$ model. From the superpotential (\ref{Wphi4zeta6}), and the BPS solutions eq.(\ref{eqbpsphi4zeta6}), we obtain
\br
\label{vvpzcomraiz}
u_+ &=& 2b +2c \tanh(x), \label{v+vpz}\\
u_- &=& -2 \tanh(x),\label{v-vpz}
\er
where we are assuming that $\a=1$, and 
\br
\label{condicaotirardaraiz}
b \geq |1 + c|,
\er
in order to simplify  the square root term. Using these results, we get the corresponding quantum mechanical potentials (see figure \ref{potphi4zeta6}),
\br
\label{U+vpz}
U_{+}(x) & = & 4 b^2 + 4 c^2 + 8 b c \,\tanh{(x)} - 2c (2c - 1) \,\sech^2{(x)},\\
\label{U-vpz}
U_{-} (x)& = & 4 - 6\, \sech^2{(x)}.
\er
They have again the form of the Rosen-Morse II potentials \cite{susyqm} (see appendix \ref{exatos}). In this case the parameters of the potential $U_+(x)$ (\ref{U+vpz}) are given by,
\br
\label{comparandopotencialvpz}
A = -2 c,\qquad B = 4 b c.
\er
Then, from the stability condition (\ref{condicaodeestabilidade}), we see that the parameters have to satisfy%
\br
c < 0, \qquad |b| < |c|, \qquad 0 \leq k < A - \sqrt{4 \: |b| \: |c|}. \label{cond118}
\er
Now, let us choose some interesting values for the parameters. From eq. (\ref{condicaotirardaraiz}), we note that if $b = 0$, then $c = -1$, and we get that
\br
\label{Uquanticovpzb2igual0}
U_+(x) = U_-(x)= 4 - 6 \sech^2(x),
\er
both potentials are equal, and stability can be guaranteed. On the other hand, when $b<0$ the condition (\ref{condicaotirardaraiz}) is not satisfied, and then stability cannot be proven in that case, at least analytically. Finally, by considering $b>0$, together with the conditions (\ref{condicaotirardaraiz}) and (\ref{cond118}), we find that
\br
\label{condicaovpzparab2positivo}
|1+c| \leq b < |c|,
\er
and then $c < - {1}/{2}$ for consistency. Furthermore, analysing possible values of number of bound states $k$, we see that if $-\frac{1}{2}\big(1+\frac{\sqrt{2}}{2}\big) <c <-\frac{1}{2}$, we will have only the zero mode $k=0$.  For values $c \leq -\frac{1}{2}\big(1+\frac{\sqrt{2}}{2}\big)$, we have the following possibilities,
\br
  k  = \left\{ \begin{array}{r c l} 0, & \quad \mbox{if} & \quad \dfrac{1}{|c|} \(c + \frac{1}{2}\)^2 \leq b <|c|, \\[0.2cm]
  \mbox{0 and  1}, & \quad \mbox{if} &  \quad |1 + c| \leq b < \dfrac{1} {|c|} \(c + \frac{1}{2}\)^2.
  \end{array}   \right.
\er  
We can also see that, since the  potential $U_-$ has eigenvalues $E_0 = 0$ and $E_1 = 3$, it will have common eigenvalues  with $U_+$ only if
\br
\label{autovalorescomuns}
b=\sqrt{\frac{(1+c)(1+2c)^2}{(1+4c)}},\qquad     \left(-1 - \frac{\sqrt{3}}{2}\right) <  c  < -1.
\er
In the table \ref{tabela1}, we have chosen some particular values for the parameters in order to illustrate our results. For all these cases, the stability of the solutions is guaranteed. 
\begin{table}[t!]
\captionsetup{font=footnotesize}
\centering 
  \begin{tabular}{ | c | c | c | c | }
    \hline
    $c$  & $b$ & $k$ & $E_k = \omega_k^2$\\ \hline
    -1 & 0.6 & 0& 0\\ \hline
    -1 & 0.125 & 0 & 0\\
     & &1 & 2.81\\ \hline
      -2 & 1.5 & 0& 0\\ \hline
      -2& 1.1 & 0& 0\\ 
      && 1& 3.24\\\hline
  \end{tabular}
  \caption{Number of bound states and their eigenvalues for different values  of the parameters $b$ and $c$.}  \label{tabela1}
 \end{table} 
%

\subsection{The extended ($\p^4$+ sG) model}

\begin{figure}[t!]
\centering
\includegraphics[scale=0.27]{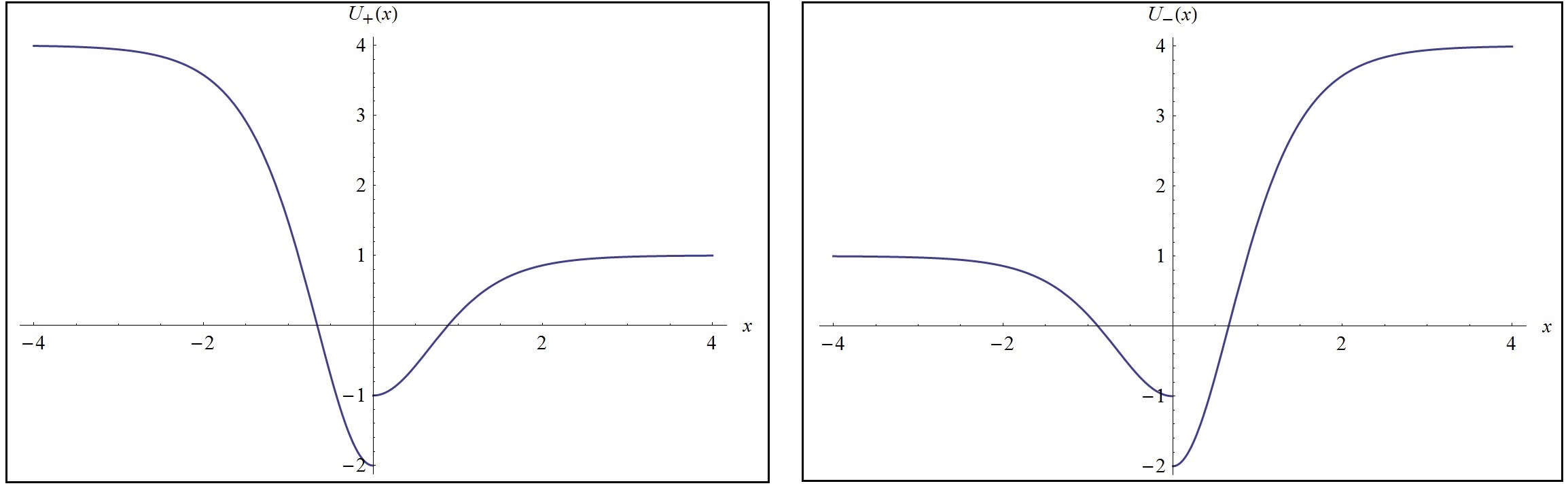}
\caption{Discontinuous quantum-mechanical potentials $U_{\pm}(x)$ associated to the extended ($\p^4$+sG) model, for $\a=1$ and $a_2=-2a_1$.} 
\label{figel2numero4}
\end{figure}

Now, we will analyze the stability of the BPS solutions (\ref{eqs3.56}) of the extended ($\p^4$+ sG) model described by the superpotential (\ref{potencialfsin}). For sake of simplification, we have chosen $a_{2} = - 2a_{1}$, and $\a = 1$,  to get 
\br
\label{vvpsGcomraiz}
u_{\pm} = -\frac{3}{2} \tanh(x)\pm \frac{1}{2}\mbox{sgn}(x) \tanh(x),
\er
where sgn($x$) is the signum function. Then, we obtain the following quantum-mechanical potentials,
\br
\label{U+vpsG}
&& U_{\pm}(x) = \frac{1}{2} (5\mp 3\, \mbox{sgn}(x))-2 (2 \mp \mbox{sgn}(x) ) \sech^2(x).
\er
In general, they are also associated to the Rosen-Morse II potential (\ref{potencialrosenmorse1}), with $B = 0$ and $\a = 1$. 
However, these quantum-mechanical potentials are discontinuous, as it can be seen from figure \ref{figel2numero4}, and that novel feature will require special attention in order to determine the eigenvalues. To do that,  we will use the procedure introduced in \cite{Glasser1, Glasser2} to determine the energy levels of composite potentials, which is based on the so-called Green function factorization theorem \cite{Garcia}. The main idea consists in decomposing the discontinuous potential into two ``pieces'', namely
\br
 U(x) = U^{(\mbox{\tiny L})} (x) \th(-x) + U^{(\mbox{\tiny R})}(x) \th(x),
\er 
where $\th(x)$ is the unit step function, and $U^{(\mbox{\tiny L/R})}(x)$ are continuous and symmetric (around the origin) potentials, for which the corresponding energy levels and wave functions for all stationary states are assumed to be known.  Then, by considering the  Green functions $G^{(\mbox{\tiny L/R})}(x,x';E)$ associated to each potentials $U^{(\mbox{\tiny L})}$ and $U^{(\mbox{\tiny R})}$,  the allowed eigenvalues $E$ of the composite potential  $U$ will be given by the solutions of the following transcendental equation \cite{Glasser1, Glasser2},
\br
\label{eqtranscendental}
G^{({\mbox{\tiny L}})}(0,0;E)+G^{({\mbox{\tiny R}})}(0,0;E)=0.
\er
In our case, both Green functions are associated to the solvable Rosen-Morse II potential, and its explicit formula can be written as follows \cite{Kleinert},
\br
\nonumber
G(x,x';E) \!&\!=\!&\! -\frac{i}{2}\Gamma(\sqrt{A^2-E}-A)\Gamma(1+A+\sqrt{A^2-E})\times\\
\label{greenrosenmorse}
&&\Big[\th(x-x')P_{A}^{-\sqrt{A^2-E}}(\tanh(x))P_{A}^{-\sqrt{A^2-E}}(-\tanh(x')) + (x\leftrightarrow x')\Big],
\er  
where  $A$ is the parameter given in (\ref{potencialrosenmorse1}),  $\G(z)$ is the Gamma function, and
\br
P_{a}^{b}(z) = \left(\frac{1+z}{1-z}\right)^{\frac{b}{2}} \frac{1}{\Gamma(1-b)}F\Big(-a,1+a;1-b;\frac{1-z}{2}\Big),
\er
are the associated Legendre polynomials, which are defined in terms of the hypergeometric function $F(a, b; c; z)$. The corresponding discrete eigenvalue spectrum satisfies
\br
\lim_{E\rightarrow E_{k}}(E-E_{k})G(x,x';E) \!&\!=\!&\! \frac{i(-1)^k}{k!}\sqrt{A^2-E_{k}}\,\Gamma(1+2A-k)\times \label{polodegreen}  \\
&&\Big[\th(x-x')P_{A}^{-\sqrt{A^2-E}}(\tanh(x))P_{A}^{-\sqrt{A^2-E}}(-\tanh(x'))+ (x\leftrightarrow x')\Big].\nonumber 
\er
Therefore, for the  $U_{+}(x)$ potential, we have  that
\br
U_{+}^{(\mbox{\tiny L})}=4-6\sech^2(x), \qquad 
U_{+}^{(\mbox{\tiny R})}=1-2\sech^2(x),
\er
and then 
\br
G_{+}^{(\mbox{\tiny L})}(0,0;E)=\frac{i}{2}\frac{(3-E)}{E\sqrt{4-E}}, \qquad G_{+}^{(\mbox{\tiny R})}(0,0;E)=\frac{i}{2}\frac{\sqrt{1-E}}{E}.
\er
From these results, we see that the transcendental equation (\ref{eqtranscendental}) becomes 
\br
\label{eqtranscendentalenergia}
E\sqrt{1-E}\sqrt{4-E}+E(3-E)=0,
\er
which only allows $E=0$ in the spectrum of the discontinuous potential $U_{+}(x)$. From the eq.(\ref{polodegreen}), we can see that the zero energy eigenvalue is  common to the decomposed potentials, which is consistent with the fact that $E = 0$ is a pole of the Green function ({\ref{greenrosenmorse}) for both cases. An identical transcendental equation will be obtained for the potential $U_-(x)$, since $U_-^{(\mbox{\tiny L/R})}=U_+^{(\mbox{\tiny R/L})}$, which again will allow only the zero energy eigenvalue. These results lead us to ensure the stability for the BPS solutions of the extended ($\vp^4$+sG) model, at least for our particular choice of parameters.


\subsection{The extended (sG+E) model}

%
\begin{figure}[t]
\centering
\includegraphics[scale=0.25]{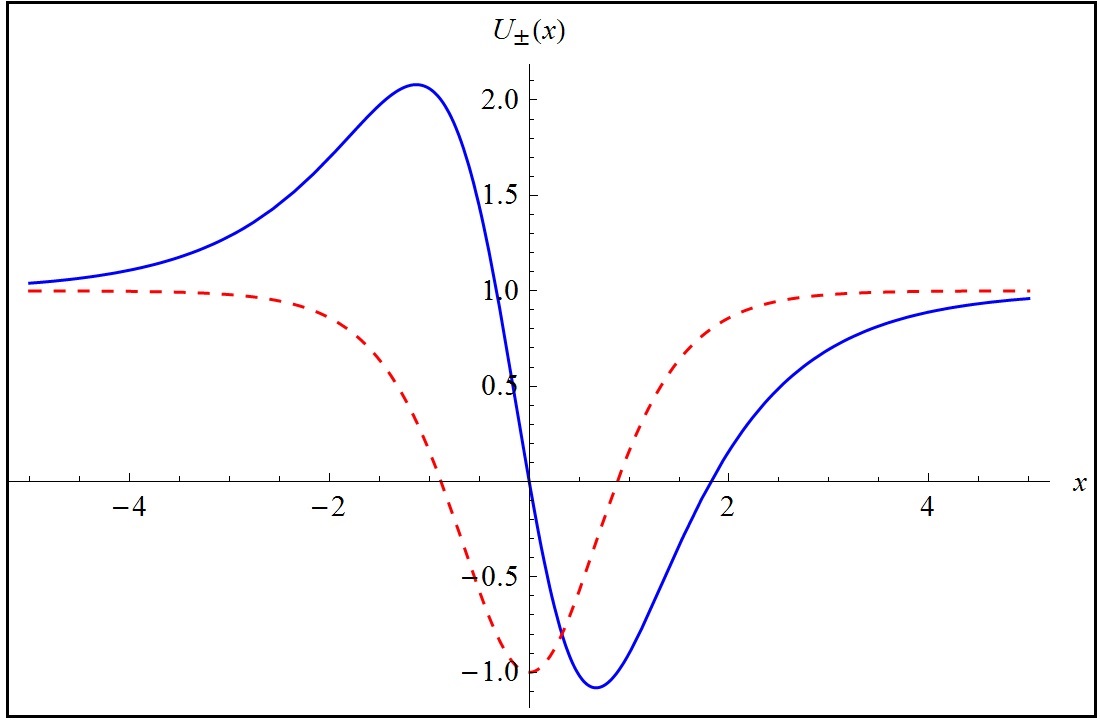}
\caption{Quantum-mechanical potentials $U_+$ (dashed line) and $U_-$ (solid line) of the extended (sG+E) model for $a_2=b_3=0$, and $\a=1$.}
\label{figel2numero5}
\end{figure}
%
%
%

Finally, let us study the stability of the BPS solutions (\ref{eqs3.78}) of the extended (sG+E) superpotential (\ref{ESGpot}). In this case, we find that
\br
\nonu
u_{\pm} &=& \frac{\a}{2} \Bigg\{\sech(\a x)+\Big(a_{2}+b_{3}-2+\frac{b_{3}}{\b^2}e^{-2\arctan(\sinh(\a x))}+a_{2}\b^2 e^{2\arctan(\sinh(\a x))}\Big)\tanh(\a x)\\
\nonu
&& \quad \pm \,2 \bigg[\Big(\frac{b_{3}}{\b}e^{-\arctan(\sinh(\a x))}+a_{2}\b e^{\arctan(\sinh(\a x))}\Big)^2\tanh^2(\a x)\\
\nonu
&& \qquad \quad  +\frac{e^{-4\arctan(\sinh(\a x))}}{4\b^4}\Big(\b^2 e^{2\arctan(\sinh(\a x))}\,\sech(\a x)\\
\label{vsGEcomraiz}
&&\qquad \quad -\big(\b^2 e^{2\arctan(\sinh(\a x))}-1\big)\big(b_{3}+a_{2}\b^2 e^{2\arctan(\sinh(\a x))}\big)\tanh(\a x)\Big)^2\bigg]^{\frac{1}{2}}\Bigg\}. 
\er
In order to simplify the root term in  eq.(\ref{vsGEcomraiz}), and study analytically the associated quantum-mechanical potentials, we could choose $a_{2} = b_{3} = 0$, obtaining
\br
\label{U+sGE}
&& U_{+}^{(0)}(x) = \a^2-\a^2 \sech^2(\a x)-3\a^2\sech(\a x)\tanh(\a x),\\
\label{U-sGE}
&& U_{-}^{(0)}(x) = \a^2-2\a^2\sech^2(\a x),
\er
which correspond to the Scarf II (\ref{potencialscarfII}) and Rosen-Morse II (\ref{potencialrosenmorse1}) potentials, respectively. However, this choice of parameters trivially decouples the fields $\eta$ and $\chi$. See their plots in figure \ref{figel2numero5}.

From the analytical point of view it is quite complicated to study these quantum-mechanical potentials for general values of the parameters $a_{2}$ and $b_{3}$. Instead, we  will perform a more qualitative and approximated analysis of the bound states for some values of the parameters. At this point, we can only guarantee that stability does exist at least for some very small values of the parameters, that is, the potentials possess the zero-mode as their fundamental bound state, and there is no negative energy eigenvalues. Of course, a more precise analysis requires a deeper numerical study. 

Before doing that, let us take a look of the potential deformations for some small values of the parameters. In the figures \ref{figura6} and \ref{figel2numero8} we have plotted the potentials for some configurations with $a_{2} = 0$, and small values of $b_3$. While, in figures \ref{figel2numero9} and \ref{figel2numero10}, we have plotted configurations with $b_{3} = 0$, and small values of $a_{2}$. \\

%
\begin{figure}[t!]
\centering
\includegraphics[width=\linewidth]{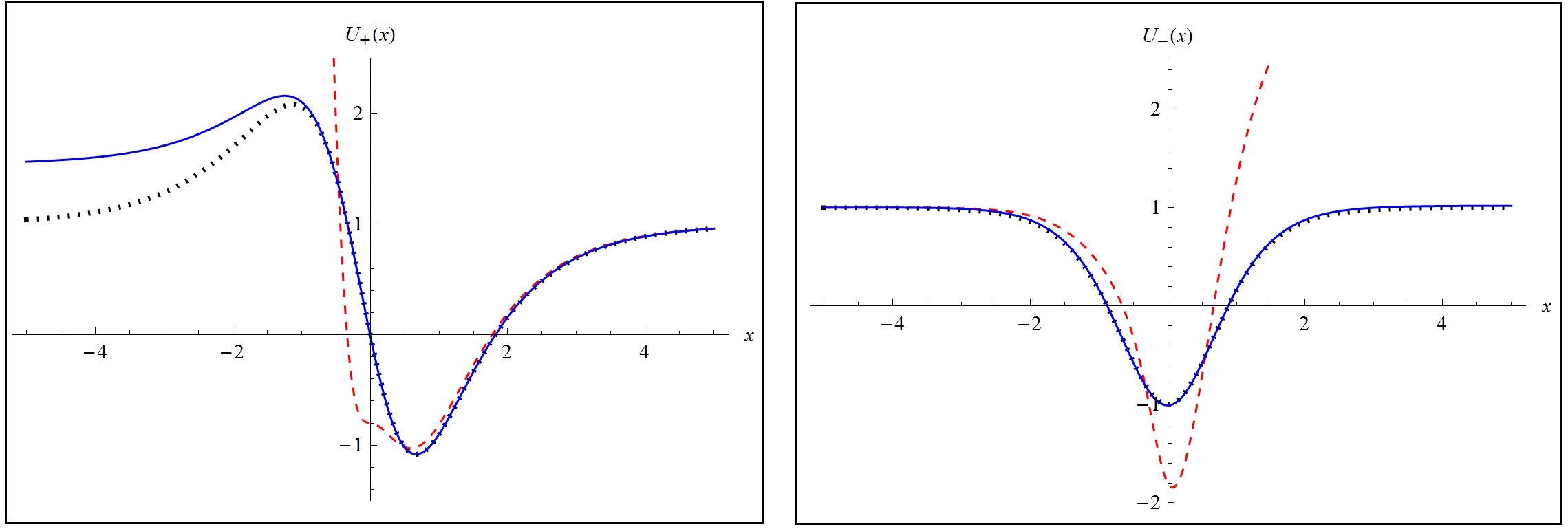}
\caption{Quantum-mechanical potentials $U_{+}$ (on the left) and $U_-$ (on the right) for $a_{2}=0$ and $\a=1$. For both, we have plotted the values  $b_{3} = 0$, $b_{3} = - 0.01$,  and $b_ {3} = - 0.8$, depicted with dotted, solid, and dashed lines, respectively.}
\label{figura6}
\end{figure}
\vspace{0.4cm}
\begin{figure}[t!]
\centering
\includegraphics[width=\linewidth]{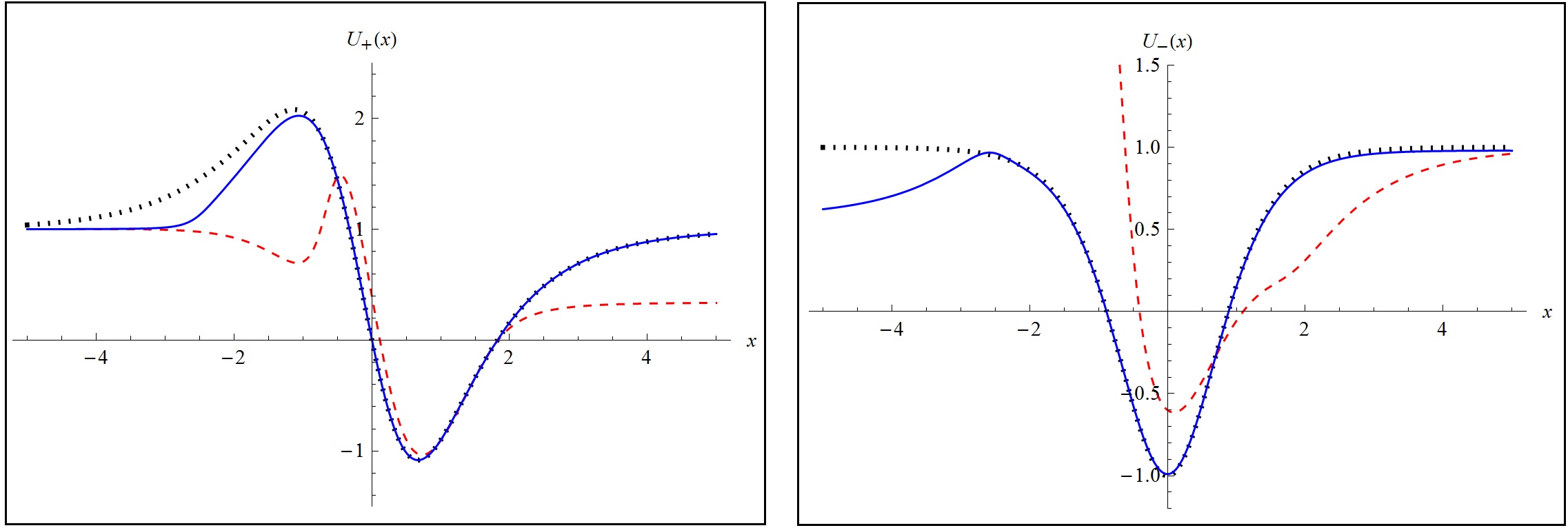}
\caption{Quantum-mechanical potentials $U_{+}$ (on the left) and $U_-$ (on the right) for $a_{2}=0$ and $\a=1$. For both, we have plotted the values  $b_{3} = 0$, $b_{3} =  0.01$,  and $b_ {3} = 0.4$, depicted with dotted, solid, and dashed lines, respectively.}
\label{figel2numero8}
\end{figure}
\vspace{0.4cm}
\begin{figure}[t!]
\centering
\includegraphics[width=\linewidth]{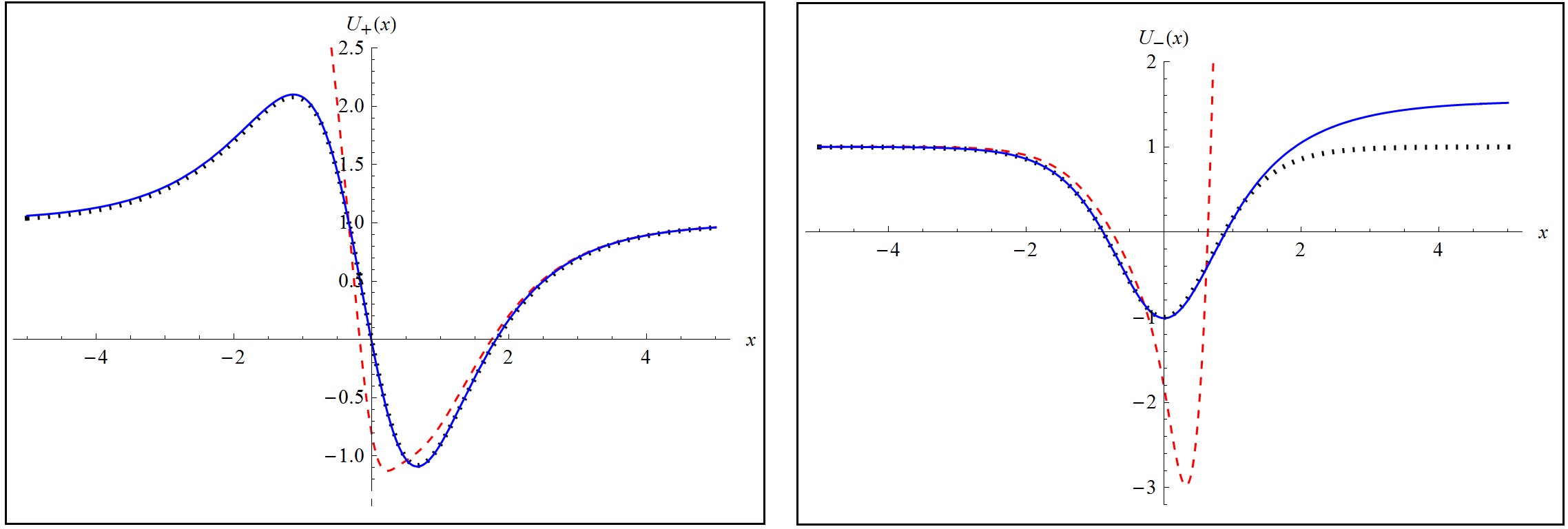}
\caption{Quantum-mechanical potentials $U_{+}$ (on the left) and $U_-$ (on the right) for $b_{3}=0$ and $\a=1$. For both, we have plotted the values  $a_{2} = 0$, $a_{2} =  -0.01$,  and $a_ {2} = -0.8$, depicted with dotted, solid, and dashed lines, respectively.}
\label{figel2numero9}
\end{figure}
%
%
\begin{figure}[h!]
\centering
\includegraphics[width=\linewidth]{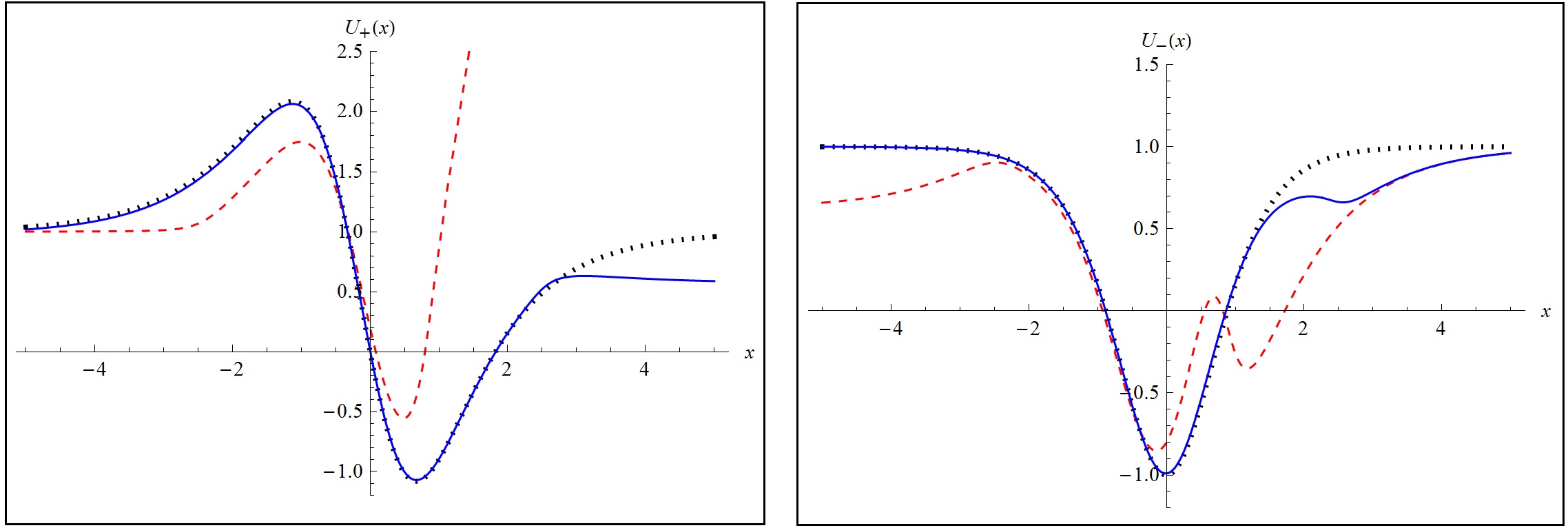}
\caption{Quantum-mechanical potentials $U_{+}$ (on the left) and $U_-$ (on the right) for $b_{3}=0$ and $\a=1$. For both, we have plotted the values  $a_{2} = 0$, $a_{2} =  0.01$,  and $a_ {2} = 0.2$, depicted with dotted, solid, and dashed lines, respectively.}
\label{figel2numero10}
\end{figure}
%
\begin{figure}[h!]
\centering
\includegraphics[width=\linewidth]{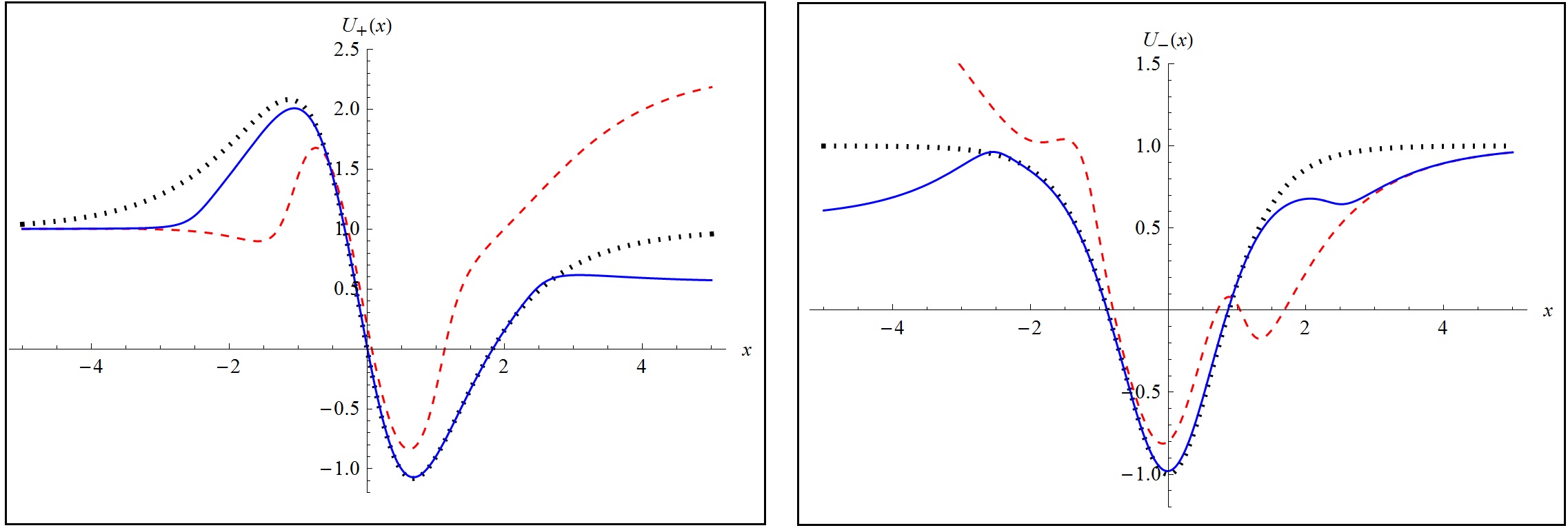}
\caption{Quantum-mechanical potentials $U_{+}$ (on the left) and $U_-$ (on the right) for  $a_{2}=0$ and $b_{3}=0$ (dotted line), $a_{2}=0.01$ and $b_{3}=0.01$ (solid line), and  $a_{2}=0.1$ and $b_{3}=0.1$ (dashed line).}
\label{figel2numero11}
\end{figure}
\mbox{}\\

Now, for very small values of the parameters ($\lesssim 10^{-2}$), it is clear that the quantum-mechanical potentials converge to exactly solvable problem, see also figure \ref{figel2numero11}. In that case it is possible to apply the time-independent perturbation theory for the calculation of the energy eigenvalues corrections.  To do that let us first consider the case when $a_{2} = 0$ and $b_{3}=\l\lesssim 10^{-2}$, so we have a perturbed Hamiltonian which consists of two parts 
\br
 H = H_0 + \l H_1,
\er
\mbox{}\\
with 
\br
 H_0 = -\frac{d^2}{dx^2} +  \left(\begin{array}{cc}
U_{+}^{(0)} &0\\
\\
0 & U_{-}^{(0)}
\end{array}\right),\qquad 
H_1 = \left(\begin{array}{cc}
U_{+}^{(1)} &0\\
\\
0 & U_{-}^{(1)}
\end{array}\right),
\er
where $U_{\pm}^{(0)}$ are the exactly solvable potentials (\ref{U+sGE}) and (\ref{U-sGE}), and the first-order corrections $U_{\pm}^{(1)}$ are given in this case by
\br 
 U_{+}^{(1)} &=& -\frac{\a^2}{\b^2} \,e^{-2 \arctan(\sinh(\a x))} \( 2 - 3 \, \sech^2(\a x) \),\\
 U_{-}^{(1)} &=& -\a^2\(2-3\, \sech^2(\a x)\).
\er
Then,  the eigenvalues $E_k$ of the perturbed problem  can be  expanded in a power series in the parameter $\l$ as follows
\br
E_{k} = E_{k}^{(0)}+\l \,E_{k}^{(1)} + \dots,\label{pertener}
\er
where $E_k^{(0)}$ are the unperturbed eigenvalues given in (\ref{energiarosenmorse1}) and (\ref{energiascarf}), respectively, and the first-order correction $E_k^{(1)}$ can be obtained from the expression 
\br 
 E_{k}^{(1)} &=&  \int_{-\infty}^{\infty}\left[\rho_{k}^{*}(x) U_{+}^{(1)}(x)\rho_{k}(x)+ \s_{k}^{*}(x) U_{-}^{(1)}(x)\s_{k}(x) \right]dx, \label{perturbacaoenegiaScII}
\er
where $\rho_k(x)$ and $\s_k(x)$ will be given in this case by the Scarf II  (\ref{ondascarf}) and  Rosen-Morse II (\ref{ondarosen}) eigenfunctions, respectively. In fact, from the explicit form of the potentials (\ref{U+sGE}) and  (\ref{U-sGE}), we see that the associated parameters are $A = \a$ in both cases,  and \mbox{$B = -\a$} for $U_{+ }^{(0)}$, and $B=0$ for $U_{-}^{(0)}$. Thus, both potentials possess only one bound state, the zero mode,
\br 
 \Psi_0 = \left( \begin{array}{c} \rho_0(x) \\ \s_0(x) \end{array} \right) = \left( \begin{array}{c} \sech(\a x) e^{\arctan(\sinh(\a x))} \\ \sech(\a x)\end{array} \right),
\er
with energy $E_0^{(0)} =0$. Now, by using eq. (\ref{perturbacaoenegiaScII}) we can show straightforwardly  that there is no correction to the zero mode energy, at least at first-order approximation, i.e. $E_{0}^{(1)} = 0$.

Let us consider now the case when $b_{3} = 0$ and $a_{2} =\l \lesssim 10^{-2}$. Similarly, we find that the first-order corrections $U_{\pm}^{(1)}$ are given in this case by,
\br
 U_{+}^{(1)}  &=& -\a^2 \(2 -3\,\sech^2(\a x)- 2\,\sech(\a x)\tanh(\a x)\), \label{b30deltaa2mais}\\
 U_{-}^{(1)}  &=& -\a^2\b^2 e^{2\arctan(\sinh(\a x))} \(2-3\,\sech^2(\a x)-2\,\sech(\a x)\tanh(\a x) \).\label{b30deltaa2menos}
\er
Then, by substituting in eq. (\ref{perturbacaoenegiaScII}) we find that the first-order correction to the zero mode also vanishes. 
Therefore,  we can ensure that in the weak coupling regime, for  $a_{2} \lesssim 10^{-2}$ and $b_3\lesssim 10^{-2}$, the stability of the extended (sG+E) BPS solutions. Of course, for greater values of the coupling parameters, we should do a more complete analytical or numerical analysis of the spectral problem. We will leave this specific issue to be explored in other future work.


%
\section{Extended three-scalar fields models}
\label{3field}

Now, we will construct some new three-scalar field extended models by applying a generalization of the extension method for three-field systems \cite{santos2018} to the one-field systems studied so far.

\subsection{$\p^4$ model coupled to the $\vp^6$-like and the inverted $\z^{4I}$ models}

Let us start by considering the coupling of the standard $\p^4$ model with the $\vp^6$-like model, and also with the so-called the inverted $\z^{4I}$-model \cite{Bazeia1}. The starting point is again the first-order equation 
\br
 \p' = W_\p^{(1)} = \a(1-\p^2), \label{eq6.1}
\er
together with the deformation functions,
\br
\p \!&\!=\!&\! f_{1}(\vp) = |\vp|-1,\\
 |\p| \!&\!=\!&\! f_{2}(\z) = \sqrt{1-\frac{\z^2}{\b^2}},\label{eqn6.2}\\
 \z \!&\!=\!&\! f_{3}(\vp) = f_{1}^{-1}(f_{2}(\vp))= \b\sqrt{1-(|\vp|-1)^2},
\er
from which we have the corresponding the first-order equations for the deformed models,
\br
\vp' \!&\!=\!&\! \widetilde{W}_\vp^{(1)} = \a\vp(2-|\vp|),\label{eq6.6}\\
\z' \!&\!=\!&\! \widehat{W}_\z^{(1)}  = -\a\omega\z\sqrt{1-\frac{\z^2}{\b^2}},\label{eq6.5}
\er
where we have defined $\omega = \mbox{sgn}(\p)$. Their corresponding static solutions are
\br
 \phi(x) = \tanh(\a x), \qquad \vp_\pm(x)=\pm(1+\tanh(\a x)),\qquad \z(x) = \b\sech(\a x). \label{eqn6.7}
\er
Now, the main idea of the method can be straightforwardly generalized to three-fields. First, we write the right-hand side of eq. (\ref{eq6.1}) now in seven different and equivalent forms by using  the deformation functions and their inverse functions, as follows
\br
\nonumber
    W^{(1)}_\p(\p)& =& \a\left(1-\p^2\right), \qquad \,\,\,   W^{(1)}_\p(\vp) =\a\left[1-(|\vp|-1)^2\right],\qquad W^{(1)}_\p(\z) = \frac{\a}{\b^2}\z^2 , \\
\nonu  
     W^{(1)}_\p(\p,\vp) &=& \a\left[1-\p (|\vp|-1)\right], \qquad W^{(1)}_\p(\p,\z)= \frac{\a}{\b}\z\sqrt{1-\p^2} ,\\
\label{fsfuncaopparapvpz} 
     W^{(1)}_\p(\vp,\z)&=& \frac{\a}{\b}\z\sqrt{1-(|\vp|-1)^2},\qquad W^{(1)}_\p(\p,\vp,\z) = \frac{\a}{\b}\z\sqrt{1-\p (|\vp|-1)}.
\er
%
Similarly, for  eq. (\ref{eq6.6})
\br
\nonumber
     \wt W^{(1)}_\vp(\z)\!&\!=\!&\! \frac{\a}{\b^2}\epsilon\z^2,  \qquad \wt W^{(1)}_\vp(\p) = \a\epsilon(1-\p^2), \qquad   \wt W^{(1)}_\vp(\vp)=\a\vp(2-|\vp|),  \\
\nonu
    \wt W^{(1)}_\vp(\p,\z) \!&\!=\!&\!  \frac{\a}{\b}\epsilon\z\sqrt{1-\p^2}, \qquad \wt W^{(1)}_\vp(\p,\vp)= \a\vp(1-\p),\\
\label{fsfuncaozparapvpz} 
     \wt W^{(1)}_\vp(\vp,\z)\!&\!=\!&\! \frac{\a}{\b}\epsilon\z\sqrt{1-(|\vp|-1)^2},\qquad \wt W^{(1)}_\vp(\p,\z,\vp)= \frac{\a}{\b}\epsilon\z\sqrt{1-\p(|\vp|-1)},
\er
and for the eq. (\ref{eq6.5}) we have
\br
\nonumber
     \wh W^{(1)}_\z(\z)\!&\!=\!&\! -\a\omega\z\sqrt{1-\frac{\z^2}{\b^2}},\qquad \wh W^{(1)}_\z(\p) = -\a\b\p\sqrt{1-\p^2}, \qquad \wh W^{(1)}_\z(\p,\z)= -\a\p\z,\\
\nonu
     \wh W^{(1)}_\z(\vp)\!&\!=\!&\! -\a\b(|\vp|-1)\sqrt{1-(|\vp|-1)^2},\qquad \wh W^{(1)}_\z(\p,\vp)= -\a\b\p\sqrt{1-(|\vp|-1)^2},\\
\label{fsfuncaovpparapvpz}
     \wh W^{(1)}_\z(\vp,\z)\!&\!=\!&\! -\a\z(|\vp|-1),\qquad \wh W^{(1)}_\z(\p ,\z ,\vp)= -\a\b\omega\sqrt{1-\p(|\vp|-1)}\sqrt{1-\frac{\z^2}{\b^2}}.
\er
%
%
%
%
Now, we will use a generalization of the ansatz used in eqs. (\ref{modeloextendidoA}) and (\ref{modeloextendidoB}) for the case of three-field systems, in the following form
\br
\nonumber
W_\p^{(3)}(\p,\vp,\z) &=& a_{1} W^{(1)}_\p(\vp)+a_{2} W^{(1)}_\p(\p,\vp)+a_{3} W^{(1)}_\p(\p)+a_{4} W^{(1)}_\p(\z)+a_{5} W^{(1)}_\p(\p,\z)\\[0.1cm]
\nonumber
&& +a_{6} W^{(1)}_\p(\vp,\z) +a_{7} W^{(1)}_\p(\p,\vp,\z) +p_{1}\,g(\vp)+p_{2}\,g(\p,\vp)+p_{3}\,g(\p)+ p_{4}\,g(\z) \\[0.1cm]
\label{fsmodeloextendidoAparapvpz}
&&+p_{5}\,g(\p,\z)  +p_{6}\,g(\vp,\z)+p_{7}\,g(\p,\vp,\z), 
\er
\br
\nonumber
W_\vp^{(3)}(\p,\vp,\z) &=& b_{1}\wt W^{(1)}_\vp(\vp)+b_{2}\wt W^{(1)}_\vp(\p,\vp)+b_{3}\wt W^{(1)}_\vp(\p)+b_{4}\wt W^{(1)}_\vp(\z)+b_{5}\wt W^{(1)}_\vp(\p,\z)\\[0.1cm]
\nonumber
&&+ b_{6}\wt W^{(1)}_\vp(\vp,\z) +b_{7}\wt W^{(1)}_\vp(\p,\vp,\z)+q_{1}\,\act g(\vp)+q_{2}\,\act g(\p,\vp)+q_{3}\,\act g(\p)+q_{4}\,\act g(\z)\\[0.1cm]
\label{fsmodeloextendidoBparapvpz}
&&+q_{5}\,\act g(\p,\z)+q_{6}\,\act g(\vp,\z)+q_{7}\,\act g(\p,\vp,\z),\\[0.3cm]
\nonumber
W_\z^{(3)}(\p,\vp,\z) &=& c_{1}\wh W^{(1)}_\z(\vp)+c_{2}\wh W^{(1)}_\z(\p,\vp)+c_{3}\wh W^{(1)}_\z(\p)+c_{4}\wh W^{(1)}_\z(\z)+c_{5}\wh W^{(1)}_\z(\p,\z)\\[0.1cm]
\nonumber
&&+c_{6}\wh W^{(1)}_\z(\vp,\z)+c_{7}\wh W^{(1)}_\z(\p,\vp,\z)+r_{1}\,\hat{g}(\vp)+r_{2}\,\hat{g}(\p,\vp)+r_{3}\,\hat{g}(\p)+r_{4}\,\hat{g}(\z)\\[0.1cm]
\label{fsmodeloextendidoCparapvpz}
&&+r_{5}\,\hat{g}(\p,\z)+r_{6}\,\hat{g}(\vp,\z)+r_{7}\,\hat{g}(\p,\vp,\z),
\er 
where the parameters must satisfy the following conditions 
\br
\sum_{i=1}^{7} a_{i}= \sum_{i=1}^{7} b_{i} = \sum_{i=1}^{7} c_{i} =1,  \qquad \mbox{and}\qquad 
\sum_{i=1}^{7} p_{i}  =\sum_{i=1}^{7} q_{i} = \sum_{i=1}^{7} r_{i} =0. 
\er
In addition, the $g$-functions are determined from the following constraints (see in appendix \ref{3campos} more details of the full derivation),  
\br
 W_{\p\vp}^{(3)} = W_{\vp\p}^{(3)}, \qquad W_{\p\z}^{(3)} = W_{\z\p}^{(3)}, \qquad W_{\vp\z}^{(3)} = W_{\z\vp}^{(3)}.
\label{fscondicaodecontinuidade}
\er
Using the explicit results for the $g$-functions into eqs.  (\ref{fsmodeloextendidoAparapvpz})--(\ref{fsmodeloextendidoCparapvpz}), yields
\begin{eqnarray}
\nonumber
W_\p^{(3)}(\p,\vp,\z)&=&\a (1-\p^2) + a_{2} \p\( 1+\p-|\vp|\)- \frac{1}{2} (1-c_{6})\left(\z^2-\b^2(1-\p^2)\right), \label{fssuperpotencialpparapvpz}\\
\nonu
W_\vp^{(3)}(\p,\vp,\z)&=& \frac{\a \epsilon}{2}\Big( (2+c_{6}\b^2) \(1- (|\vp|-1)^2\)-a_{2} \(\p^2- (|\vp|-1)^2\) -c_{6} \z^2\Big),
\label{fssuperpotencialparapvpz}\\
W_\z^{(3)}(\p,\vp,\z)&=& -\a\z \left( \p - c_{6} \left(1+\p - |\vp|\)\right).
\label{fssuperpotencialvpparapvpz}
\end{eqnarray}
After integrating, we finally obtain the following three-field superpotential
\br
\nonumber
W^{(3)}(\p,\vp,\z)&=& \a\left(1-a_{2}+\frac{\b^2}{2}(1-c_{6})\right)\p\left(1-\frac{\p^2}{3}\right) -\frac{\a}{2}(1-c_{6})\p\vp^2 -\frac{\a}{2}c_{6}\vp^2(|\z|-1)\\
\nonumber
&& +\a a_{2}\p\left(1-\frac{\p}{2}\left(|\z|-1\right)\right)+\frac{\a a_{2}}{2}|\z| +\frac{\a}{2}\left(2-a_{2}+\b^2 c_{6}\)\z^2\left(1- \frac{|\z|}{3}\)\\
&& +\frac{\a}{12}\left(2b_{4}+{c_{6}\b^2}\).\label{fspotencialcompletopvpz}
\er
From now on we will named this three field model as the extended ($\p^4+\vp^6_l+\z^{4I}$) model, for which possesses the static configurations given in eq.(\ref{eqn6.7}) are BPS solutions, with energy
\br 
 E_{\mbox{\tiny BPS}} = \frac{2\a}{3}\(4+\b^2\).
\er
The issue that arises from these results concerns the linear stability of the solutions for this superpotential. Although the stability analysis for three-field systems follows the same steps that the one presented in section \ref{linear},  it is actually further more complicated mostly because of the diagonalization of the Schr\"odinger-type operator, which is the key point in order to find the normal mode fluctuations. In this case, we will have
\br
\phi(x,t)&=&\p_s(x)+\sum_k\rho_k(x)\cos(w_kt), \nonu\\
\vp(x,t)&=&\vp_s(x)+\sum_k\s_k(x)\cos(w_kt),\nonu\\
\label{camposcomflutuacoes3campos}
\z(x,t)&=&\z_s(x) +\sum_k\xi_k(x)\cos(w_kt),
\er
where $\rho_k$, $\s_k$, and $\xi_k$ are the fluctuations around the static solutions $\p_s(x), \vp_s(x)$, and $\z_s(x) $. Considering the dynamics of these three time-dependent fields up to first-order, we will obtain a corresponding Schr\"odinger-like equation $H\Psi_k(x) = w^2_k\Psi_k(x)$, 
\br
&&H=-\frac{d^2}{dx^2}+\left ( \begin{array}{ccc}
V_{\p\p} & V_{\p\vp} & V_{\p\z}\\
V_{\vp\p} & V_{\vp\vp} & V_{\vp\z}\\
V_{\z\p} & V_{\z\vp} & V_{\z\z}\\
\end{array}\right),\qquad 
\label{elHV3campos}
\Psi_k(x)=\left(\begin{array}{c}
\rho_k(x)
\\
\s_k(x)
\\
\xi_k(x)
\end{array}\right).
\er
For the case of BPS potentials, we can write this Hamiltonian in terms of linear operators, namely $H = A^\dag_-A_-=A_+A_-$, where \br
\label{eldiagS3campos}
&&A_{\pm}=\pm\frac{d}{dx}+ \mathbf{W}, \qquad \mathbf{W}\,=\,\left ( \begin{array}{ccc}
W_{\p\p} & W_{\p\vp} & W_{\p\z}\\
W_{\vp\p} & W_{\vp\vp} & W_{\vp\z}\\
W_{\z\p} & W_{\z\vp} & W_{\z\z}\\
\end{array}\right).
\er
Our strategy again will be trying to diagonalize the matrix ${\bf W}$, and then the Schr\"odinger-type equation will  be split into three equations, which will be analysed separately.

In the case of the BPS solutions (\ref{eqn6.7}) of the ($\p^4 +\vp_l^6 + \z^{4I} $) model (\ref{fspotencialcompletopvpz}), this matrix takes the following form 
\br
\label{matrizp4vp4Iz6}
\bw =\left ( \begin{array}{ccc}
-2 \a \tanh(a x) & 0 & 0\\
0 &-\a (2+\b^2) \tanh(\a x) & -\a\b\e\, \sech{(\a x)}\\
0 & -\a\b\e \,\sech{(\a x)} &  -\a\tanh(\a x)\\
\end{array} \right ),
\er
where we have chosen the parameters being $a_{2} = 0$ e $c_{6} = 1$ for simplicity. By computing its corresponding eigenvalues, we find
\br
\label{p4vp4Iz6u+}
u_{0} = -2\a\tanh(\a x),\qquad  u_{\pm } = -\frac{\a}{2}\left((3+\b^2)\tanh(\a x)\pm \sqrt{4\b^2+(\b^2-1)^2\tanh^2(\a x)}\right).
\er
Now, by setting  $\b = 1$, we will find that the quantum mechanical potentials are given by,
\br
\label{potencialp4vp4Iz6U+}
U_{0} = 4\a^2-6\a^2\sech^2(\a x),\qquad U_{\pm} = 5\a^2-6\a^2\sech^2(\a x) \pm 4\a^2\tanh(\a x),
\er
which are again Rosen-Morse II potentials (\ref{potencialrosenmorse1}). The $U_0$ potential has parameters $A = 2 \a$ and $B = 0$,  and possesses eigenvalues $E_0 = 0$ and $E_1 = 3 \a^2$. The other two potentials $U_\pm$ have parameters $A = 2 \a$ and $B = \pm 2 \a^2$ respectively, and only have the ground state $E_0 = 0$. Therefore, for these choice of parameters, we will have stability guaranteed. Another possible choice would be $a_{2} = 0$ and $c_{6} = 0$, however we will get  essentially the same results.


\subsection{$\p^4$ model coupled to sine-Gordon model and the E-model}

Let us now construct a model obtained by the coupling of the standard $\p^4$ model with sine-Gordon, and the E-model. The first-order equation for each one of these models are given by eqs. (\ref{eq2.17}), (\ref{eq2.28}), and (\ref{eq2.30}), namely 
\br
 &&\p' = W_\p^{(1)} = \a(1-\p^2),\\
 &&\chi' = \wt{W}_\chi ^{(1)}= \frac{\a}{\b}\cos(\bc),\\
 &&\eta' = \wh{W}_\eta ^{(1)}= \a (1+\eta)\cos\Big(\frac{1}{2} \ln (1+\eta)^2\Big),
\er
with static solutions  
\br 
\p(x) =  \tanh(\a x),\qquad 
\chi(x) = \frac{1}{\b}\arctan{\big(\sinh(\a x)\big)},\qquad 
\eta(x) = \exp \left( \arctan(\sinh(\a x)) \right) -1. \quad \mbox{}\label{eqn6.31}
\er
The deformation functions connecting the three models have the following forms,
\br
\label{f1}
 &&\p = f_{1}(\c) = \sin(\bc),\\
\label{f2} 
 &&\eta = f_{2}(\chi) = e^{\bc}-1,\\
\label{f3}
 &&\p = f_{3}(\eta) = f_{1}\left(f_{2}^{-1}(\eta)\right) = \sin\Big(\frac{1}{2} \ln (1+\eta)^2\Big).
\er
By using these deformation functions and their inverse functions, we get the following expressions,
\br
\nonumber
     W_\p^{(1)}(\p) \!&=&\!\a(1-\p^2),\qquad  W_\p^{(1)}(\chi)=\a\cos^2(\bc), \qquad   W_\p^{(1)}(\eta)=\a\cos^2\Big(\frac{1}{2} \ln (1+\eta)^2\Big),    \\
 \nonumber    
      W_\p^{(1)}(\p,\chi)\!&=&\! \a\big(1-\p\sin(\bc)\big),   \qquad W_\p^{(1)}(\p ,\eta)=\a\left[1-\p\sin\Big(\frac{1}{2} \ln (1+\eta)^2\Big)\right], \\
 \nonumber     
    W_\p^{(1)}(\eta,\chi) \!&=&\!\a\left[1-\sin(\bc)\sin\Big(\frac{1}{2} \ln (1+\eta)^2\Big)\right], \\
        \quad 
\label{7potencialvp} 
 W_\p^{(1)}(\p ,\eta ,\chi) \!&=&\! \a\left[1-2\p\sin\Big(\frac{\bc}{2}\Big)\cos\Big(\frac{1}{4} \ln (1+\eta)^2\Big)\right]. \mbox{\!\!\!\!}
\er
Similarly, we have 
\br
\nonu
     \wt W_\c^{(1)}(\c) \!\!&=&\!\!\frac{\a}{\b}\cos(\bc),\qquad  \wt W_\c^{(1)}(\p) =\frac{\a}{\b}\sqrt{1-\p^2}, \qquad   \wt W_\c^{(1)}(\eta)=     \frac{\a}{\b}\cos\Big(\frac{1}{2} \ln (1+\eta)^2\Big),\\
\nonu  
  \wt W_\c^{(1)}(\p,\c) \!\!&=&\!\! \frac{\a}{\b}\left[\big(1-2\p^2\big)\cos(\bc)+2\p\sqrt{1-\p^2}\sin(\bc)\right],\\
\nonu
  \wt W_\c^{(1)}(\eta ,\c) \!\!&=&\!\! \frac{\a}{\b}\left[2\cos\Big(\frac{\bc}{2}\Big)\cos\Big(\frac{1}{4} \ln (1+\eta)^2\Big)-1\right],\\
\nonu 
    \wt W_\c^{(1)}(\p,\eta)\!\!&=&\!\! \frac{\a}{\b}\left[\big(1-2\p^2\big)\cos\Big(\frac{1}{2} \ln (1+\eta)^2\Big)+2\p\sqrt{1-\p^2}\sin\Big(\frac{1}{2} \ln (1+\eta)^2\Big)\right],\\
\nonu
     \wt W_\c^{(1)}(\p ,\eta ,\c)\!\!&=&\!\!\frac{\a}{\b}\bigg[1-2\p\sin\Big(\frac{\bc}{2}\Big)\cos\Big(\frac{1}{4} \ln (1+\eta)^2\Big)+2\sqrt{1-\p^2}\sin\Big(\frac{\bc}{2}\Big)\sin\Big(\frac{1}{4} \ln (1+\eta)^2\Big)\bigg],\\\label{7potencialchi}
\er
and 
\br
\nonumber
   \wh W_\eta^{(1)}(\eta)\!\!\! &=&\!\!\!\a(1+\eta)\cos\!\Big(\frac{1}{2} \ln (1+\eta)^2\Big),\qquad\quad \,  \wh W_\eta^{(1)}(\eta, \c)= \a(1+\eta)\cos(\bc),    \\
\nonumber     
\wh W_\eta^{(1)}(\c) \!\!\! &=&\!\! \!\a \,e^{\bc}\cos(\bc), \qquad \qquad \qquad \qquad \quad \! \wh W_\eta^{(1)}(\p,\c) = \a \, e^{\bc}\sqrt{1-\p^2}, \\
 \nonumber
   \wh W_\eta^{(1)}(\p)\!\!\! &=&\!\!\!\a\, e^{\arcsin(\p)}\sqrt{1-\p^2}, \qquad \qquad \qquad \quad \!\! \wh W_\eta^{(1)}(\p,\eta) = \a\sqrt{1-\p^2}(1+\eta),\\
 \wh W_\eta^{(1)}\!(\p ,\eta ,\c)\!\! \!&=&\!\! \!\a(1+\eta)\!\bigg[1- 2\sin\Big(\!\frac{\bc}{2}\Big)\!\!\left(\p \!\cos\!\Big(\frac{1}{4} \ln (1+\eta)^2\Big)  \! +\!\sqrt{1-\p^2}\sin\!\Big(\frac{1}{4} \ln (1+\eta)^2\Big)\!\right)\!\!\bigg]. \qquad \quad \mbox{}\label{7potencialp}   
\er
Now, from above parametrizations we can derive explicitly the functions $g$, $\tilde{g}$, and $\hat{g}$ (see  \mbox{appendix \ref{3campos}}  for more details of the full derivation). After doing that, we find
\begin{eqnarray}
\nonumber
W_\p^{(3)}(\p,\eta,\c)&=&\a(1-a_{4}-a_{5})(1-\p^2)+\a a_{4}\cos^2\Big(\frac{1}{2} \ln (1+\eta)^2\Big) \nonu\\
&&+\a a_{5}\left[1-\p\sin\Big(\frac{1}{2} \ln (1+\eta)^2\Big)\right], \label{w3vp}\\ 
W_\c^{(3)}(\p,\eta,\c) &=& \frac{\a}{\b}\cos(\bc) +\a\b c_{1}\,e^{\bc}(1+\eta-e^{\bc})\big(\cos(\bc)-\sin(\bc)\big),  \nonu\\
&&-\frac{\a\b c_{6}}{2}\sin(\bc)\left((1+\eta)^2 - e^{2\bc}\right), \label{w3c}\\
W_\eta^{(3)}(\p,\eta,\c) &=& \a c_{1}\,e^{\bc}\cos(\bc)+\a(1-c_{1}-c_{6})(1+\eta)\cos\Big(\frac{1}{2} \ln (1+\eta)^2\Big)\nonu\\
&&+\a c_{6}(1+\eta)\cos(\bc)+\frac{\a a_{4}}{(1+\eta)}\sin\big( \ln (1+\eta)^2\big)\left[\sin\Big(\frac{1}{2} \ln (1+\eta)^2\Big)-\p\right]\nonu\\
&&+\frac{\a a_{5}}{2(1+\eta)}\cos\Big(\frac{1}{2} \ln (1+\eta)^2\Big)\left[\sin^2\Big(\frac{1}{2} \ln (1+\eta)^2\Big)-\p^2\right],
\label{w3p}
\end{eqnarray}
which after being integrated lead us to the three-field superpotential 
\br
\nonumber
W^{(3)}(\p,\eta,\c) &=& \a(1-a_{4}-a_{5})\left(\p-\frac{\p^3}{3}\right) +\frac{\a}{\b^2}\sin(\bc) \\
\nonumber
&&+\frac{\a}{5}(1-c_{1}-c_{6})(1+\eta)^2\left[2\cos\Big(\frac{1}{2} \ln (1+\eta)^2\Big)+\sin\Big(\frac{1}{2} \ln (1+\eta)^2\Big)\right]\nonu\\
&& + \a a_{4}\left[\p\cos^2\Big(\frac{1}{2} \ln (1+\eta)^2\Big)+\frac{2}{3}\sin^3\Big(\frac{1}{2} \ln (1+\eta)^2\Big)\right] \nonu\\
&&+\a a_{5}\left[\p-\frac{\p^2}{2}\sin\Big(\frac{1}{2} \ln (1+\eta)^2\Big)+\frac{1}{6}\sin^3\Big(\frac{1}{2} \ln (1+\eta)^2\Big)\right]\nonu\\
&&+\frac{\a c_{1}}{5}\,e^{\bc}\left[\big(5(1+\eta)-3e^{\bc}\big)\cos(\bc)+e^{\bc}\sin(\bc)\right]\nonu\\
&&+\frac{\a c_{6}}{10}\left[\big(5(1+\eta)^2-e^{2\bc}\big)\cos(\bc)+2e^{2\bc}\sin(\bc)\right].
\label{potencialvppc}
\er
This extended three-field superpotential describes the coupling of $\p^4$, sine-Gordon, and the E-model, where the static configurations (\ref{eqn6.31}) are BPS solutions of this superpotential connecting the minima $m_1=(-1,-\frac{\pi}{2\b}, -1 + e^{-\pi/2})$  and $m_2 =(+1,+\frac{\pi}{2\b}, -1 + e^{+\pi/2})$, with BPS energy 
\br
\label{evppc}
E_{\mbox{\tiny BPS}}= {2\a}\left(\frac{2}{3}+\frac{1}{\b^2}+\frac{1}{5}\cosh(\pi)\right).
\er
Now, regarding the linear stability of these BPS solutions, we compute the eigenvalues of the matrix ${\bw}$ by choosing $a_5= -2a_4$ and $c_6 =-c_1$,  leading us to 
\br
\label{p4sGEu+}
u_{0} &=& -2\a\tanh(\a x),\\
\nonu
u_\pm &=& -\frac{\a}{2}\sech(\a x)\Big[2\sinh(\a x)-1+c_{1} \left(1+\b^2 e^{2\sin^{-1}(\tanh(\a x))}\right)\\
\label{p4sGEu0}
&& \pm \sqrt{1+2c_{1}\left(\b^2 e^{2\sin^{-1}(\tanh(\a x))}-1\right)+c_{1}^2\left(1+\b^2 e^{-2\sin^{-1}(\tanh(\a x))}\right)^2}\Big].
\er
We see that the first quantum-mechanical potential derived from eq. (\ref{p4sGEu+}) is simply given by
\br
\label{potencialp4sGEU+}
U_{0}(x) = 4\a^2 - 6\a^2\sech^2(\a x),
\er
whose energy eigenvalues are $E_0 = 0$ and $E_1 = 3 \a^2$, which partially guarantees stability. However, the others two potentials $U_\pm$ have complicated forms (see figure \ref{figel3pGEU0}), which as a consequence arises some difficulties in obtaining analytical results, except for the $c_1=0$ case which decouples the sine-Gordon field. Instead of that,  we will perform an approximated analysis for those cases. We see from the plots of the potentials in figure \ref{figel3U0a31a36menor1} that for small values of $c_1$ these potentials approximate to Rosen-Morse II and Scarf II profiles, respectively.

\begin{figure}[t!]
\centering
\includegraphics[width=\linewidth]{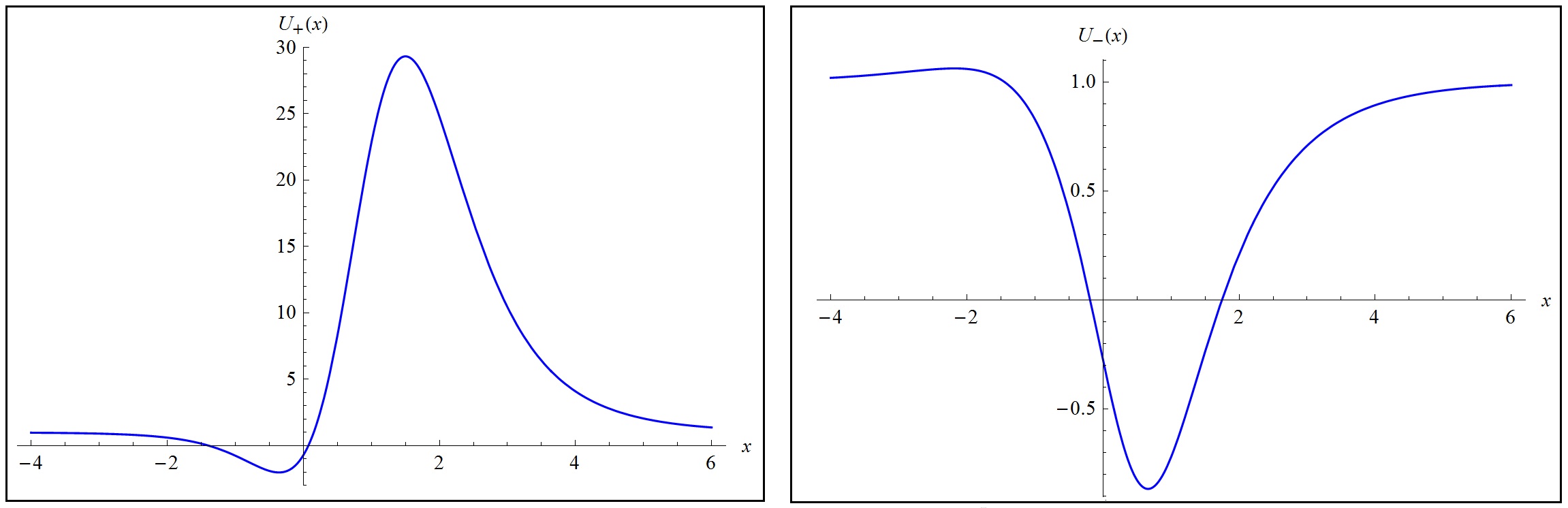}
\caption{Quantum-mechanical potentials $U_{+}$ (on the left) and $U_-$ (on the right) for $\a=\b=c_1=1$.}
\label{figel3pGEU0}
\end{figure}
Let us consider small values of the parameter, that is $c_1=\l \lesssim 10^{-2}$, so we obtain the approximated potentials up to first-order,
\br 
 U_\pm(x) = U_\pm^{(0)}(x) +\l\,  U_\pm^{(1)}(x),
\er
where the unperturbed potentials are given by
\br
 U_+^{(0)}(x)= \a^2 - 2\a^2 \sech^2(\a x), \qquad U_-^{(0)}(x) = \a^2 - \a^2 \sech^2(\a x) - 3\a^2 \sech(\a x)\tanh(\a x),
\er
while the first-order corrections are,
\br
U_+^{(1)}(x) &=& \a^2\b^2 e^{2\sin^{-1}(\tanh(\a x))} \big(3\sech(\a x)\tanh(\a x)-2\sech^2(\a x)\big),\nonu \\
\nonu
U_{-}^{(1)}(x) &=& \a^2 \big(3\sech(\a x)\tanh(\a x)-2\sech^2(\a x)\big).\label{psGEUopertubacaoU-}
\er
We notice that the unperturbed potential $U_+^{(0)}$ is described by the potential Rosen-Morse II with $A = \a$ and $B = 0$, while $U_{-}^{(0)}$ is described by the potential Scarf II with $A = \a$ and $B = -\a$.  Both potentials only possess one bound state, the zero mode $E_0^{(0)}=0$. Therefore, the corresponding first-order correction $E_0^{(1)}$ to the zero energy  will be obtained from
\br
\label{perturbadopsGE}
E_{0}^{(1)}=\int_{-\infty}^{\infty}\Big[\s^{*}_{0}(x)U_{+}^{(0)}(x)\s_{0}(x)+\xi^{*}_{0}(x)U_{-}^{(0)}(x)\xi_{0}(x)\Big]dx,
\er
where the zero mode eigenfunctions  $\s_0 (x)$ and $\xi_{0}(x)$ can be computed from eqs. (\ref{ondarosen}) and (\ref{ondascarf}), respectively. Computing explicitly the integral we can easily verified that $E_{0}^{(1)} = 0$,  and then we can ensure the stability of the solutions up to first-order approximation.
\begin{figure}[t]
\centering
\includegraphics[width=\linewidth]{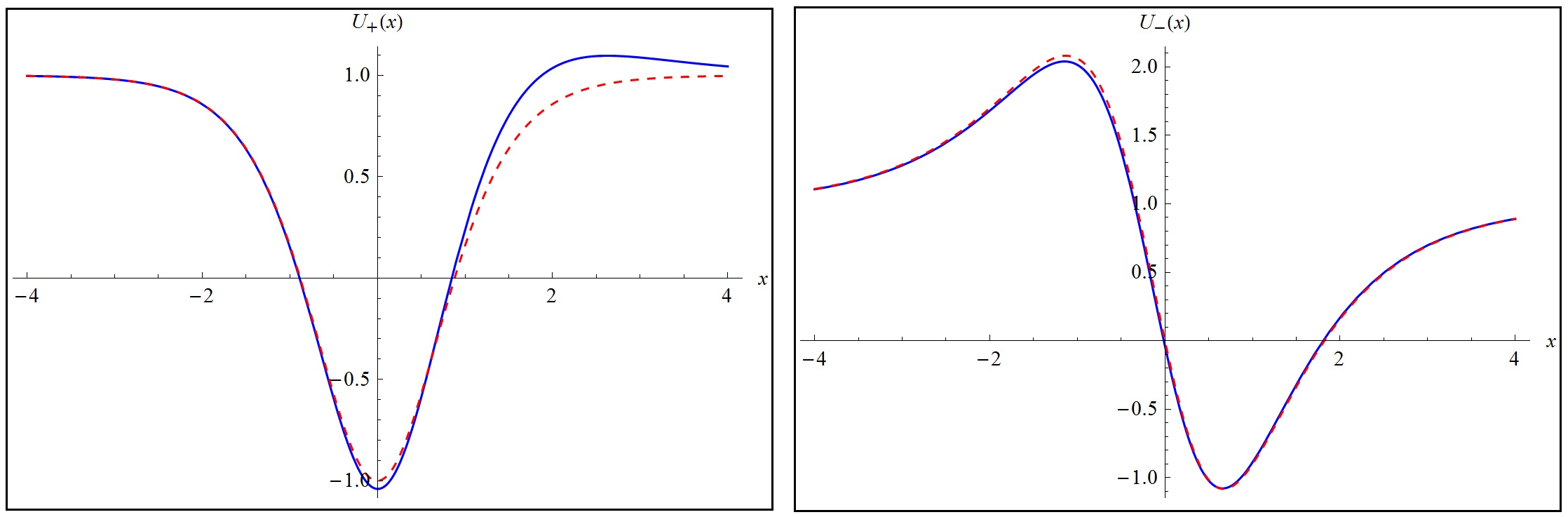}
\caption{Quantum-mechanical potentials $U_{+}$ (on the left) and $U_-$ (on the right) for $\a =\b=1$. For both, we have plotted the values  $c_{1} = 0$ (dashed red lines) and  $c_{1} =  0.02$ (blue solid lines).}
\label{figel3U0a31a36menor1}
\end{figure}
%

On the other hand, it is clear that there is enough room for several different topological sectors depending on the values of the four arbitrary parameters. In particular, if we chose $c_{1}=c_{6}=0$, we get the following two BPS solutions,
\begin{eqnarray}
 \eta^{(\pm)} (x) &=& -1 + e^{\pm \pi/2},\qquad \chi(x) = \frac{1}{\b}\arctan(\sinh(\a x)), \nonu
\end{eqnarray}
\begin{equation}
 \phi^{(\pm)}(x)= \frac{(1-a_4) \pm e^{(\pm 2(1-a_4)\mp a_5)\a x}  }{\pm(a_5- (1-a_4)) +  e^{(\pm 2(1-a_4)\mp a_5)\a x} }, \label{eq6.50}
\end{equation}
providing that $a_5\neq 2(1-a_4)$, and also that $a_5>(1-a_4)$ for $\phi^{(+)}(x)$, and $a_5<(1-a_4)$ for $\phi^{(-)}(x)$.
The kink solutions for the $\phi^4$ field in eq. (\ref{eq6.50}) have the very same form as the ones previously found eqs. (\ref{sol3.58}) and (\ref{sol3.60}) for the extended ($\phi^4$+sG) model, and also interpolates between the values $\pm 1$ and $ \frac{\pm(1-a_4)}{a_5-(1-a_4)}$, respectively. Depending on the values of the parameters, we will have two topological sectors with the corresponding BPS energies given by 
\br 
 E_{\mbox{\tiny{BPS}}}^{(\pm)}= \left|\frac{2\a}{\b^2} \pm \frac{\a}{6}\frac{\(a_5 -2(1-a_4)\)^3}{\(a_5 - (1-a_4)\)^2}  \right|.
\er
By analysing the stability of the $\p^{(-)}$ solution, we will find that the associated $\bw$ matrix reads,
\br
 \label{matriznovasolucoes}
\bw = \left ( \begin{array}{ccc}
2\a(a_{4}-1)\p^{(-)} & 0 & 0\\
0 & -\a \sin(\bc) & 0\\
0 & 0 & 2\a e^{\pi}a_{4}(1+\p^{(-)})+\a\\
\end{array} \right ),
\er
where we have chosen $a_5=0$ for simplicity, and therefore we have that $a_4<1$. The quantum-mechanical potentials will be given by
\br
\label{potencialU+novasolucoes}
&&U_{+}(x)=4\a^2(a_{4}-1)^2\frac{\left[(a_{4}-1)^2+4(a_{4}-1)e^{2\a(a_{14}-1)x}+e^{4\a(a_{4}-1)x}\right]}{\left[1-a_{4}+e^{2\a(a_{4}-1)x}\right]^2},\\
\label{potencialU0novasolucoes}
&&U_{0}(x)=\a^2-2\a^2\sech^2(\a x),\\
\label{potencialU-novasolucoes}
&&U_{-}(x)=\frac{8\a^2 a_{4}(a_{4}-1)^2 e^{\pi + 2\a (a_{4}-1) x}}{\left[1-a_{4}+e^{2\a(a_{4}-1)x}\right]^2}+\left(\a-\frac{4\a e^{\pi}a_{4}(a_{4}-1)}{1-a_{4}+e^{2\a(a_{4}-1)x}}\right)^2,
\er
from where we immediately see that the potential $U_0$ only possesses the eigenvalue $E_0=0$. In its turn, we can verify that the potentials $U_\pm$ can be described by shifted Rosen-Morse II potentials, namely
\br 
\label{potencialRMnovasolucoes}
&&U_\pm(x) = A_\pm^2+\frac{B_\pm^2}{A_\pm^2}-A_\pm(A_\pm +\kappa)\sech^2[\kappa (x-x_{0}^\pm)]+2B_\pm\tanh[\kappa(x-x_{0}^\pm)],
\er
where $\kappa = \a (1-a_4)$, and the parameters are
\br
\label{parametrosparanovasolucoesU+}
A_+=2\a(1-a_{4}),\qquad B_+=0,\qquad x_{0}^+=\frac{1}{{2\a(a_{4}-1)}}{\ln{\left(1-a_{4}\right)}},
\er
and 
\br
\nonu
A_-&=&-2\a e^{\pi}a_{4},\qquad B_-=-2\a^2 e^{\pi}a_{4}(1+2e^{\pi}a_{4}),\qquad \\[0.15cm]
\label{parametrosparanovasolucoesU-}
 x_{0}^- &=&\frac{1}{2\a (a_4-1)}\ln{\left(\frac{(1+4e^{\pi})^2a_{4}^2(1-a_{4})}{(2-a_4)^2}\right)}.
\er
We find that the potential $U_+$ possesses the eigenvalues  $E_{0} = 0$ and $E_{1} = 3 \a^2(1-a_{4})^2$, whereas the potential $U_-$ only possesses the eigenvalue $E_0=0$, if we have that
\br
\label{intervaloa14}
-\frac{e^{-\pi}}{2} < a_{4} < -\frac{e^{-\pi}}{4}.
\er
Thus, this particular solution is stable only if the parameter $a_4$ satisfies the constraint (\ref{intervaloa14}), at least for our choice of parameters. Following an analogous procedure, the stability analysis of the solution $\p^{(+)}$ will lead us to similar results.


\subsection{$\p^4$ model coupled to two sine-Gordon models}
\label{se6.3}
In this last example, we will construct a three-field system that couples the $\phi^4$ field with two different sine-Gordon fields $\chi$ and $\psi$. The  first-order equations are  
\br
 \phi'(x) \!\!&=&\!\! W_\p^{(1)} = \,\a (1-\phi^2),\\
 \chi'(x) \!\!&=&\!\! \wt{W}^{(1)}_\chi =\, \frac{\a}{\b} \cos(\b \chi),\\
 \psi'(x) \!\!&=&\!\! \wh{W}^{(1)}_\psi = \,\frac{\a}{\g} \cos(\g \psi),
\er
and their corresponding static solutions are
\br
 \phi(x) = \tanh(\a x), \qquad \chi(x) = \frac{1}{\b}\arctan(\sinh(\a x), \qquad \psi(x) = \frac{1}{\g}\arctan(\sinh(\a x). \label{eq6.42}
\er
The deformation functions are, 
\br 
 \phi \!\!\!&=&\!\! \! f_1(\chi) \,=\, \sin(\b \chi), \\
 \phi \!\!\!&=&\!\! \! f_2(\psi) \,=\, \sin(\g \psi), \\
 \chi \!\!\!&=&\!\! \! f_3(\psi)\,=\, \frac{\g}{\b} \psi.
\er
As it was already done in the previous models, we use these functions to write the following equivalent expressions,
\br
 W_\p^{(1)} (\p) &=& \a(1-\p^2), \qquad \qquad \,\, \, W_\p^{(1)} (\p,\c)= \a\(1-\phi \sin(\bc)\),\nonu\\
  {W}_\p^{(1)}(\c)  &=& \a \cos^2(\bc),\qquad \qquad W_\p^{(1)} (\chi,\psi)= \a\(1- \sin(\bc)\sin(\gp)\), \nonu\\
  {W}_\p^{(1)}(\psi) &=& \a \cos^2(\gp), \qquad  \qquad W_\p^{(1)} (\p,\psi) = \a\(1- \phi\sin(\gp)\), \nonu\\
   {W}_\p^{(1)}(\p,\c,\psi)&=& \a\left(1-2\p\sin\Big(\frac{\bc}{2}\Big)\cos\Big(\frac{\gamma \psi}{2}\Big)\right),\label{eqn6.63}
\er
and%
\br
 \wt{W}^{(1)}_\c (\chi) &=& \frac{\a}{\b}\cos(\bc),\qquad\qquad  \wt{W}^{(1)}_\c (\c,\p) = \frac{\a}{\b} \sqrt{1-\p \sin(\bc)}, \nonu\\
 \wt{W}^{(1)}_\c (\p) &=& \frac{\a}{\b}\sqrt{1-\p^2}, \qquad \qquad \!\! \wt{W}^{(1)}_\c (\p,\psi) = \frac{\a}{\b} \sqrt{1-\p\sin(\gp)},\nonu \\
  \wt{W}^{(1)}_\c (\psi) &=& \frac{\a}{\b}\cos(\gp), \qquad \qquad \! \wt{W}^{(1)}_\c (\c,\psi) = \frac{\a}{\b} \sqrt{1-\sin(\bc)\sin(\gp)},\nonu \\
  \wt{W}^{(1)}_\c (\p,\c,\psi) &=&\frac{\a}{\b} \big(\cos(\bc)\cos^2(\gp) - \p^2\cos(\bc) +2\p \sin(\bc)\cos(\gp)\big),
\er
and also
\br
 \wh{W}^{(1)}_\psi (\psi ) &=& \frac{\a}{\g}\cos(\gp),\qquad\qquad  \wh{W}^{(1)}_\psi  (\psi,\p) = \frac{\a}{\g} \sqrt{1-\p \sin(\gp)}, \nonu\\
 \wh{W}^{(1)}_\psi  (\p) &=& \frac{\a}{\g}\sqrt{1-\p^2}, \qquad \qquad \!\! \wh{W}^{(1)}_\psi  (\p,\c) = \frac{\a}{\g} \sqrt{1-\p\sin(\bc)},\nonu \\
  \wh{W}^{(1)}_\psi  (\chi) &=& \frac{\a}{\g}\cos(\bc), \qquad \qquad \! \wh{W}^{(1)}_\psi  (\c,\psi) = \frac{\a}{\g} \sqrt{1-\sin(\bc)\sin(\gp)},\nonu \\
  \wh{W}^{(1)}_\psi  (\p,\c,\psi) &=&\frac{\a}{\g} \big(\cos(\gp)\cos^2(\bc) - \p^2\cos(\gp) +2\p \sin(\gp)\cos(\bc)\big).\label{eqn6.65}
\er
As before, we use all of these expressions to obtain the corresponding $g$-functions (see  details in \mbox{appendix \ref{3campos}}), and then by substituting the results in eqs. (\ref{fsmodeloextendidoAparapvpz}) -- (\ref{fsmodeloextendidoCparapvpz}), we have
\br 
 W_\p^{(3)} (\p,\c,\psi) &=&  \a(1-\phi^2) +\a a_1 \left(\phi^2-\sin^2(\bc) \) +\a a_2 \p\left(\p-\sin(\bc)\) +\a a_4\left(\p^2-\sin^2(\gp)\) \nonu\\ 
 && +\a a_5 \p\left(\p - \sin(\gp)\)                  ,\\[0.1cm]
  W_\chi^{(3)} (\p,\c,\psi) &=& \frac{\a}{\b}(1-b_4) \cos(\bc) +\a\b \left(2a_1 +\frac{a_2}{2}\right)\sin^2(\bc) \cos(\bc)    + \frac{\a b_4}{\b} \cos(\gp)\nonu \\
  && -\a\b a_1 \phi \sin(2\bc) -\frac{\a\b a_2}{2} \phi^2\cos(\bc),\\[0.1cm]
W_\psi^{(3)} (\p,\c,\psi) &=&   \frac{\a}{\g}\cos(\gp) +\a\g \left(2a_4+\frac{a_5}{2}\)\sin^2(\gp)\cos(\gp) - \a\g a_4 \phi \sin(2\gp)\nonu \\
&& -\frac{\a \g a_5}{2} \phi^2 \cos(\gp) + \frac{\a \g b_4}{\b^2 } \left(\g\psi - \b\chi  \)\sin(\gp),
\er
with the corresponding superpotential given by
\br
W^{(3)}  (\p,\c,\psi)  &=&  \a\p - \a \left(1-a_1-a_2-a_4-a_5\)\frac{\p^3}{3} - \a a_1 \p \sin^2(\bc) - \frac{\a a_2}{2} \p^2\sin(\bc) \nonu\\
&& -\a a_4 \phi \sin^2(\gp) -\frac{\a a_5}{2} \phi^2 \sin(\gp) +\frac{\a}{\b^2} (1-b_4) \sin(\bc) \nonu \\
&& + \frac{\a}{3} \left(2a_1+\frac{a_2}{2}\)\sin^3(\bc) +\frac{\a b_4}{\b^2}\left(\bc - \gp \) \cos(\gp) + \frac{\a}{\g^2}\left(1+\frac{b_4 \g^2}{\b^2}\) \sin(\gp)  \nonu\\
&& +\frac{\a}{3}\left(2a_4 +\frac{a_5}{2}\)\sin^3(\gp). \label{eq6.52}
\er
This new extended three-field superpotential describes the coupling of the $\p^4$  field with two different sine-Gordon fields, and will be named as the {extended} ($\phi^4$+sG$_1$+sG$_2$) {model}. The static solutions (\ref{eq6.42}) are BPS solutions of its first-order equations, connecting the minima $m_1=(-1,-\frac{\pi}{2\b},-\frac{\pi}{2\g})$  and $m_2 =(1,\frac{\pi}{2\b}, \frac{\pi}{2\g})$, with BPS energy given by
\br
E_{\mbox{\tiny BPS}}=  {2\a}\left(\frac{2}{3}+\frac{1}{\b^2}+\frac{1}{\g^2}\right).
\er
Now, in order to analyse linear stability of the BPS solutions (\ref{eq6.42}), we find that in this case the corresponding matrix $\bw$ takes the following form,
\br
\label{matrizp4sGsG}
\bw = \left ( \begin{array}{ccc}
-2 \a \tanh(a x) & 0 & 0\\
0 & \a(b_{4}-1)\tanh(\a x) & -\frac{\a\gamma b_{4}}{\b}\tanh(\a x)\\
0 & -\frac{\a\gamma b_{4}}{\b}\tanh(\a x) & \a\left(\frac{\gamma^2 b_{4}}{\b^2}-1\right)\tanh(\a x)\\
\end{array} \right ),
\er
where we have chosen $a_5=-2a_4$ and $a_2=-2a_1$, for simplicity. By diagonalizing this matrix, we will find the following eigenvalues
\br
\label{p4sGsGu+}
u_{0} = -2\a\tanh(\a x),\qquad 
u_{+} = -\a\tanh(\a x),\qquad 
u_{-} = -\mu \tanh(\a x),\quad \mbox{}
\er
where in this case the parameter $\mu = \a - \a b_4 (1+\frac{\g^2}{\b^2})$.  Then, the corresponding quantum-mechanical potentials will be given as follows,
\br
U_{0} = 4\a^2-6\a^2\sech^2(\a x),\quad 
U_{+} =\a^2-2\a^2\sech^2(\a x),\quad 
U_{-} = \mu^2- \mu(\mu +\a)\sech^2(\a x).\quad \mbox{}\label{potencialp4sGsGU-}
\er
which are again Rosen-Morse II potentials. We see that the potential $U_0$ is the same as the one in eq.(\ref{potencialp4vp4Iz6U+}), and has the eigenvalues $E_0=0$ and $E_1 = 3 \a^2$.  The parameters for the potential $U_+$ are $A = \a$ and $B = 0$, and has only one eigenvalue, $E_0 = 0$. For the potential $U_-$ the parameters are $A = \mu$ and $B = 0$. In this case, the  number of eigenvalues will be now constrained by $0 \leq k<1- b_4 (1+\frac{\g^2}{\b^2})$, which requires that  $b_{4} <\frac{\b^2}{(\b^2 + \gamma^2)}$ in order to guarantee stability. Therefore, when $0<b_{4}<\frac{\b^2}{(\b^2 + \gamma^2)}$ there exists only one eigenvalue $E_0 = 0$. For $b_{4} <0$, we note that the number of bound states increases for  decreasing $b_4$, and then the potential could have more than one non-negative eigenvalue, guaranteeing in this way the stability of the BPS solutions.
\begin{figure}[t!]
\centering
\includegraphics[scale=0.3]{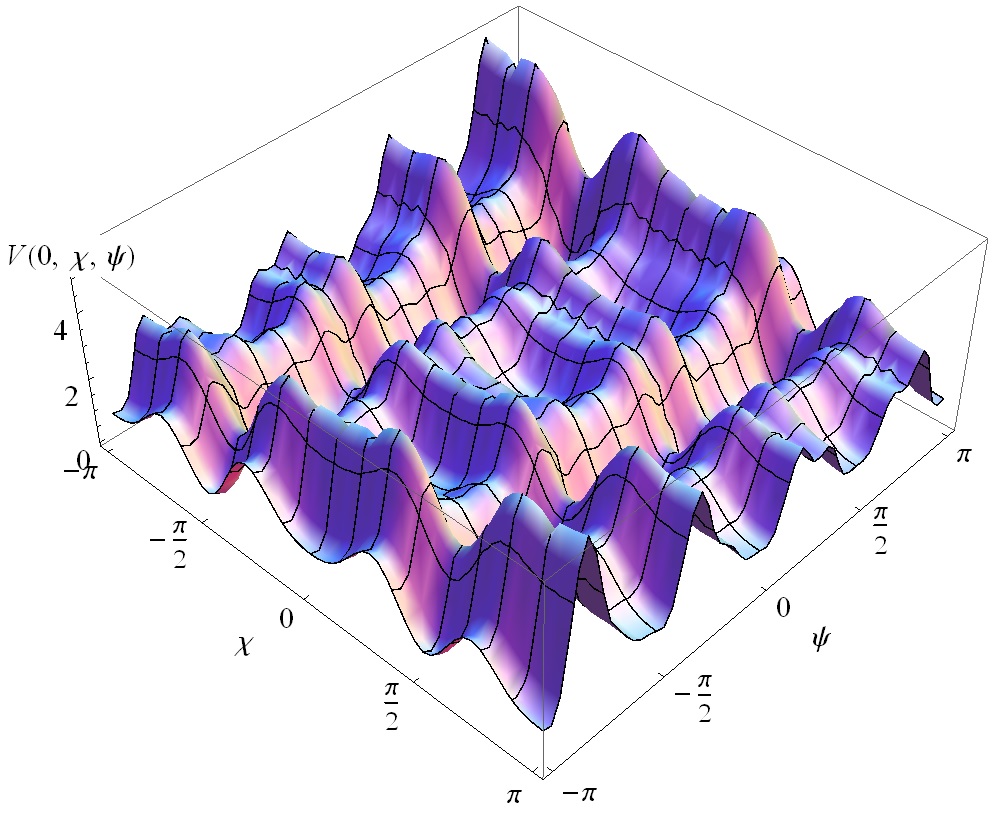}
\caption{Plot of the projection of the potential $V(0,\chi,\psi)$ for the values $\a=1$, $\b=1$, $\g=2$, $a_1 =2$ and  $b_4=0.1$.}
\label{figpropotsine}
\end{figure}

\mbox{}\\

\noindent There are also several others interesting features that can be mentioned about this new model. For instance, the projection of the corresponding potential in the $(\chi,\psi)$ plane gives the following,
\br
 V(0,\chi,\psi) &=&  \frac{\a^2}{2}\left[  \frac{1}{\b^2}\(1-b_4+2a_1 \b^2\)^2 \cos^2(\bc)  + \(\frac{b_4^2}{\b^2}+\frac{\(1+2\g^2(1-a_1)\)^2}{\g^2} \)   \cos^2(\gp)     \right. \nonu\\
 &&\left.  \qquad + \frac{2b_4}{\b^2}\(1-b_4 +2a_1\b^2\)\cos(\bc)\cos(\gp) +2a_1 (1-a_1) \cos^2(\bc)\cos^2(\gp) \right. \nonu\\
 &&\left. \qquad  -4a_1b_4 \cos^3(\bc)\cos(\gp) + a_1\(a_1+4(b_4 -1) -8a_1\b^2\)\cos^4(\bc) \right. \nonu\\
 &&\left. \qquad  + (1-a_1)\((1-a_1)(1-8\g^2) -4\) \cos^4(\gp) + 4a_1^2 \b^2 \cos^6(\bc)  \right. \nonu\\
 &&\left.  \qquad   + 4 \g^2(1-a_1)^2\cos^6(\gp) - \frac{b_4}{\b^2}\big(\b\chi-\g\psi\big) \sin(2\gp)\(1+2(1-a_1)\g^2 \sin^2(\gp)\) \right. \nonu\\
 &&\left. \qquad  +\frac{\g^2b_4^2}{\b^4}\big(\bc-\gp\big)^2 \sin^2(\gp)
 \right],\label{eqn6.64}
\er
where we have considered $a_2 = a_5 =0$, and $a_4=1-a_1$, without loss of generality. It is worth pointing out that this potential is not BPS, and even though its minima are located at 
\br 
m_k = \left(\frac{\pi}{2\b}(2k-1), \frac{\pi}{2\g}(2k-1)\), \qquad k \in \mathbb{Z},
\er
the static sine-Gordon kinks are no longer solutions of its field equations. Despite of being an interesting potential (see figure \ref{figpropotsine}), we have not been able to find any explicit analytical solutions for it.  It would be interesting to look for at least numerical solutions and also further explore this potential. That could be addressed in more detail in another work. 
%
%
%

On the other hand, when substituting $\phi= \pm 1$ directly in (\ref{eq6.52}), and setting $a_1=a_2=a_4=a_5=0$, we  end up with a different effective two-fields superpotential,  and its corresponding potential, given by
\br 
W^{(2)}_{\mbox{\tiny eff}}(\chi,\psi)& =& \frac{\alpha  (1-b_4)}{\b^2} \sin (\beta  \chi ) +{\alpha} \left(\frac{1}{\g^2}+\frac{b_4}{\beta ^2}\right) \sin (\gamma  \psi )+\frac{\alpha  b_4}{\b^2}  (\beta  \chi -\gamma  \psi )\cos (\gamma  \psi ), \qquad \mbox{}\label{eq6.54}\\
V_{\mbox{\tiny eff}}(\chi,\psi)& =&  \frac{\a^2}{2}\left[  \frac{1}{\b^2}\(1-b_4\)^2 \cos^2(\bc)  + \(\frac{b_4^2}{\b^2}+\frac{1}{\g^2} \)   \cos^2(\gp)    + \frac{2b_4}{\b^2}\(1-b_4 \)\cos(\bc)\cos(\gp)  \right. \nonu\\
 &&\left.  \qquad   - \frac{b_4}{\b^2}\big(\b\chi-\g\psi\big) \sin(2\gp)+\frac{\g^2b_4^2}{\b^4}\big(\bc-\gp\big)^2 \sin^2(\gp) \right]. \label{eqn6.67}
\er 
Note that although this potential  is somehow contained within the projection $V(0,\chi,\psi)$,  they are  actually different even if we set $a_1=0$ in eq. (\ref{eqn6.64}), and in this case the static solutions for the sine-Gordon fields given in (\ref{eq6.42}) are BPS solutions of the first-order equation for the effective  superpotential (\ref{eq6.54}). It is worth also noting that the simple coupling  between the two sine-Gordon fields contained in the last term of eq. (\ref{eq6.54}) differs from some models previously constructed  in  the literature. In particular, if we eliminate the coupling term by setting $b_4=0$,  then our potential will take the form of the non-integrable two-frequency sine-Gordon model considered in \cite{Delfino}, where the authors studied how the particle spectrum of the model changes by considering the second interaction as a perturbation of the original integrable sine-Gordon model. In addition,  after proper redefinitions
our superpotential  (\ref{eq6.54}), also with $b_4=0$, can be also seen as a limit case of the FKZ (Ferreira, Klimas, and Zakrewski) pre-potential based on the $SU(3)$ Lie algebra\footnote{In fact, the model contains three parameters $\g_1, \g_2,$ and $\g_3$, and then the exact equivalence will requires that the latter one vanishes.}  \cite{Luiz}. However, their potential $V$ will be quite different since a constant, real and positive-definite matrix $\eta_{ab}$, which is basically a modified version of the associated Cartan matrix, is directly involved in the definition of the FKZ models. Despite of these differences, it would be interesting to analyse if there exist any common points between the two methods of constructing multi-scalar field theories. This issue will be addressed in future investigations.
\begin{figure}[t!]
\centering
\includegraphics[width=\linewidth]{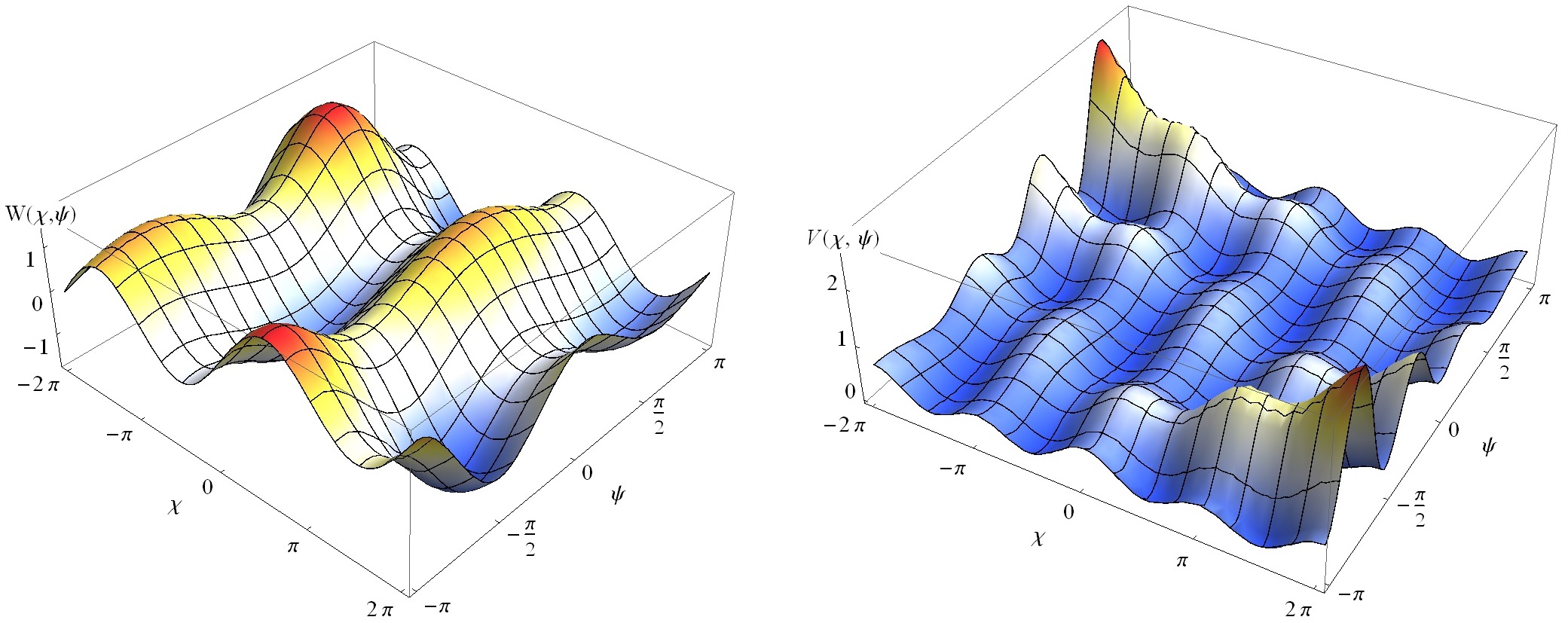}
\caption{Effective two-fields superpotential (on the left) and the  associated potential (on the right) for the two coupled sine-Gordon fields. For both, we have plotted the values $\a=1$, $\b=1$, $\g=2$, and $b_4=0.1$.}
\label{figel2numero7}
\end{figure}


\section{Final remarks}
\label{remarks}

In this paper, we have presented the explicit construction of several interesting new models described by two and three real scalar fields theories in ($1+1$)-dimensions supporting BPS states. The way of constructing  such field theories is called the extension method, which was introduced originally in \cite{santos13, santos2018}. This method requires considering initially several (not necessarily different) one-field systems which are known to support BPS states, and that are also connected through some mappings called deformation functions. Then, the corresponding first-order equations are rewritten in several different but  equivalent non-trivial ways by  using such functions and their inverse functions. Doing that, the fields are then coupled by introducing an ansatz for the first-order equations for the resulting two-field model eqs. (\ref{modeloextendidoA}) and (\ref{modeloextendidoB}), and  respectively for three-field model eqs. (\ref{fsmodeloextendidoAparapvpz})--(\ref{fsmodeloextendidoCparapvpz}). To finish the procedure, some functions, called here as $g$-functions, are then introduced in order to guarantee smoothness of the superpotential, which are properly derived from consistency constraints (\ref{fscondicaodecontinuidade}).

The constructed theories were obtained by coupling basically some known one-scalar field BPS models, namely $\phi^4$ model,  the $\vp^6$-like model, the sine-Gordon model, the E-model, and finally the inverse $\z^{4I}$ model. One of the most important advantages of this method of constructing multi-fields models is that it maintains the BPS solutions of the original one-field systems. However, they are not the only possible BPS solutions for the multi-field superpotential. In fact, in some cases we have been able to find analytically (or numerically) other BPS solutions for the resulting model. We have also studied in some details the linear stability of the BPS states for the resulting multi-scalar superpotential. In general, these studies lead us with two very-well known exactly solvable quantum-mechanical problems, the Rosen-Morse II and the Scarf II potentials. For several choices of the potential parameters we have been able to perform analytically such analysis, and have found that they are stable with respect of small perturbations. However, in some cases the problem is somehow complicated and we have only been able to study in a qualitative and approximated way, with no full guarantee of the stability. Of course, such analysis could be improved by performing proper numerical simulations. Those investigations represent the next step in our studies on multi-scalar field theories and will be done in future works.

There are several other interesting issues that can be addressed in next investigations from our results. For instance, a more complete numerical study of the solutions and their stability, specially two-solitons solutions, would provide a good scenario for investigating the behaviour during kink collisions. In addition, that kind of analysis also could bring some additional information that allow us identify possible quasi-integrable multi-scalar models \cite{quasi1, quasi2}. In particular, we are interested in the two coupled sine-Gordon model obtained in section \ref{se6.3}, which are slightly  related to the FKZ models. We believe that a more detailed analysis would give some interesting connections between the two methods and probably help us answer some unsolved problems from both sides.

Finally, one more question of interest involves the investigations of possible supersymmetric generalization of the extension method. As it is well-known the interest on the study of supersymmetric kinks has a long history, and essentially concerns with the calculations of quantum corrections to the kink mass,  and the central charge \cite{susykink}, \cite{Nastase}--\cite{Shifman2}. Therefore, it will be interesting to construct general supersymmetric field theories by using the extension method, especially for the ones that possess intrinsically infinite  number of degenerate vacua, as it is the case of sine-Gordon and the E-model. These issues are also currently under investigations.\\[0.3cm]


\newpage
\noindent
{\bf Acknowledgements} \\
\vskip .1cm \noindent
{Authors would like to thank to CAPES-Brazil for financial support. Authors are also grateful to the Directorate of Innovation and Research of the Federal University of Itajub\'a (DIP-UNIFEI) for partial financial support at the very initial stage of this project. }

%
%
\appendix 
\section{Associated exactly solvable potentials}
\label{exatos}
The very well-known exactly solvable Rosen-Morse II potential (or modified Po\"schl-Teller potential) can be written in the following form \cite{susyqm},
\br
\label{potencialrosenmorse1}
U(x) = A^2 + \frac{B^2}{A^2}  - A(A+\a) \sech^2(\a x)+ 2B\tanh(\a x),
\er
where $\a>0$, and $A$ and $B$ are arbitrary real parameters. The bound states have the following eigenvalues,
\br
\label{energiarosenmorse1}
E_k& =& A^2 + \frac{B^2}{A^2} - (A - k\a)^2 - \frac{B^2}{(A - k\a)^2},  \qquad 0 \leq k < \frac{(A - \sqrt{|B|})}{\a}.
\er
By imposing the stability condition,  we find that  
\br 
A > 0, \qquad \mbox{and}\qquad |B| < A^2. \label{condicaodeestabilidade}
\er
In addition, the corresponding wave eigenfunctions are given by
\br
\psi_k (x) &=& (1-\tanh(\a x))^{(s+t-k)/2} (1+\tanh(\a x))^{(s-t-k)/2} \,P_k^{(s+t-k,s-t-k)}(\tanh(\a x)), \label{ondarosen}
\er
where $P_k^{(\a,\b)}$ are the Jacobi polynomials, and
\br 
s = A/\a, \qquad t=\frac{B}{\a^2 (s-k)}.
\er
Now, let us consider another very well-known exactly solvable potential, namely the Scarf II potential \cite{susyqm},
\br
U(x) = A^2 + \big(B^2-A(A+\a)\big) \sech^2(\a x) + B(2A+\a) \sech(\a x) \tanh(\a x),
\label{potencialscarfII}
\er
where $\a$, $A$, and $B$ are real parameters. Its corresponding bound states possess energy eigenvalues given by
\br
E_k = A^2 - (A-k\a)^2, \qquad 0 \leq k < \frac{A}{\a}. \label{energiascarf}
\er
Their associated eigenfunctions can be written as follows,
\br
\psi_{k}(x) = i^k(\sech(\a x))^{s}\, e^{-u \arctan(\sinh(\a x))}P_k^{(iu-s-1/2,-i u-s-1/2)}(i\sinh(\a x)), \label{ondascarf}
\er
where $P_k^{(\a,\b)}$ are again the Jacobi polynomials, $s = A/\a$, and $u={B}/{\a}$. It is clear that both potentials (\ref{potencialrosenmorse1}) and (\ref{potencialscarfII}) coincides when $B=0$.

\newpage
\section{Calculation of the $g$-functions for three-fields systems}
\label{3campos}

Here, we will present the explicit derivations of the $g$-function for the three-field model constructed in section \ref{3field}. In principle they are arbitrary function constructed in a similar way as the superpotential, by using the deformation functions and the corresponding inverse functions. The specific form will come out of the following constraints
\br
 W_{\p\vp}^{(3)} = W_{\vp\p}^{(3)}, \qquad W_{\p\z}^{(3)} = W_{\z\p}^{(3)}, \qquad W_{\vp\z}^{(3)} = W_{\z\vp}^{(3)},
\label{fscondicaodecontinuidadeparapvpz}
\er
which are basically consistency conditions for the existence of a well-defined continuous superpotential function given by the ansatz (\ref{fsmodeloextendidoAparapvpz}) -- (\ref{fsmodeloextendidoCparapvpz}).


\subsection{The extended ($\p^4+\vp^6_l+\z^{4I}$) model}
Let us start with the derivation of the $g$-functions for the extended ($\p^4+\vp^6_l+\z^{4I}$) model.  From the consistency conditions  eq. (\ref{fscondicaodecontinuidadeparapvpz}), and by using eqs.(\ref{fsfuncaopparapvpz})--(\ref{fsmodeloextendidoCparapvpz}), we obtain the following constraints
\begin{eqnarray}
\nonumber
0\!\!\!&=&\!\!\! p_{1}g_{\vp}(\vp)+p_{2}g_{\vp}(\p,\vp)+ p_{6}g_{\vp}(\vp,\z)+ p_{7}g_{\vp}(\p,\vp,\z)-q_{5}\act{g}_{\p}(\p,\z)- q_{3}\act{g}_{\p}(\p)-q_{2}\act{g}_{\p}(\p,\vp)\\
\nonu
&&\!\!\! -q_{7}\act{g}_{\p}(\p,\vp,\z) -\a\epsilon (a_{2}-2b_{3})\p + \frac{\a \epsilon b_{5}}{\b}\frac{\p\z}{\sqrt{1-\p^2}}+\a b_{2}\vp +2\a\epsilon a_{1}(1-|\vp|)\\
\label{fseqcontinuidadegeral1parapvpz}
&&\!\!\! +\frac{\a \epsilon a_{6}}{\b}\frac{\z (1-|\vp|)}{\sqrt{1-(|\vp|-1)^2}}- \frac{\a \epsilon b_{7}}{2\b}\frac{\z(1-|\vp|)}{\sqrt{1-\p(|\vp|-1)}} -\frac{\a\epsilon a_{7} }{2\b}\frac{\p\z}{\sqrt{1-\p(|\vp|-1)}},
\end{eqnarray}
and
\begin{eqnarray}
\nonumber
0\!\!\! &=&\!\!\! p_{4}g_{\z}(\z)+p_{5}g_{\z}(\p,\z)+p_{6}g_{\z}(\vp,\z) + p_{7}g_{\z}(\p,\vp,\z)-r_{5}\hat g_{\p}(\p,\z)-r_{3}\hat g_{\p}(\p)-r_{2}\hat g_{\p}(\p,\vp)\\
\nonumber
&&\!\!\! -r_{7}\hat g_{\p}(\p,\vp,\z) +\a\left(\frac{2 a_{4}}{\b^2}+c_{5}\right)\z+\frac{\a}{\b} \left(a_{5}+\b^2 c_{3}\right)\sqrt{1-\p^2}+\frac{\a a_7}{\b}\sqrt{1-\p(|\vp|-1)}\\
\label{fseqcontinuidadegeral2parapvpz}
&&\!\!\!-\a\b c_{3}\frac{\p^2}{\sqrt{1-\p^2}}+\frac{\a}{\b}(a_{6}+\b^2 c_{2})\sqrt{1-(|\vp|-1)^2}-\frac{\a\b \omega c_{7} }{2}\frac{(|\vp|-1)}{\sqrt{1-\p(|\vp|-1)}}\sqrt{1-\frac{\z^2}{\b^2}}, \qquad \mbox{}
\end{eqnarray}
and also,
\begin{eqnarray}
\nonumber
0 \!\!\! &=& \!\!\! r_{1}\hat g_{\vp}(\vp)+r_{2}\hat g_{\vp}(\p,\vp)+r_{6}\hat g_{\vp}(\vp,\z) + r_{7}\hat g_{\vp}(\p,\vp,\z) - q_{4}\act{g}_{\z}(\z)-q_{5}\act{g}_{\z}(\p,\z)-q_{6}\act{g}_{\z}(\vp,\z)\\
\nonumber
&&\!\!\! -q_{7}\act{g}_{\z}(\p,\vp,\z) - \a\epsilon \left(c_{6}+\frac{2}{\b^2}b_{4}\right)\z -\frac{\a\epsilon b_7}{\b}\sqrt{1-\p (|\vp|-1)}\\
\nonu
&&\!\!\!  +\a\b\epsilon c_{2}\frac{\p(|\vp|-1)}{\sqrt{1-(|\vp|-1)^2}} -\frac{\a \epsilon}{\b} (b_{6}+\b^2 c_{1})\sqrt{1-(|\vp|-1)^2}+\a\b\epsilon c_{1}\frac{(|\vp|-1)^2}{\sqrt{1-(|\vp|-1)^2}}\\
\label{fseqcontinuidadegeral3parapvpz}
&& \!\!\!  -\frac{\a \epsilon b_{5}}{\b}\sqrt{1-\p^2}+\frac{\a\b\omega\epsilon c_{7} }{2}\frac{\p}{\sqrt{1-\p(|\vp|-1)}}\sqrt{1-\frac{\z^2}{\b^2}}.
\end{eqnarray}
%
%
%
%
%
In order to solve the system of  eqs.(\ref{fseqcontinuidadegeral1parapvpz})-(\ref{fseqcontinuidadegeral3parapvpz}), we will choose $p_{2}=p_{5}=p_{6}=p_{7}=q_{2}=q_{5}=q_{6}=q_{7}=r_{1}=r_{3}=r_{5}=r_{6}=r_{7}=0$, and $a_{6}=a_{7}=b_{7}=c_{2}=c_{7}=0$. Doing that, we 	get
\begin{subequations}
\br
\label{fsgzderivadaz}
p_{1} g_\vp(\vp) &=& 2\a\left(a_{1}-\frac{b_{2}}{2}\right)\vp-2\a\epsilon a_{1},\\
\label{fsgvpderivadavp}
p_{4} g_\z(\z)&=&-\a\left(\frac{2a_{4}}{\b^2}+c_{5}\right)\z,
\er
and
\br
\label{fshpderivadap}
q_{3} \act{g}_\p(\p) &=& \a\epsilon (2 b_{3}-a_{2})\p,\\
\label{fshvpderivadavp}
q_{4} \act{g}_\z(\z) &=&-\a\epsilon \left(c_{6}+\frac{2}{\b^2}b_{4}\right)\z,
\er
and also
\br
\label{fsgtiopzderivadap}
r_{2} \hat g_\p(\p,\vp)&=& \frac{\a}{\b} (a_{5}+\b^2 c_{3})\sqrt{1-\p^2}-\a\b c_{3}\frac{\p^2}{\sqrt{1-\p^2}}+\a\b c_{2}\sqrt{1-(|\vp|-1)^2},\\
\nonu
r_{2} \hat g_\vp(\p,\vp)&=& -\a\b\epsilon c_{2}\frac{\p (|\vp|-1)}{\sqrt{1-(|\vp|-1)^2}}+\frac{\a}{\b}\epsilon(b_{6}+\b^2 c_{1})\sqrt{1-(|\vp|-1)^2}\\
\label{fsgtiopzderivadaz}
&& - \a\b\epsilon c_{1}\frac{(|\vp|-1)^2}{\sqrt{1-(|\vp|-1)^2}}.
\er
\end{subequations}
Then, by  integrating these expressions respectively, we will find
\begin{subequations}
\br
\label{fsgz}
p_{1} g(\vp) &=& \a\left(a_{1}-\frac{b_{2}}{2}\right)\vp^2-2\a\epsilon a_{1}\vp,\\
\label{fsgvp}
p_{4} g(\z) &=&-\frac{\a}{2}\left(\frac{2a_{4}}{\b^2}+c_{5}\right)\z^2,\\
\label{fshp}
q_{3}\act{g}(\p) &=& \frac{\a}{2}\epsilon (2b_{3}-a_{2})\p^2,\\
\label{fshvp}
q_{4}\act{g}(\z) &=&-\frac{\a}{2}\epsilon\left(c_{6}+\frac{2}{\b^2}b_{4}\right)\z^2+C,\\
\label{fsgtiopz}
r_{2} \hat g(\p,\vp) &=& \frac{\a}{2\b}\left((a_{5}+2\b^2 c_{3})\p\sqrt{1-\p^2}+ a_{5}\arcsin(\p)\right)+ \a\b c_{2}\p\sqrt{1-(|\vp|-1)^2},\qquad \mbox{}\\
\nonu
r_{2}\hat g(\p,\vp) &=& \frac{\a}{2\b}\left((b_{6}+2\b^2 c_{1})(|\vp|-1)\sqrt{1-(|\vp|-1)^2}+ b_{6}\arcsin(|\vp|-1)\right)\\
\label{fsgtiopz2}
&& +\a\b c_{2}\p\sqrt{1-(|\vp|-1)^2},
\er
\end{subequations}
where the parameters satisfy the following constraints,
\begin{subequations}
\br
a_{1}&=&\frac{p_{1}}{p_{4}}\left(a_{4}+\frac{\b^2}{2}c_{5}\right),\qquad a_{5}=b_{6}, \qquad b_{2}=0,\qquad c_{1}=c_{3},\\
\left(c_{6}+\frac{2}{\b^2}b_{4}\right) &=&\frac{q_{4}}{q_{3}}\left(\frac{2 b_{3}-a_{2}}{\b^2}\right), \qquad \,\,\, \,C=\frac{\a}{2}\frac{q_{4}}{q_{3}}\epsilon(2b_{3}-a_{2}).
\er
\end{subequations}
Now, the deformation functions allow us to write the $g$-functions as follows,  
\begin{subequations}
\br
\label{fsgp}
p_{3} g(\p) & =&-\frac{\a p_{3}}{p_{4}}\left(a_{4}+\frac{\b^2}{2} c_{5}\right)(1-\p^2),\\
\label{fshz}
q_{1} \act{g}(\vp) &=& \frac{\a \epsilon}{2} (2b_{3}-a_{2})(|\vp|-1)^2,\\
\label{fsgtiovp}
r_{4} \hat g(\z) &=& -\a\epsilon\left[\left(c_{2}+c_{3}+\frac{a_{5}}{2\b^2}\right)\z\sqrt{1-\frac{\z^2}{\b^2}}+\frac{a_{5}}{2\b}\arcsin\left(\sqrt{1-\frac{\z^2}{\b^2}}\right)\right].
\er
\end{subequations}
By using the above results and eqs. (\ref{fsfuncaopparapvpz})-(\ref{fsmodeloextendidoCparapvpz}), we have
\begin{eqnarray}
\nonumber
W_\p^{(3)}(\p,\vp,\z)&=&-\frac{\a c_{5}}{2}\z^2+\frac{\a}{\b}a_{5}\z\sqrt{1-\p^2} +\a\left(1-a_{2}+\frac{\b^2}{2} c_{5}\right)(1-\p^2)+ \a a_{1}\left(1-(|\vp|-1)^2\right)\\
&&+\a a_{2}\left(1-\p(|\vp|-1)\right)+\frac{\a p_{1}}{p_{4}}\left(a_{4}+\frac{\b^2}{2}c_{5}\right)\left[(1-\p^2)+\vp^2-2|\vp|\right],\\
\nonu
W_\vp^{(3)}(\p,\vp,\z)&=&\frac{\a \epsilon b_{4}}{\b^2}\z^2+\frac{\a \epsilon\b^2}{2} \frac{q_{4}}{q_{3}}\left(\frac{2b_{3}-a_{2}}{\b^2}\right)\left(1-\frac{\z^2}{\b^2}\right)+\a\epsilon b_{3}(1-\p^2) +\a b_{1}\vp(2-|\vp|)\\
&&+\frac{\a \epsilon a_{5}}{\b}\z\sqrt{1-(|\vp|-1)^2}+\frac{\a \epsilon}{2} (2b_{3}-a_{2})(\p^2+(|\vp|-1)^2),\\
\nonu
W_\z^{(3)}(\p,\vp,\z) &=&-\a\epsilon\left(c_{2}+c_{3}+c_{4}+\frac{a_{5}}{2\b^2}\right)\z \sqrt{1-\frac{\z^2}{\b^2}}-\frac{\a \epsilon a_{5}}{2\b}\arcsin\left(\sqrt{1-\frac{\z^2}{\b^2}}\right)-\a c_{5}\p\z\\
\nonu
&&+\frac{\a a_{5}}{2\b} \left(\p\sqrt{1-\p^2}+\arcsin(\p)\right)-\a\b c_{3} (|\vp|-1)\sqrt{1-(|\vp|-1)^2}\\
&&-\a c_{6}\z(|\vp|-1).
\end{eqnarray}
%
%
By integrating and comparing, we see that we should also  have $a_{1}=a_{5}=b_{6}=c_{1}=c_{3}=0$, and $p_{1}=0$. In addition, this will require that $q_{4}=-2q_{3}$. Finally, we can write the superpotential as follows
\br
\nonumber
W^{(3)}(\p,\vp,\z)&=&\frac{\a\b^2}{3}\epsilon\left(1-c_{5}-c_{6}\right)\left(1-\frac{\z^2}{\b^2}\right)^{\frac{3}{2}}-\frac{\a c_{5}}{2}\p\z^2+\a\left(1-a_{2}+\frac{\b^2}{2}c_{5}\right)\p\left(1-\frac{\p^2}{3}\right)\\
\nonumber
&&+\a a_{2}\p\left[1-\frac{\p}{2}\left(|\vp|-1\right)\right]+\a\left(1-b_{1}+\frac{\b^2}{2}c_{6}\right)|\vp|+\a b_{1}\vp^2\left(1-\frac{|\vp|}{3}\right)\\
&&-\frac{\a}{6}\Big(b_{4}+\frac{\b^2}{2}c_{6}\Big)(|\vp|-1)^3-\frac{\a c_{6}}{2}\z^2(|\vp|-1).
\er
Here, in order to avoid rational exponents that will bring some additional difficulties in analysing the linear stability of the model, we will also choose $c_{5}=1-c_{6}$.  Putting all these results back, we will finally get
\begin{eqnarray}
\nonumber
W_\p^{(3)}(\p,\vp,\z)&=&\a (1-\p^2) + a_{2} \p\( 1+\p-|\vp|\)- \frac{1}{2} (1-c_{6})\left(\z^2-\b^2(1-\p^2)\right), \\
\nonu
W_\vp^{(3)}(\p,\vp,\z)&=& \frac{\a \epsilon}{2}\Big( (2+c_{6}\b^2) \(1- (|\vp|-1)^2\)-a_{2} \(\p^2- (|\vp|-1)^2\) -c_{6} \z^2\Big),\\
\nonu
W_\z^{(3)}(\p,\vp,\z)&=& -\a\z \left( \p - c_{6} \left(1+\p - |\vp|\)\right).
\end{eqnarray}
After integrating, we finally obtain the following three-field superpotential
\br
\nonumber
W^{(3)}(\p,\vp,\z)&=& \a\left(1-a_{2}+\frac{\b^2}{2}(1-c_{6})\right)\p\left(1-\frac{\p^2}{3}\right) -\frac{\a}{2}(1-c_{6})\p\vp^2 -\frac{\a}{2}c_{6}\vp^2(|\z|-1)\\
\nonumber
&& +\a a_{2}\p\left(1-\frac{\p}{2}\left(|\z|-1\right)\right)+\frac{\a a_{2}}{2}|\z| +\frac{\a}{2}\left(2-a_{2}+\b^2 c_{6}\)\z^2\left(1- \frac{|\z|}{3}\)\\
&& +\frac{\a}{12}\left(2b_{4}+{c_{6}\b^2}\).
\er

\newpage
\subsection{The extended ($\p^4$+sG+E) model}
Here, we present the explicit derivation of the three-field superpotential for the coupling of  $\p^4$, sine-Gordon and the E-model. 
From eq. (\ref{fscondicaodecontinuidadeparapvpz}), and by using eqs.(\ref{7potencialvp})--(\ref{7potencialp}), we obtain respectively
\begin{eqnarray}
\nonumber
0\!\!\! &=& \!\!\! p_{1}g_{\c}(\c)+p_{2}g_{\c}(\p,\c)+p_{6}g_{\c}(\eta,\c) +p_{7}g_{\c}(\p,\eta,\c) -q_{2}\act{g}_{\p}(\p,\c)  -q_{3}\act{g}_{\p}(\p) -q_{5}\act{g}_{\p}(\p,\eta)\\
\nonumber
&&\!\!\! -q_{7}\act{g}_{\p}(\p,\eta,\c) -2\a\b a_{1}\cos(\bc)\sin(\bc)+ \frac{\a}{\b}b_{3}\frac{\p}{\sqrt{1-\p^2}} - \a\b a_{2}\p\cos(\bc)\\
\nonumber
&& \!\!\! -\a\b a_{6}\cos(\bc)\sin\Big(\frac{1}{2} \ln (1+\eta)^2\Big)  - \a\b a_{7}\p\cos\Big(\frac{\bc}{2}\Big)\cos\Big(\frac{1}{4} \ln (1+\eta)^2\Big)\nonu\\
&&\!\!\! +\frac{2\a b_2}{\b}\left[2\p\cos(\bc)+\frac{\p^2}{\sqrt{1-\p^2}}\sin(\bc)-\sqrt{1-\p^2}\sin(\bc)\right]\nonu\\
&&\!\!\! +\frac{2\a b_{5}}{\b^2}\left[2\p\cos\Big(\frac{1}{2} \ln (1+\eta)^2\Big)+\frac{\p^2}{\sqrt{1-\p^2}}\sin\Big(\frac{1}{2} \ln (1+\eta)^2\Big)-\sqrt{1-\p^2}\sin\Big(\frac{1}{2} \ln (1+\eta)^2\Big)\right]\nonu\\
&& \!\!\! + \frac{2\a b_{7}}{\b}\left[\sin\Big(\frac{\bc}{2}\Big)\cos\Big(\frac{1}{4} \ln (1+\eta)^2\Big)+\frac{\p}{\sqrt{1-\p^2}}\sin\Big(\frac{\bc}{2}\Big)\sin\Big(\frac{1}{4} \ln (1+\eta)^2\Big)\right],
\label{condWx1}
\end{eqnarray}
and 
\begin{eqnarray}
\nonumber
0 \!\!\!&=&\!\!\!   p_{4}g_{\eta}(\eta)+ p_{5}g_{\eta}(\p,\eta)+p_{6}g_{\eta}(\eta,\c)+p_{7}g_{\eta}(\p,\eta,\c)-r_{2}\hat{g}_{\p}(\p,\c) -r_{3}\hat{g}_{\p}(\p)-r_{5}\hat{g}_{\p}(\p,\eta)\\
\nonumber
&& \!\!\!   -r_{7}\hat{g}_{\p}(\p,\eta,\c) - 2\a a_{4}\frac{1}{(1+\eta)}\cos\Big(\frac{1}{2} \ln (1+\eta)^2\Big)\sin\Big(\frac{1}{2} \ln (1+\eta)^2\Big) +\a c_{5}\frac{\p}{\sqrt{1-\p^2}}(1+\eta)\\
\nonumber
&&\!\!\! - \a c_{3}\left(1-\frac{\p}{\sqrt{1-\p^2}}\right)e^{\arcsin{(\p)}}-\a a_{6}\frac{1}{(1+\eta)}\sin(\bc)\cos\Big(\frac{1}{2} \ln (1+\eta)^2\Big)+\a c_{2}\frac{\p}{\sqrt{1-\p^2}}e^{\bc}\\
&&\!\!\! - \a a_{5}\frac{\p}{(1+\eta)}\cos\Big(\frac{1}{2} \ln (1+\eta)^2\Big)+\a a_{7}\frac{\p}{(1+\eta)}\sin\Big(\frac{\bc}{2}\Big)\sin\Big(\frac{1}{4} \ln (1+\eta)^2\Big)\nonu\\
&&\!\!\! + 2\a c_{7}(1+\eta)\left[\sin\Big(\frac{\bc}{2}\Big)\cos\Big(\frac{1}{4} \ln (1+\eta)^2\Big)+\frac{\p}{\sqrt{1-\p^2}}\sin\Big(\frac{\bc}{2}\Big)\sin\Big(\frac{1}{4} \ln (1+\eta)^2\Big)\right],
\label{condWx2}
\end{eqnarray}
and also,
\begin{eqnarray}
\nonumber
0\!\!\!&=& \!\!\! q_{4}\act{g}_\eta (\eta)+q_{5}\act{g}_\eta (\p,\eta)+q_{6}\act{g}_{\eta}(\eta,\c)+ q_{7}\act g_{\eta}(\p,\eta,\c)    -r_{1}\hat{g}_{\c}(\c)-r_{2}\hat{g}_\c (\p,\c)-r_{6}\hat{g}_{\c}(\eta,\c)\\
\nonumber
&& \!\!\! -r_{7}\hat{g}_{\c}(\p,\eta,\c) - \frac{\a b_{4}}{\b}\frac{1}{(1+\eta)}\sin\Big(\frac{1}{2} \ln (1+\eta)^2\Big) -\a\b c_{1}e^{\bc}\left(\cos(\bc)-\sin(\bc)\right)\\
\nonumber
&&\!\!\! -\frac{\a b_{6}}{\b}\frac{1}{(1+\eta)}\cos\Big(\frac{\bc}{2}\Big)\sin\Big(\frac{1}{4} \ln (1+\eta)^2\Big)+\a\b c_{6}(1+\eta)\sin(\bc)-\a\b c_{2}\sqrt{1-\p^2}e^{\bc}\\
&&\!\!\!   + \frac{\a b_{5}}{\b}\left[\frac{(2\p^2-1)}{(1+\eta)}\sin\Big(\frac{1}{2} \ln (1+\eta)^2\Big)+\frac{2\p\sqrt{1-\p^2}}{(1+\eta)}\cos\Big(\frac{1}{2} \ln (1+\eta)^2\Big)\right]\nonu\\
&& \!\!\!  + \frac{\a b_{7}}{\b}\left[\frac{\p}{(1+\eta)}\sin\Big(\frac{\bc}{2}\Big)\sin\Big(\frac{1}{4} \ln (1+\eta)^2\Big)+\frac{\sqrt{1-\p^2}}{(1+\eta)}\sin\Big(\frac{\bc}{2}\Big)\cos\Big(\frac{1}{4} \ln (1+\eta)^2\Big)\right]\nonu\\
&&\!\!\! - \a\b c_{7}(1+\eta)\left[\sqrt{1-\p^2}\cos\Big(\frac{\bc}{2}\Big)\sin\Big(\frac{1}{4} \ln (1+\eta)^2\Big)-\p\cos\Big(\frac{\bc}{2}\Big)\cos\Big(\frac{1}{4} \ln (1+\eta)^2\Big)\right]\!.\qquad\,\,\, \mbox{}
\label{condWx3}
\end{eqnarray}
By choosing $p_{1}=p_{4}=p_{5}=p_{6}=p_{7}=q_{2}=q_{3}=q_{4}=q_{5}=q_{7}=r_{1}=r_{2}=r_{3}=r_{6}=r_{7}=0$ and $a_{6}=a_{7}=b_{5}=b_{7}=c_{2}=c_{7}=0$, we get 
\begin{subequations}
\br
p_{2}g_\c (\p,\c) \!&\!=\!&\! \a\b a_{1}\sin(2\bc)-\frac{2\a b_{2}}{\b}\left[2\p\cos(\bc) +\frac{\p^2}{\sqrt{1-\p^2}}\sin(\bc)-\sqrt{1-\p^2}\sin(\bc)\right]\nonu\\
\label{gc}
&&+\a\b a_{2}\p\cos(\bc)-\frac{\a b_{3}}{\b}\frac{\p}{\sqrt{1-\p^2}},\\[0.1cm]
q_{6}\act{g}_\eta(\eta,\c) \!&\!=\!&\! \frac{\a b_{4}}{\b}\frac{1}{(1+\eta)}\sin\Big(\frac{1}{2} \ln (1+\eta)^2\Big)+\a\b c_{1}e^{\bc}\left(\cos(\bc)-\sin(\bc)\right)-\a\b c_{6}(1+\eta)\sin(\bc)\nonu\\
\label{tgp}
&&+\frac{\a b_{6}}{\b}\frac{1}{(1+\eta)}\cos\Big(\frac{\bc}{2}\Big)\sin\Big(\frac{1}{4} \ln (1+\eta)^2\Big),\\[0.2cm]
r_{5}\hat{g}_\p(\p,\eta) \!&\!=\!&\! -2\a a_{4}\frac{1}{(1+\eta)}\cos\Big(\frac{1}{2} \ln (1+\eta)^2\Big)\sin\Big(\frac{1}{2} \ln (1+\eta)^2\Big)-\a a_{5}\frac{\p}{(1+\eta)}\cos\Big(\frac{1}{2} \ln (1+\eta)^2\Big)\nonu\\
&&-\a c_{3}\left(1-\frac{\p}{\sqrt{1-\p^2}}\right)e^{\arcsin(\p)}+\a c_{5}\frac{\p}{\sqrt{1-\p^2}}(1+\eta).
\er
\end{subequations}
Now, by performing the integrations we find 
\begin{subequations}
\br
g(\p,\c) \!&\!=\!&\! -\frac{\a}{2}\frac{a_{1}}{p_{2}}\cos(2\bc)-\frac{\a}{\b}\frac{b_{3}}{p_{2}}\frac{\p\c}{\sqrt{1-\p^2}}+\a\frac{a_{2}}{p_{2}}\p\sin(\bc)\nonu\\
\label{c12g}
&& -\frac{2\a}{\b^2}\frac{b_{2}}{p_{2}}\left[2\p\sin(\bc) +\sqrt{1-\p^2}\cos(\bc)-\frac{\p^2}{\sqrt{1-\p^2}}\cos(\bc)\right],\\
\act{g}(\eta,\c) \!&\!=\!&\! -\frac{\a}{\b}\frac{b_{4}}{q_{6}}\cos\Big(\frac{1}{2} \ln (1+\eta)^2\Big)+\a\b\frac{c_{1}}{q_{6}}\eta e^{\bc}\big(\cos(\bc)-\sin(\bc)\big)\nonu\\[0.1cm]
\label{c26tg}
&&-\frac{2\a}{\b}\frac{b_{6}}{q_{6}}\cos\Big(\frac{\bc}{2}\Big)\cos\Big(\frac{1}{4} \ln (1+\eta)^2\Big)-\a\b\frac{c_{6}}{q_{6}}\left(\eta+\frac{\eta^2}{2}\right)\sin(\bc),\\[0.1cm]
\hat{g}(\p,\eta) \!&\!=\!&\! -2\a\frac{a_{4}}{r_{5}}\frac{\p}{(1+\eta)}\cos\Big(\frac{1}{2} \ln (1+\eta)^2\Big)\sin\Big(\frac{1}{2}\ln (1+\eta)^2\Big)-\a\frac{c_{3}}{r_{5}}\sqrt{1-\p^2}e^{\arcsin(\p)}\nonu\\
\label{c35h}
&&-\frac{\a}{2}\frac{a_{5}}{r_{5}}\frac{\p^2}{(1+\eta)}\cos\Big(\frac{1}{2} \ln (1+\eta)^2\Big)-\a\frac{c_{5}}{r_{5}}\sqrt{1-\p^2}(1+\eta).
\er
\end{subequations}
In addition, we can use the deformation functions, as well as their inverse functions, to write  
\begin{subequations}
\br
\label{c13g}
g(\p) \!&\!=\!&\! -\frac{\a}{2}\frac{a_{1}}{p_{2}}\cos\big(2\arcsin(\p)\big)-\frac{\a}{\b^2}\frac{b_{3}}{p_{2}}\frac{\p}{\sqrt{1-\p^2}}\arcsin(\p)+\a\frac{a_{2}}{p_{2}}\p^2-\frac{2\a}{\b^2}\frac{b_{2}}{p_{2}},\\[0.1cm]
\act{g}(\c) \!&\!=\!&\! -\frac{\a}{\b}\frac{b_{4}}{q_{6}}\cos(\bc)-\frac{2\a}{\b}\frac{b_{6}}{q_{6}}\cos^2\Big(\frac{\bc}{2}\Big)-\frac{\a\b}{2}\frac{c_{6}}{q_{6}}\big(e^{2\bc}-1\big)\sin(\bc)\nonu\\
\label{c21tg}
&&+\a\b\frac{c_{1}}{q_{6}}e^{\bc}\big(e^{\bc}-1\big)\big(\cos(\bc)-\sin(\bc)\big),\\[0.1cm]
\hat{g}(\eta) \!&\!=\!&\! -\frac{\a}{2}\frac{(4a_{4}+a_{5})}{r_{5}}\frac{1}{(1+\eta)}\cos\Big(\frac{1}{2} \ln (1+\eta)^2\Big)\sin^2\Big(\frac{1}{2} \ln (1+\eta)^2\Big)\nonu\\
&&-\a\frac{(c_{3}+c_{5})}{r_{5}}(1+\eta)\cos\Big(\frac{1}{2} \ln (1+\eta)^2\Big).
\er
\end{subequations}
Here, in order to avoid possible divergences  in the first-order equations at the minima of the  $\p^4$ field, we chose to set $b_{2} = b_{3}=0$. It is also  worth pointing that the apparent divergence at the value $\eta=-1$ is and inherent issue of the E-model superpotential, and as far as the kink solutions (\ref{eqn6.31}) are concerned, this value will be never reached. Putting together all these results, we finally get
\begin{eqnarray}
\nonumber
W_\p^{(3)}(\p,\eta,\c)&=&\a(1-a_{4}-a_{5})(1-\p^2)+\a a_{4}\cos^2\Big(\frac{1}{2} \ln (1+\eta)^2\Big) \nonu\\
&&+\a a_{5}\left[1-\p\sin\Big(\frac{1}{2} \ln (1+\eta)^2\Big)\right], \\ 
W_\c^{(3)}(\p,\eta,\c) &=& \frac{\a}{\b}\cos(\bc) +\a\b c_{1}\,e^{\bc}(1+\eta-e^{\bc})\big(\cos(\bc)-\sin(\bc)\big),  \nonu\\
&&-\frac{\a\b c_{6}}{2}\sin(\bc)\left((1+\eta)^2 - e^{2\bc}\right), \\
W_\eta^{(3)}(\p,\eta,\c) &=& \a c_{1}\,e^{\bc}\cos(\bc)+\a(1-c_{1}-c_{6})(1+\eta)\cos\Big(\frac{1}{2} \ln (1+\eta)^2\Big)\nonu\\
&&+\a c_{6}(1+\eta)\cos(\bc)+\frac{\a a_{4}}{(1+\eta)}\sin\big( \ln (1+\eta)^2\big)\left[\sin\Big(\frac{1}{2} \ln (1+\eta)^2\Big)-\p\right]\nonu\\
&&+\frac{\a a_{5}}{2(1+\eta)}\cos\Big(\frac{1}{2} \ln (1+\eta)^2\Big)\left[\sin^2\Big(\frac{1}{2} \ln (1+\eta)^2\Big)-\p^2\right],
\end{eqnarray}
which after being integrated lead us to the three-field superpotential 
\br
\nonumber
W^{(3)}(\p,\eta,\c) &=& \a(1-a_{4}-a_{5})\left(\p-\frac{\p^3}{3}\right) +\frac{\a}{\b^2}\sin(\bc) \\
\nonumber
&&+\frac{\a}{5}(1-c_{1}-c_{6})(1+\eta)^2\left[2\cos\Big(\frac{1}{2} \ln (1+\eta)^2\Big)+\sin\Big(\frac{1}{2} \ln (1+\eta)^2\Big)\right]\nonu\\
&& + \a a_{4}\left[\p\cos^2\Big(\frac{1}{2} \ln (1+\eta)^2\Big)+\frac{2}{3}\sin^3\Big(\frac{1}{2} \ln (1+\eta)^2\Big)\right] \nonu\\
&&+\a a_{5}\left[\p-\frac{\p^2}{2}\sin\Big(\frac{1}{2} \ln (1+\eta)^2\Big)+\frac{1}{6}\sin^3\Big(\frac{1}{2} \ln (1+\eta)^2\Big)\right]\nonu\\
&&+\frac{\a c_{1}}{5}\,e^{\bc}\left[\big(5(1+\eta)-3e^{\bc}\big)\cos(\bc)+e^{\bc}\sin(\bc)\right]\nonu\\
&&+\frac{\a c_{6}}{10}\left[\big(5(1+\eta)^2-e^{2\bc}\big)\cos(\bc)+2e^{2\bc}\sin(\bc)\right].
\er


\subsection{The extended ($\phi^4$+sG$_1$+sG$_2$) model}

Finally, let us present the explicit derivation of the three-field superpotential for the coupling of  the $\p^4$ model with two different sine-Gordon models.
From eq. (\ref{fscondicaodecontinuidadeparapvpz}), and by using eqs. eqs. (\ref{eqn6.63})--(\ref{eqn6.65}), we obtain respectively
\begin{eqnarray}
\nonumber
0 \!\!\! &=& \!\!\! p_{1}g_{\c}(\c)+p_{2}g_{\c}(\p,\c)+p_{6}g_{\c}(\c,\psi)+p_{7}g_{\c}(\p,\c,\psi) -q_{2}\act{g}_{\p}(\p,\c)-q_{3}\act{g}_{\p}(\p)-q_{5}\act{g}_{\p}(\p,\psi)\\
\nonumber
&& \!\!\!  -q_{7}\act{g}_{\p}(\p,\c,\psi) -2\a\b a_{1}\cos(\bc)\sin(\bc) -\a\b\left(a_{2}-\frac{2b_{7}}{\b^2}\right)\p\cos(\bc)  +\frac{\a b_{3}}{\b}\frac{\p}{\sqrt{1-\p^2}}\\
\nonumber
&& \!\!\! + \frac{\a b_{2}}{2\b} \frac{\sin(\bc)}{\sqrt{1-\p\sin(\bc)}} +\frac{\a b_{5}}{2\b}\frac{\sin(\gamma\psi)}{\sqrt{1-\p\sin(\gamma\psi)}} -\a\b a_{6}\cos(\bc)\sin(\gamma\psi) - \frac{2\a b_{7}}{\b}\sin(\bc)\cos(\gamma\psi)\\
\nonu
&& \!\!\! \\
\label{condWx1sG}
&& \!\!\! +\a\b a_{7}\p\left(2\p\sin(\bc)\cos(\gamma\psi)+\cos(\bc)\cos^2(\gamma\psi)-2\sin(\gamma\psi)\sin(\bc)\cos(\bc)\right),
\end{eqnarray}
and,
\begin{eqnarray}
\nonumber
0 \!\!\! &=& \!\!\! p_{4}g_{\psi}(\psi)+p_{5}g_{\psi}(\p,\psi) +p_{6}g_{\psi}(\c,\psi)+ p_{7}g_{\psi}(\p,\c,\psi)-r_{2}\hat{g}_{\p}(\p,\c)-r_{3}\hat{g}_{\p}(\p)-r_{5}\hat{g}_{\p}(\p,\psi)\\
\nonumber
&&\!\!\! -r_{7}\hat{g}_{\p}(\p,\c,\psi) +\frac{\a c_{2}}{2\gamma}\frac{\sin(\bc)}{\sqrt{1-\p\sin(\bc)}}+\frac{2\a c_{7}}{\gamma}\p\cos(\bc) -\a\gamma\left(a_{6}+\frac{2c_{7}}{\gamma^2}\right)\sin(\bc)\cos(\gamma\psi)\\
\nonumber
&&\!\!\!  + \frac{\a c_{3}}{\gamma}\frac{\p}{\sqrt{1-\p^2}}-2\a\gamma a_{4}\sin(\gamma\psi)\cos(\gamma\psi)+\frac{\a c_{5}}{2\gamma}\frac{\sin(\gamma\psi)}{\sqrt{1-\p\sin(\gamma\psi)}} -\a\gamma a_{5}\p\cos(\gamma\psi)\\
\label{condWx2sG}
&&\!\!\!  +\a\gamma a_{7}\p\left(2\p\cos(\bc)\sin(\gamma\psi)-2\sin(\bc)\sin(\gamma\psi)\cos(\gamma\psi)-\cos(\gamma\psi)\sin^2(\bc)\right),
\end{eqnarray}
and also,
\begin{eqnarray}
\nonumber
0 \!\!\! &=&\!\!\! q_{4}\act{g}_{\psi}(\psi)+q_{5}\act{g}_{\psi}(\p,\psi)+q_{6}\act{g}_\psi(\c,\psi)+q_{7}\act{g}_{\psi}(\p,\c,\psi) - r_{1}\hat{g}_\c(\c)-r_{2}\hat{g}_{\c}(\p,\c)-r_{6}\hat{g}_\c(\c,\psi)\\
\nonumber
&&\!\!\! -r_{7}\hat{g}_{\c}(\p,\c,\psi) +\frac{\a\b c_{1}}{\gamma}\sin(\bc) + \frac{\a\b c_{2}}{2\gamma}\frac{\p\cos(\bc)}{\sqrt{1-\p\sin(\bc)}} -\frac{\a\b c_{7}}{\gamma}\p^2\sin(\bc) - \frac{\a\gamma b_{4}}{\b}\sin(\gamma\psi)\\
\nonumber
&&\!\!\! - \frac{\a\gamma b_{5}}{2\b}\frac{\p\cos(\gamma\psi)}{\sqrt{1-\p\sin(\gamma\psi)}} +\frac{\a\b c_{6}}{2\gamma}\frac{\cos(\bc)\sin(\gamma\psi)}{\sqrt{1-\sin(\bc)\sin(\gamma\psi)}} - \frac{\a\gamma b_{6}}{2\b}\frac{\sin(\bc)\cos(\gamma\psi)}{\sqrt{1-\sin(\bc)\sin(\gamma\psi)}}\\
\nonu
&&\!\!\!\\
\nonu
&&\!\!\! + \frac{\a\b c_{7}}{\gamma}\sin(\bc)\cos^2(\gamma\psi) - \frac{2\a\gamma b_{7}}{\b}\sin(\gamma\psi)\cos(\gamma\psi)\cos(\bc) - \frac{2\a\b c_{7} }{\gamma}\p\cos(\bc)\cos(\gamma\psi) \\
\label{condWx3sG}
&&\!\!\! - \frac{2\a\gamma b_{7}}{\b}\p\sin(\gamma\psi)\sin(\bc).
\end{eqnarray}
Now, by choosing $p_{1}=p_{2}=p_{4}=p_{5}=p_{7}=q_{3}=q_{4}=q_{5}=q_{6}=q_{7}=r_{1}=r_{2}=r_{3}=r_{5}=r_{6}=0$, and $a_{7}=b_{2}=b_{3}=b_{5}=b_{6}=c_{2}=c_{3}=c_{5}=c_{6}=0$, we obtain
\begin{subequations}
\br
\label{gDccpsi1}
p_{6}g_\c (\c,\psi) &=& \a\b a_{6}\cos(\bc)\sin(\gamma \psi)+\frac{2\a b_{7}}{\b}\sin(\bc)\cos(\gamma\psi),\\
\label{gDccpsi2}
p_{6}g_\c (\c,\psi) &=& \a\gamma \left(a_{6}+\frac{2c_{7}}{\gamma^2}\right)\cos(\gamma\psi)\sin(\bc),\\
\label{tgDffc}
q_{2}\act{g}_\p (\p,\c) &=& -2\a\b a_{1}\sin(\bc)\cos(\bc)-\a\b\left(a_{2}-\frac{2b_{7}}{\b^2}\right)\p\cos(\bc),\\
\label{hDffcpsi}
r_{7}\hat{g}_\p(\p,\c,\psi) &=& \frac{2\a c_{7}}{\gamma}\p\cos(\bc)-2\a\gamma a_{4}\sin(\gamma\psi)\cos(\gamma\psi)-\a\gamma a_{5}\p\cos(\gamma\psi),\\
\nonu
r_{7}\hat{g}_\c(\p,\c,\psi) &=& \frac{\a\b c_{1}}{\gamma}\sin(\bc)-\frac{\a\b c_{7}}{\gamma}\p^2\sin(\bc)+\frac{\a\b c_{7}}{\gamma}\sin(\bc)\cos^2(\gamma\psi)-\frac{\a\gamma b_{4}}{\b}\sin(\gamma\psi)\\
\nonu
&& -\frac{2\a\gamma b_{7}}{\b}\sin(\gamma\psi)\cos(\gamma\psi)\cos(\bc)-\frac{2\a\b c_{7}}{\gamma}\p\cos(\bc)\cos(\gamma\psi)\\
\label{hDcfcpsi}
&& -\frac{2\a\gamma b_{7}}{\b}\p\sin(\gamma\psi)\sin(\bc).
\er
\end{subequations}
Then, by performing the integrations we find 
\begin{subequations}
\br
\label{gcpsi}
p_{6}g(\c,\psi) &=& \a a_{6}\sin(\bc)\sin(\gamma \psi),\\
\label{tgfc}
q_{2}\act{g} (\p,\c) &=& -2\a\b a_{1}\p\sin(\bc)\cos(\bc)-\frac{\a\b}{2}a_{2}\p^2\cos(\bc),\\
r_{7}\hat{g}(\p,\c,\psi) &=& -\frac{\a c_{1}}{\gamma}\cos(\bc)-2\a\gamma a_{4}\p\sin(\gamma\psi)\cos(\gamma\psi) -\frac{\a\gamma a_{5}}{2}\p^2\cos(\gamma\psi) \nonu\\
&&-\frac{\a\gamma b_{4}}{\b}\c\sin(\gamma\psi) \label{hfcpsi},
\er
\end{subequations}
where it is necessary to have $b_{7}=c_{7}=0$ for consistency. In addition, by using the deformation functions and their inverse functions, we get   
\begin{subequations}
\br
\label{gp}
p_{3}g(\p)&=&-\a a_{6}\p^2,\\
\label{gc}
q_{1}\act{g}(\c)&=&\a\b\left(2a_{1}+\frac{a_{2}}{2}\right)\sin^2(\bc)\cos(\bc),\\
r_{4}\hat{g}(\psi)&=&\frac{\a c_{1}}{\gamma}\cos(\gamma\psi)+\a\gamma\left(2a_{4}+\frac{a_{5}}{2}\right)\sin^2(\gamma\psi)\cos(\gamma\psi)+\frac{\a\gamma^2 b_{4}}{\b^2}\psi\sin(\gamma\psi).\label{hpsi}
\er
\end{subequations}
Putting together all these results, we obtain 
\br 
 W_\p^{(3)} (\p,\c,\psi) &=&  \a(1-\phi^2) +\a a_1 \left(\phi^2-\sin^2(\bc) \) +\a a_2 \p\left(\p-\sin(\bc)\) +\a a_4\left(\p^2-\sin^2(\gp)\) \nonu\\ 
 && +\a a_5 \p\left(\p - \sin(\gp)\)                  ,\\[0.1cm]
  W_\chi^{(3)} (\p,\c,\psi) &=& \frac{\a}{\b}(1-b_4) \cos(\bc) +\a\b \left(2a_1 +\frac{a_2}{2}\right)\sin^2(\bc) \cos(\bc)    + \frac{\a b_4}{\b} \cos(\gp)\nonu \\
  && -\a\b a_1 \phi \sin(2\bc) -\frac{\a\b a_2}{2} \phi^2\cos(\bc),\\[0.1cm]
W_\psi^{(3)} (\p,\c,\psi) &=&   \frac{\a}{\g}\cos(\gp) +\a\g \left(2a_4+\frac{a_5}{2}\)\sin^2(\gp)\cos(\gp) - \a\g a_4 \phi \sin(2\gp)\nonu \\
&& -\frac{\a \g a_5}{2} \phi^2 \cos(\gp) + \frac{\a \g b_4}{\b^2 } \left(\g\psi - \b\chi  \)\sin(\gp),
\er
which upon being integrated results in the following superpotential
\br
W^{(3)}  (\p,\c,\psi)  &=&  \a\p - \a \left(1-a_1-a_2-a_4-a_5\)\frac{\p^3}{3} - \a a_1 \p \sin^2(\bc) - \frac{\a a_2}{2} \p^2\sin(\bc) \nonu\\
&& -\a a_4 \phi \sin^2(\gp) -\frac{\a a_5}{2} \phi^2 \sin(\gp) +\frac{\a}{\b^2} (1-b_4) \sin(\bc) \nonu \\
&& + \frac{\a}{3} \left(2a_1+\frac{a_2}{2}\)\sin^3(\bc) +\frac{\a b_4}{\b^2}\left(\bc - \gp \) \cos(\gp) + \frac{\a}{\g^2}\left(1+\frac{b_4 \g^2}{\b^2}\) \sin(\gp)  \nonu\\
&& +\frac{\a}{3}\left(2a_4 +\frac{a_5}{2}\)\sin^3(\gp). 
\er


\end{document}